\documentclass[twocolumn]{aastex631}

\shorttitle{COS-Holes Survey}
\shortauthors{Garza et al.}
\DeclareRobustCommand{\ion}[2]{%
\relax\ifmmode
\ifx\testbx\f@series
{\mathbf{#1\,\mathsc{#2}}}\else
{\mathrm{#1\,\mathsc{#2}}}\fi
\else\textup{#1\,{\mdseries\textsc{#2}}}%
\fi}

\usepackage{longtable}
\usepackage{textcomp}
\usepackage{graphicx}	% Including figure files
\usepackage{amssymb,amsmath} % Extra maths symbols, % Advanced maths commands
\usepackage{hyperref}
\usepackage{booktabs}
\usepackage{gensymb}
\usepackage{listings}
\usepackage{lipsum}
\usepackage{float}
\usepackage{tabularx}
\newcolumntype{Y}{>{\centering\arraybackslash}X}
\usepackage{xcolor}
\usepackage{soul}
\begin{document}

\title{The COS-Holes Survey: Connecting Galaxy Black Hole Mass with the State of the CGM}

\correspondingauthor{Samantha L. Garza}
\email{samgarza@uw.edu}

\author[0000-0003-4521-2421]{Samantha L. Garza}
\affiliation{Department of Astronomy, University of Washington, Seattle, WA, 98195}

\author[0000-0002-0355-0134]{Jessica K. Werk}
\affiliation{Department of Astronomy, University of Washington, Seattle, WA, 98195}

\author[0000-0002-3391-2116]{Benjamin D. Oppenheimer}
\affiliation{Center for Astrophysics and Space Astronomy, 389 UCB, Boulder, CO, 80309, USA}

\author[0000-0003-0789-9939]{Kirill Tchernyshyov}
\affiliation{Department of Astronomy, University of Washington, Seattle, WA, 98195}

\author[0000-0001-7589-6188]{N. Nicole Sanchez}
\affiliation{Carnegie Observatories, 813 Santa Barbara Street, Pasadena, California, 91101 USA}
\affiliation{California Institute of Technology, TAPIR 350-17, 1200 E. California Boulevard, Pasadena, California, 91125-0001 USA}

\author[0000-0003-3520-6503]{Yakov Faerman}
\affiliation{Department of Astronomy, University of Washington, Seattle, WA, 98195}

\author[0000-0001-6248-1864]{Kate H. R. Rubin}
\affiliation{Department of Astronomy, San Diego State University, San Diego, CA 92182 USA}
\affiliation{Department of Astronomy $\&$ Astrophysics, University of California, San Diego, La Jolla, CA 92093, USA}

\author[0000-0002-2816-5398]{Misty C. Bentz}
\affiliation{Department of Physics and Astronomy, Georgia State University, Atlanta, GA 30303, USA}

\author[0000-0002-8337-3659]{Jonathan J. Davies}
\affiliation{Department of Physics and Astronomy, University College London, Gower Street, London WC1E 6BT, UK}

\author[0000-0002-1979-2197]{Joseph N. Burchett}
\affiliation{Department of Astronomy, New Mexico State University, Las Cruces, NW, 88003}

\author[0000-0001-6258-0344]{Robert A. Crain}
\affiliation{Astrophysics Research Institute, Liverpool John Moores University, 146 Brownlow Hill, Liverpool, L3 5RF}

\author[0000-0002-7738-6875]{J. Xavier Prochaska}
\affiliation{Department of Astronomy and Astrophysics, University of California, Santa Cruz, CA 95064, USA}
\affiliation{Kavli Institute for the Physics and Mathematics of the Universe (Kavli IPMU), 5-1-5 Kashiwanoha, Kashiwa, 277-8583, Japan}
\affiliation{Division of Science, National Astronomical Observatory of Japan, 2-21-1 Osawa, Mitaka, Tokyo 181-8588, Japan}

%% Note that the \and command from previous versions of AASTeX is now
%% depreciated in this version as it is no longer necessary. AASTeX 
%% automatically takes care of all commas and "and"s between authors names.

%% AASTeX 6.31 has the new \collaboration and \nocollaboration commands to
%% provide the collaboration status of a group of authors. These commands 
%% can be used either before or after the list of corresponding authors. The
%% argument for \collaboration is the collaboration identifier. Authors are
%% encouraged to surround collaboration identifiers with ()s. The 
%% \nocollaboration command takes no argument and exists to indicate that
%% the nearby authors are not part of surrounding collaborations.

%% Mark off the abstract in the ``abstract'' environment. 
\begin{abstract}

We present an analysis of \textit{HST}/COS/G160M observations of \ion{C}{IV} in the inner circumgalactic medium (CGM) of a novel sample of eight z$\sim$0, L$\approx$L$^{\star}$ galaxies, paired with UV-bright QSOs at impact parameters ($R_\mathrm{proj}$) between 25-130 kpc.
The galaxies in this stellar-mass-controlled sample (log$_{10}$M$_{\star}$/M$_{\odot}$ $\sim$ 10.2-10.9 M$_{\odot}$) host super-massive black holes (SMBHs) with dynamically-measured masses spanning log$_{10}$M$_\mathrm{BH}$/M$_{\odot}$ $\sim$ 6.8-8.4; 
this allows us to compare our results with models of galaxy formation where the integrated feedback history from the SMBH alters the CGM over long timescales. We find that the \ion{C}{IV} column density measurements (N$_{\rm C IV}$) (average log$_{10}$N$_{\rm C IV, CH}$ = 13.94$\pm$0.09 cm$^{-2}$) are largely consistent with existing measurements from other surveys of N$_{\rm C IV}$ in the CGM (average log$_{10}$N$_{\rm C IV, Lit}$ = 13.90$\pm$0.08 cm$^{-2}$), but do not show obvious variation as a function of the SMBH mass. In contrast, specific star-formation rate (sSFR) is highly correlated with the ionized content of the CGM. We find a large spread in sSFR for galaxies with log$_{10}$M$_\mathrm{BH}$/M$_{\odot}$ $>$  7.0, where the CGM \ion{C}{IV} content shows clear dependence on galaxy sSFR but not M$_\mathrm{BH}$. Our results do not indicate an obvious causal link between CGM \ion{C}{IV} and the mass of the galaxy's SMBH; however through comparisons to the EAGLE, Romulus25, $\&$ IllustrisTNG simulations, we find that our sample is likely too small to constrain such causality.

%This example manuscript is intended to serve as a tutorial and template forauthors to use when writing their own AAS Journal articles. The manuscriptincludes a history of \aastex\ and includes figure and table examples to illustrate these features. Information on features not explicitly mentioned in the article can be viewed in the manuscript comments or more extensive online documentation. Authors are welcome replace the text, tables, figures, and bibliography with their own and submit the resulting manuscript to the AAS Journals peer review system.  The first lesson in the tutorial is to remind authors that the AAS Journals, the Astrophysical Journal (ApJ), the Astrophysical Journal Letters (ApJL), the Astronomical Journal (AJ), and the Planetary Science Journal (PSJ) all have a 250 word limit for the  abstract\footnote{Abstracts for Research Notes of the American Astronomical Society (RNAAS) are limited to 150 words}.  If you exceed this length the Editorial office will ask you to shorten it. This abstract has 161 words.

\end{abstract}

%% Keywords should appear after the \end{abstract} command. 
%% The AAS Journals now uses Unified Astronomy Thesaurus concepts:
%% https://astrothesaurus.org
%% You will be asked to selected these concepts during the submission process
%% but this old "keyword" functionality is maintained in case authors want
%% to include these concepts in their preprints.

%% From the front matter, we move on to the body of the paper.
%% Sections are demarcated by \section and \subsection, respectively.
%% Observe the use of the LaTeX \label
%% command after the \subsection to give a symbolic KEY to the
%% subsection for cross-referencing in a \ref command.
%% You can use LaTeX's \ref and \label commands to keep track of
%% cross-references to sections, equations, tables, and figures.
%% That way, if you change the order of any elements, LaTeX will
%% automatically renumber them.
%%
%% We recommend that authors also use the natbib \citep
%% and \citet commands to identify citations.  The citations are
%% tied to the reference list via symbolic KEYs. The KEY corresponds
%% to the KEY in the \bibitem in the reference list below. 

\section{Introduction} \label{sec:intro}

For decades, absorption-line experiments using bright background quasars (QSOs) have been recognized as an efficient way of studying diffuse gaseous atmospheres of the Milky Way and other galaxies \citep[e.g.][]{Bahcall_Spitzer_1969, bergeron1986, werk_2013}. With the more recent addition of results from \textit{HST}/COS, astronomers have established that this diffuse outer part of galaxies, called the circumgalactic medium (CGM), is a highly-ionized, massive, spatially extended reservoir of both fuel for future star formation and the byproducts of stellar evolution \citep{Lehner_howk_2011, Peeples_2014, Werk_2014, tumlinson_2017}. The properties of the CGM, particularly the highly ionized CGM traced by \ion{O}{VI}, are linked to the star-forming properties of host galaxies \citep[e.g.][]{Tumlinson11, Tchernyshyov2022}. For this reason, the CGM can serve as an excellent testing ground for astrophysical models of galaxy-scale feedback. In this work, we focus on testing the cumulative effect of feedback from supermassive black holes (SMBHs) on the content of the cool CGM.

It is well-known that the properties of galactic SMBHs correlate with their parent galaxy properties. For example, \cite{kormendy_richstone_1995} found that black hole masses scale linearly with the absolute luminosity of the host bulge (or elliptical galaxy). This result inspired many investigations into other scaling relationships between these galaxy properties and their corresponding central black hole properties, which found an indirect link between galaxy formation and the growth of their SMBHs \citep{Haehnelt_Natarajan_Rees1998, magorrian_1998, saglia_2016}. In particular, both the relation between the mass of the central SMBH and the stellar dispersion of its host galaxy's bulge, M$_\mathrm{BH}$-$\rm \sigma$, and the bulge mass-M$_\mathrm{BH}$ correlation \citep{Silk_rees_1998, Ferrarese_Merrit_2000, Gebhardt_2000, Haring_Rix_2004, reines_2015}, reflect the assertion that the mass of the SMBH is a fundamental property of a galaxy, reflective of its history \citep{K_ho_2013, van_den_Bosch_2016}. To extend this further, we posit that there can be a significant (and observable) alteration of the CGM content of galaxies due to black hole activity over time due to the cumulative effect of the processes associated with black hole growth such as those envisioned in most kinetic-mode feedback scenarios \citep[e.g.,][]{best_2012}. 

Recently, analytical studies have shown that maintaining the large observed column densities of highly ionized gas in the CGM, traced by far ultraviolet (FUV) transitions like \ion{O}{VI}, for longer than a Gyr requires a significant source of energy that cannot be supplied by galactic supernovae and stellar winds alone \citep[][but see \citet{Faerman_2020, Faerman_2022}]{mathews_prochaska_2017}. If these are the only energy sources, \cite{mcquinn_werk_2018} asserts that much of a galaxy's energy budget must be expended in the CGM (rather than the ISM). SMBHs may provide a promising source of far-reaching intermittent feedback shocks that can keep the gas in the CGM warm and highly ionized (T $> 10^{5}$). More specifically, the energy released from building a SMBH not only exceeds the binding energy of the gas in the bulge (by orders of magnitude), but can easily exceed the binding energy of the entire gaseous halo \citep{oppy_2018_solo}. Therefore, even with a low efficiency of the SMBH rest mass energy being imparted to the gaseous halo over its history, the mass and the energetics of the CGM can be significantly affected.

In combination with established black hole scaling relations, these arguments imply that the mass of a SMBH may be a key determinant for the content and kinematics of the CGM around L$^{\star}$ galaxies. There are already established physical links (i.e. scaling relationships) for galaxies on black hole scales (sub-pc) and stellar-disk scales (kpc). If these relationships are combined with the expectation that the extended gaseous halos of galaxies fuel their star formation, it is possible that the evolution of a galaxy's central black hole likewise physically links to the properties of galactic gas on CGM scales (tens to hundreds of kpc). Our present survey, which we call COS-Holes, seeks to examine whether such a correlation exists between the pc-scale physics of black hole growth and the global, kpc-scale gas flows of the CGM that fuels star formation \citep{Oppenheimer_2020, Nelson_2018, sanchez_2019}. 

In the last five years, simulation work has already suggested that feedback from a galaxy's SMBH impacts the content and ionization state of its CGM, but they have differing views on the role the SMBH ultimately plays. 
%this sentence is true for all of them (double ccheck the into for r25 for first time)
Results from studies using the cosmological simulations TNG \citep{Nelson_2018_oxygen}, EAGLE \citep{Davies_2019, Oppenheimer_2020}, and {\sc Romulus} \citep{tremmel_2017} suggest that the SMBH at the center of galaxies enriches the CGM by driving metals out of the disk and into the halo. However in EAGLE and TNG, galaxies that host more massive BHs can provide a significant amount of energy over time which transport baryons beyond the virial radius, ultimately reducing gas accretion, overall star formation, and the total density of the CGM. Using \ion{O}{VI}, demonstrated to be a sensitive probe of SMBH feedback, \citet{sanchez_2019} reports a contrasting view to the role of the SMBH outlined above. Results from {\sc Romulus} do not show evacuation of CGM gas into the IGM, but rather suggest that galaxies with more massive BHs are more likely to have a more-metal enriched (higher ion column density) CGM. Due to these opposing simulation predictions, it is imperative to empirically test the role SMBH feedback plays (if any) in setting the content of the CGM. 

This work examines the observed relationship, if any, between black hole growth over long timescales (parameterized by a dynamically-measured SMBH mass) and the gas content and kinematics within the extended halos of galaxies. In addition, we compare these observations to predictions from cosmological simulations. Our novel sample of stellar- and halo-mass controlled nearby galaxies ($z <$ 0.005; Figure \ref{fig: galaxy_sample}) host a wide range of dynamically resolved SMBHs (log$_{10}$ M$_\mathrm{BH}$ $\sim$ 6.8-8.4) and FUV bright QSOs at impact parameters between 25 $<$ R$_\mathrm{proj}$ $<$ 130 kpc. 
%In this paper, we present a controlled observational test of the role of black hole growth in the regulation of the warm, ionized (T $\approx$ 10$^{4-5}$ K) baryonic content of extended gaseous halos as traced by \ion{C}{4}.

%BRING THIS BACK ONCE ALL THE SECTIONS ARE IN 

This paper proceeds as follows: \S\ref{Obs_and_Data} describes the sample selection \S\ref{sec: sample_def}, FUV spectroscopy \S\ref{sec: sfr}, and data reduction and analysis for the COS-Holes sample \S\ref{line_meas}; \S\ref{sec: additional_lit_data} presents the BH mass estimates for archival data collected to increase the sample size; \S\ref{sec:Results} presents general trends for the COS-Holes sample (\S\ref{sec: gen_trends}), the radial profile for the COS-Holes+Literature sample (\S\ref{sec: cos+lit data and stats}), multivariate analysis and statistics done on the sample (\S\ref{sec: multivariate}), and the minimum mass of carbon seen in the CGM of the sample (\S\ref{sec: carbon mass}); \S\ref{sec: sim_results} describes the three simulations used in this paper (\S\ref{sec: sim_descriptions}) and presents the results of the simulated values compared to the results of the combined sample (\S\ref{sec: comp_to_sim}); \S\ref{sec:Discuss} presents a discussion of sSFR dependence in \ion{C}{IV} ionization (\S\ref{sec: sSFR dom}) and whether BHs evacuate their CGMs or not (\S\ref{sec: bh_evacuate}); lastly we present a summary of our conclusions in \S\ref{sec:End}. In this work, we assume a flat-universe $\Lambda$CDM cosmology with $H_{0}$ = 67.8 km s$^{-1}$ Mpc$^{-1}$ and $\Omega_{m}$ = 0.308 \citep{Plank2016}.

\section{Observation and Data Analysis} \label{Obs_and_Data}

\subsection{Sample Selection} \label{sec: sample_def}

\begin{figure*}
    \centering
    \includegraphics[width = 0.9\textwidth]{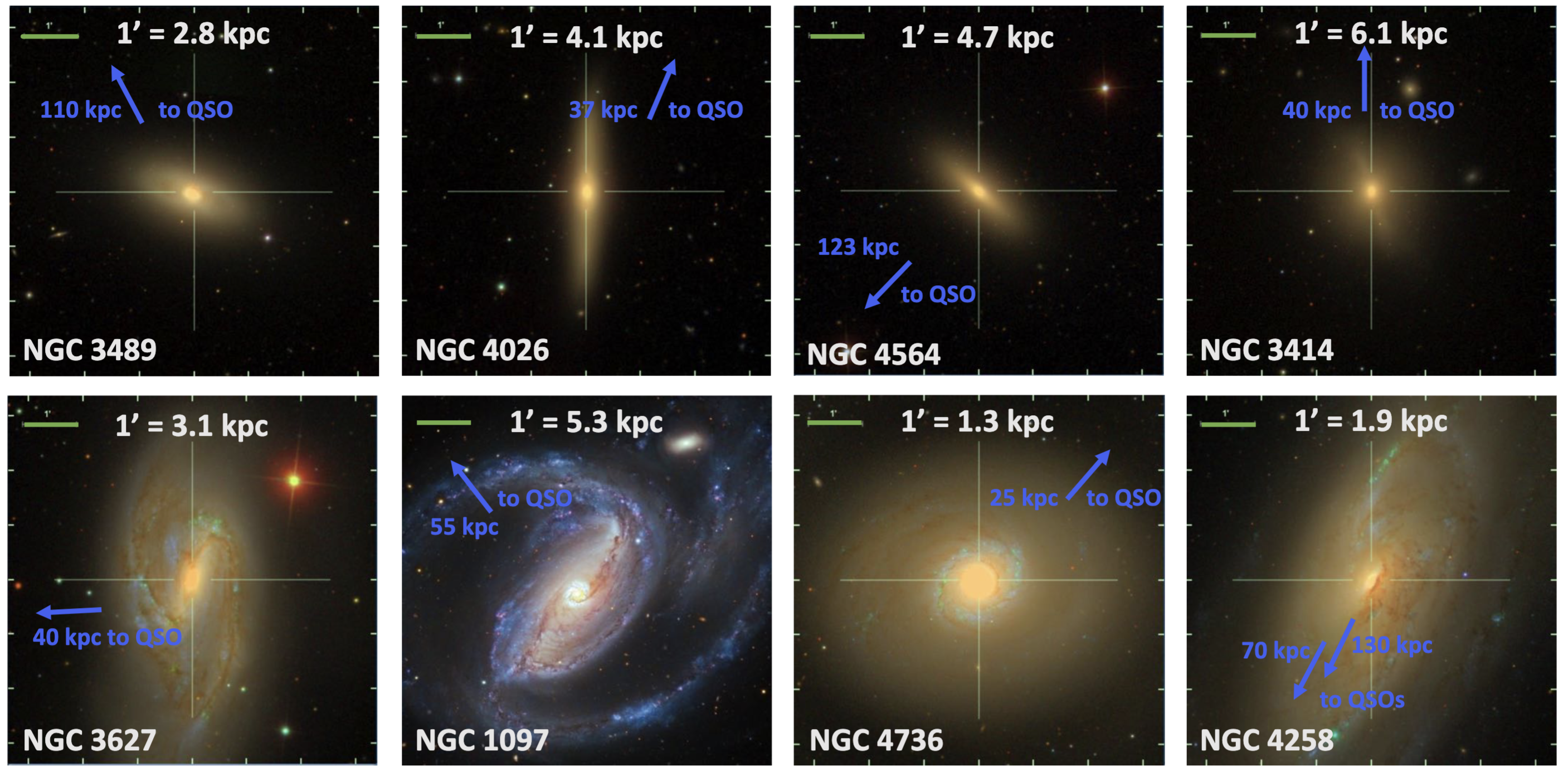}
    \qquad
    \includegraphics[width = 0.42\textwidth]{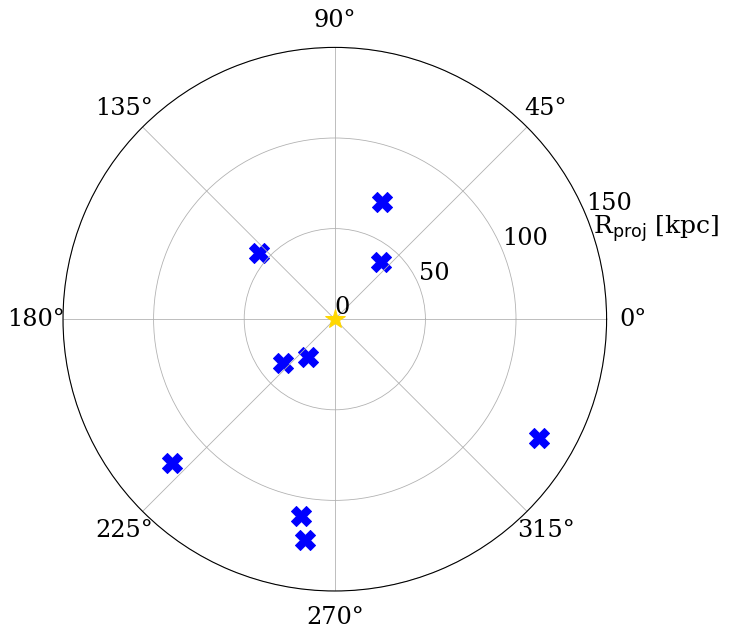}
    \qquad
    \includegraphics[width = 0.42\textwidth]{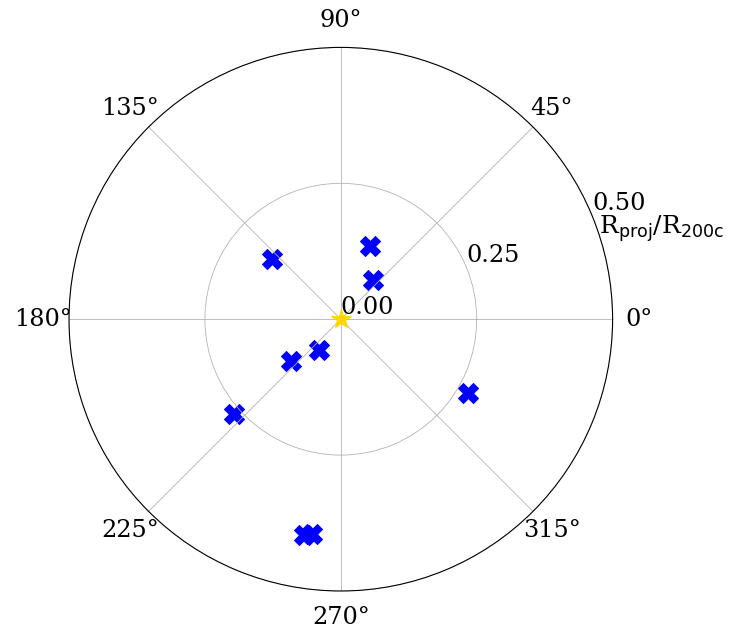}
    \caption{Top: SDSS images of each of the 8 target galaxies in the sample, with exception of NGC 1097, with the image from ESO. The physical scale (in kpc) in each galaxy's rest-frame is shown at the top of its 6'$\times$6' stamp. The targeted QSOs lie outside of the shown FOV; blue arrows and text display the direction, and distance to each FUV-bright QSO. One galaxy, NGC 4258, has two UV-bright QSOs at $<$130 kpc. Bottom: Target figure showing the distribution of QSO position angles (blue x's) on the sky with respect to the target galaxies (shifted to the center, yellow star). On the left, the radial coordinate (R$_\mathrm{proj}$) is in physical kpc at the galaxy redshift, and on the right, this coordinate is translated to the fraction of galaxy’s virial radius, R$_\mathrm{proj}$/R$_{\rm 200c}$, at which the sightline intercepts the halo. No knowledge of galaxy disk orientation or inclination with respect to the sightline is implied here.
    }
    \label{fig: galaxy_sample}
\end{figure*}

%Galaxy Information Table 
\begin{table*}
\centering
\caption{Galaxy Sample Properties}
\begin{tabular}{cccccccccc}
\hline
Galaxy   & RA        & Dec        & $z$        & Morph & D & sSFR           & M$_{*}$         & M$_{200c}$      & M$_\mathrm{BH}$                          \\
         & (deg)     & (deg)      &          &            & (Mpc)    & (log$_{10}$ yr$^{-1}$) & (log$_{10}$M$_{\odot}$) & (log$_{10}$M$_{\odot}$) & (log$_{10}$M$_{\odot}$)  
\\
(1)      & (2)                     & (3)       & (4)        & (5)   & (6)       & (7)       & (8)   & (9)   & (10) \\ \hline

NGC 1097 & 41.579 & -30.275 & 0.0042  & SB(s)b     & 14.50 a    & -9.7        & 10.5       & 11.75   & 8.14$\pm$0.090   \\
NGC 3414 & 162.818 & 27.975  & 0.0049   & S0         & 25.20$\pm$2.74 b     & -11.8        & 10.8      & 12.29  & 8.40 $\pm$ 0.07   \\
NGC 3489 & 165.078  & 13.901  & 0.0022 & SABa       & 11.98 c    & -11.2        & 10.2       & 11.45  & 6.77$\pm$0.065  \\
NGC 3627 & 170.063 & 12.991  & 0.0024  & SAB(s)b    & 10.05$\pm$1.09 b    & -10.3         & 10.8      & 12.41  & 6.92$\pm$0.048  \\
NGC 4026 & 179.855   & 50.962  & 0.0033 & S0         & 13.35 c    & -12.2        & 10.4      & 11.66  & 8.26$\pm$0.120 \\
NGC 4258 & 184.740 & 47.304  & 0.0015 & SABbc      & 7.27$\pm$0.50 b     & -10.9        & 10.9      & 12.51  & 7.58$\pm$0.030  \\
NGC 4564 & 189.113  & 11.439  & 0.0038 & S0         & 15.94 c    & -12.4        & 10.4       & 11.63  & 7.94$\pm$0.140  \\ 
NGC 4736 & 192.721 & 41.121  & 0.0010 & Sab        & 5.00$\pm$0.79 b        & -10.7        & 10.6      & 11.94  & 6.83$\pm$0.120  \\
\hline
\end{tabular}
\label{tab: galaxy_properties}
\tablecomments{Comments on columns: (1) galaxy name; (2-3) RA and Dec for the galaxy; (4) galaxy redshift; (5) Morphology; (6) assumed distance to galaxy where the letter beside the distance corresponds to one of the following references: (a) \citet{van_den_Bosch_2016}, (b) \citet{saglia_2016}, (c) \citet{Tonry_2001} SBF corrected via Eq A1 in \citet{Blakeslee_2010}. ; (7) specific star formation rate: typical errors on SFRs derived from infrared photometry are 0.2 - 0.3 dex \citep{Reike_2009,Terrazas_2017} while stellar masses are accurate to about $\sim50$\%. On average, for galaxies of these masses, sSFR errors will be on the order of a few - several tenths of a dex; (8) Stellar Mass; (9) Halo Mass; (10) SMBH Mass.}
\end{table*}

%QSO/Galaxy Pair Information Table 
%Table that has the numbers for each column 
\begin{table*}
\centering
\caption{Background QSO Properties}
\begin{tabular}{cccccccccc}
\hline
QSO                     & RA        & Dec        & z     & R$_\mathrm{proj}$     & R$_\mathrm{proj}$/R$_{200c}$      & Mag   & Mag   & N$_\mathrm{ORB}$ \\
                         & (deg)     & (deg)      &       & (kpc)     &      & (FUV) & (NUV) &      \\
(1)      & (2)                     & (3)       & (4)        & (5)   & (6)       & (7)       & (8)   & (9) \\ \hline
UVQSJ024649.87-300741.5 & 41.707792 & -30.128194 & 0.524 & 55.66 & 0.32 & 18.46 & 17.9  & 4    \\
SDSSJ105115.75+280527.1      & 162.81564 & 28.090865  & 0.423 & 40.50  & 0.15 & 18.2  & 17.75 & 4    \\
SDSSJ110139.76+142953.4      & 165.4157  & 14.498172  & 0.635 & 110.00 & 0.80 & 18.99 & 18.70  & 7    \\
SDSSJ112304.91+125748.0      & 170.77049 & 12.963349  & 0.315 & 120.00   & 0.42 & 18.76 & 18.34 & 6    \\
SDSSJ115901.72+510630.7      & 179.75718 & 51.108554  & 0.524 & 37.38  & 0.23 & 18.72 & 18.36 & 4    \\
SDSSJ122046.61+464347.5      & 185.19421 & 46.729881  & 0.707 & 69.78 & 0.22 & 18.82 & 18.21 & 6    \\
UVQSJ122208.10+461250.1 & 185.53375 & 46.213917  & 0.111 & 130.30 & 0.42 & 18.30  & 18.11 & 4    \\
LBQS-1235+1123          & 189.43571 & 11.116143  & 0.949 & 123.10  & 0.77 & 18.99 & 17.93 & 7    \\
SDSSJ124939.06+412243.5      & 192.41277 & 41.378773  & 0.368 & 25.52 & 0.13 & 18.62 & 18.55 & 6    \\

\hline
\end{tabular}
\label{tab: qso_properties}
\tablecomments{Comments on columns: (1) QSO Identification; (2-3) RA and Dec for the QSO; (4) QSO redshift; (5) QSO impact parameter; (6) impact parameter normalized by virial radius; (7) FUV Magnitude; (8) NUV Magnitude; (9) Number of orbits.}
\end{table*}

The COS-Holes Survey consists of nine UV-bright QSOs, z $<$ 0.005, probing the halos of eight galaxies at impact parameters \textit{R$_\mathrm{proj}$} $\sim$ 25-130 kpc as seen in Figure \ref{fig: galaxy_sample}. To build the survey, we cross-matched the SDSS DR14 QSO catalog \citep{Paris_2018} and the UVQS \citep{monroe_2016} QSO catalogs with several published catalogs of galaxy BH masses \citep{K_ho_2013, mcconnell_2013, bentz_and_katz_2015, Lasker_2016, van_den_Bosch_2016, Terrazas_2017} to search for FUV bright QSOs (GALEX M$_\mathrm{FUV} <$ 19) within 150 kpc projected distance from the galaxies in their rest frames. By design, the resulting sample contains galaxies that have dynamically determined SMBH masses (e.g. through stellar dynamics, ionized gas dynamics, CO molecular gas disk dynamics, maser disk dynamics etc) determined. We note that we did not select our galaxies based on assembly history or morphology. We acknowledge that recent results have shown that disruptions in the disk (either by merger or similar event) can be an important factor for how BHs grow and affect the CGM \citep{davies_2022, Davies_2024}; however testing for these morphological differences in the galaxies and how that affects the properties of the CGM is beyond the scope of this work. 
%somewhere here is where I need to address bullet point 2 from the referee: 
The property cuts implemented for our sample, as described below, were strategically made to match similar cuts made for previous surveys searching for highly ionized gas \citep[COS-Halos;][]{Tumlinson11, Werk_2012} within the cool and intermediate temperature phase of the CGM \citep{tumlinson_2017}; what sets our sample apart, however, is our focus on galaxies that have accurately measured BH masses in order to determine how they impact the state of the CGM.

To start, we eliminate from the sample any galaxies in dense cluster environments (e.g. Virgo) which have already been shown to possess significantly less gas than galaxies in more isolated environments \citep{yoon_2012, Burchett_2018}. We check the GALEX NUV magnitudes of our targets for large values of NUV-FUV colors, which would potentially indicate the presence of a strong Lyman Limit system (N$_\mathrm{HI} \geqslant 10^{17}$ cm$^{-2}$; LLS) along the line of sight that may have contaminated our transitions of interest. We note that this process does not introduce a bias to the sample selection since the LLS absorption would be at unrelated higher redshifts than our targets. 

It is known that AGN feedback can be highly directional and not necessarily aligned with the spin axis of the galaxy \citep{Bentz_2023}. However, we choose not to include QSOs that probe the halos of galaxies with black holes that are currently accreting as Seyferts or quasars themselves, as done in \citet{berg_2018}. Instead, we are more interested in the long term effects the black hole has on the halo and and thus selected galaxies based on their stellar mass, redshift, and having a dynamically-measured black-hole mass available. None of the galaxies in our sample is classified as an AGN in the mid-infrared photometrically-selected sample of \cite{Asmus_2020}, using a method that has a 90\% reliability in selecting AGN \citep{Assef_2018}. We note however, that many of our galaxies do exhibit  LINER-like emission in their central regions, possibly indicating a low-luminosity, accretion-powered active nucleus \citep[e.g.][]{Molina_2018}. LINER emission is quite common among nearby, L$^{\star}$ spiral galaxies, and it can be related to AGN phenomena, although this relation is uncertain and poorly-quantified \citep{Ho_1997}. 

In addition, there can be an azimuthal dependence of ion absorption in disk dominated galaxies. For example, it has been shown that there is a strong azimuthal dependence with \ion{Mg}{II} \citep{Bordoloi_2011}, but for \ion{O}{VI} the correlation along the major and minor axes are less clear \citep{Kacprazak_2019}. These dependencies have not been demonstrated for \ion{C}{IV} and investigating them in the COS-Holes sample is beyond the scope of this work. Moreover, as seen in Figure \ref{fig: galaxy_sample}, some of our galaxies are too face on to report accurate azimuthal angles.  The remaining galaxies have azimuthal angles consistent with a random distribution, and thus any azimuthal angle dependence of CGM CIV will not play a significant role in driving the trends (or lack thereof) we observe. 

Finally, the nearby galaxy NGC 4258 which has a highly accurate BH-mass measurement from megamaser kinematics \citep{miyoshi_1995}, is serendipitously intersected by two inner-CGM QSO sightlines at 70 and 130 kpc. We include both QSO targets in our final sample because it offers a rare opportunity to study subtle variations (e.g. column density, kinematics, etc.) within a single halo. 
%Such variations in the CGM can provide important constraints on the scatter in the trends (or lack thereof) in the sample as a whole \citep{Bowen_2016}. 

%need to expand upon the stellar mass range question from the referee here 
We selected a sample of galaxies with stellar masses spanning a narrow range around M* ($\approx 10^{10.5}$ M$_{\odot}$), since stellar mass has been found to correlate with ionized CGM content (e.g. \citet{Tchernyshyov2022}). We estimate the halo masses of our sample by following the same method as outlined in CGM$^{2}$ by \cite{Tchernyshyov2022}. Using the stellar mass-halo relation, as defined in Tab J1 of \cite{Behrooz_2019}, in combination with the approach laid out in \cite{Hu_2003}, we convert the halo masses to match the convention where the average mass density within the halo radius is 200 times the critical density of the universe. We denote these halo masses and the corresponding virial radii as M$_{200c}$ and R$_{200c}$. The final range of stellar and halo masses for the sample are log$_{10}$M$_{\star}$/M$_{\odot}$ $\sim$ 10.2-10.9 and log$_{10}$M$_\mathrm{halo}$/M$_{\odot}$ $\sim$ 11.45-12.51 respectively. We note that a stellar mass of $\approx$10$^{10.5}$ M$_{\odot}$, is representative of L$^{\star}$ galaxies, but also can also be a transitional stellar mass in terms of sSFR which is known to correlate with CGM properties in intermediate ionization states \citep{Tchernyshyov_2023}.  However, by keeping the range of stellar and halo masses relatively small we minimize the scatter due to these properties and enable a controlled examination of the role SMBHs and SFR play in shaping the properties of the CGM. The galaxy and QSO properties for the sample can be found in Tables \ref{tab: galaxy_properties} and \ref{tab: qso_properties} respectively. 

\subsubsection{Star Formation Rates}\label{sec: sfr}

We obtain star formation rates (SFRs) for the COS-Holes sample from \cite{Terrazas_2017} in \S\ref{sec:Results}. For the three galaxies in our survey not included in their sample (NGC 3489, NGC4026, and NGC 4564), we calculate their corresponding SFRs using the same methodology; we summarize the procedure here but a detailed description can be found in \cite{bell_2003} and \cite{Terrazas_2016}. We calculate far-infrared (FIR) SFRs by using Eq. (A1) in \cite{bell_2003} which uses 60 and 100 $\rm \mu$m \textit{IRAS} fluxes to estimate the FIR flux. We then estimate the total infrared (TIR) flux via TIR = 2 $\times$ FIR \citep{bell_2003}. Finally, the TIR-derived SFR is calculated using Eq. 12 in \cite{kennicutt_evans_2012}: 
\begin{equation}
    \log_{10} \rm SFR_\mathrm{TIR} ( M_{\odot} \mathrm{yr}^{-1}) = \log_{10} L_\mathrm{TIR} - 43.41
\end{equation}
where L$_\mathrm{TIR}$ is the TIR luminosity 
calculated from the TIR flux estimates and distances to the galaxy (for consistency we use the same distances presented in \citep{Terrazas_2017}; for more detailed information, see Table \ref{tab: galaxy_properties}). We note that for NGC 3489, only 65 and 90 $\mu$m fluxes from \textit{AKARI} were available on NED\footnote{The NASA/IPAC Extragalactic Database (NED) is funded by the National Aeronautics and Space Administration and operated by the California Institute of Technology.} and we use those values to calculate its respective SFR. To present the calculated SFRs as log$_{10}$sSFRs (which range between -12.4 and -9.7) in Table \ref{tab: galaxy_properties}, we divide them by the stellar mass of the galaxy. Typical stellar mass uncertainties derived from SDSS photometry are $\pm$50$\%$ \citep{Blanton_2007, Kauffmann_2003} (approximately 0.2 dex) and the SFR errors derived from infrared photemetry are approximately 0.2-0.3 dex  \citep{Reike_2009, Terrazas_2017}. 

\subsection{COS Spectroscopy}\label{sec: cos_spec}

The quasar spectra for the COS-Holes survey were taken using the G160M grating on the Cosmic Origins Spectrograph \citep[COS;][]{Froning_and_Green_2009, Green_2012} on the Hubble Space Telescope as a part of a 55-orbit Cycle 29 HST Program (PID$\#$16650). The primary spectral features of interest were absorption lines from the doublets \ion{C}{IV} ($\lambda\lambda$1548, 1550) and \ion{Si}{IV} ($\lambda\lambda$1393, 1402). 

\ion{C}{IV} is the highest ionization state transition available at these low redshifts (z $<$ 0.005) where dynamical black hole masses are available, and is easily detectable in the UV. We note that \ion{C}{IV} is an ``intermediate" ion with a potential energy of 47.89 eV required to ionize \ion{C}{III} into \ion{C}{IV}. In collisional ionization equilibrium (CIE) \ion{C}{IV} reaches a peak ion fraction at a temperature of 1.2$\times 10^{5}$ K (10$^{5.1}$) and falls rapidly at higher temperatures, with less than 10$\%$ at 1.6$\times 10^{5}$ K (10$^{5.2}$) \citep{Gnat_2007}. In photoionization equilibrium (PIE), it peaks at a density of n$_{H}$ $\approx$ 2$\times 10^{-5}$ cm$^{-3}$ at $z =$ 0 \citep{haardt_manau_2012, Khaire_19}. Thus, in CGM conditions, it can form either through photo-ionization or collisional ionization \citep{tumlinson_2017}.
In EAGLE, \citet{Oppenheimer_2020} found that \ion{C}{IV} is a very good tracer of CGM gas between T$=10{^4}-10{^5}$ K and n$_H=10^{-5}-10^{-3}$ cm$^{-3}$. While it is beyond the scope of this work to constrain the precise phase of \ion{C}{IV}, we highlight that we are explicitly avoiding characterization of the hot CGM (T $\approx$ 10$^{6}$ K). 
%"overall baryonic content" in a specific temperature and density range seems contradictory to me. I assume it is the latter, and then suggest to say "This ions tracks well CGM gas at temperatures XX-XX and densities XX-XX in the EAGLE cosmological simulations (Oppenheimer...)"

The COS-Holes QSOs have FUV magnitudes of 18.2 - 19.0 and redshifts ranging from 0.3 $-$ 0.9, and we observed each target QSO for between 4 and 7 orbits in G160M with a central wavelength of 1577 \AA. Our exposure times were calculated to detect a 40 m\AA~ feature at 2$\sigma$, consistent with detected \ion{C}{IV} around $\sim$ 0.1 - 1 L$^{\star}$ galaxies in the literature \citep{Borthakur_2013, Bordoloi_2014, burchett_2016}. All spectra achieved a S/N of 10 - 12 per resel at the wavelengths of \ion{C}{IV}. 

We combine the CALCOS-generated x1D files using v3.1.1 of the COADD\_X1D routine provided by the COS-GTO team \citep{danforth_2016}, which properly treats the error arrays of the input files using Poisson statistics. The code aligns the different exposures by determining a constant offset determined by cross-correlating strong ISM lines in a 10\AA~ wide region of the spectrum. The COS line-spread function (LSF) is well described by
a Gaussian convolved with a power law that extends to many
tens of pixels beyond the line center \citep{Green_2012}. 
These broad wings affect both the precision of our equivalent width measurements and complicate assessments of line saturation. We mediate these effects by fitting absorption lines with Voigt profiles that incorporate the COS LSF. Each COS resolution element at R  $\sim$18,000 covers 16  km s$^{-1}$ and is sampled by six raw pixels. We perform our analysis on the data binned by three native spectral pixels to a dispersion of $\Delta\lambda$ $\approx$ 0.0367 \AA. The resulting science-grade spectra are characterized by  a FWHM $\approx$ 16 km s$^{-1}$. We perform continuum fitting with the {\tt linetools} package\footnote{https://doi.org/10.5281/zenodo.1036773}, an open-source code for analysis of 1D spectra.

%not sure if this is the correct paper to cite here for \citep{coslsf}. Need to check with Jess 

\subsection{Absorption Line Measurements} \label{line_meas}

This section describes the methods used to measure and calculate key observational properties, presented in Table \ref{tab: cos_holes_analysis}. \S \ref{sec:Results} discusses the column densities versus key galaxy parameters (Figure \ref{fig: gal_properties}). 

\subsubsection{Line identification with PyIGM}\label{sec: pyigm}

We manually assign line identifications and redshifts to all absorption features in the spectra using the {\tt\string PyIGM IGMGuesses} GUI\footnote{https://doi.org/10.5281/zenodo.1045480}. To make sure that we correctly attribute absorption to a COS-Holes galaxy's CGM rather than another absorber at a different redshift, we implement the following methodology. First, we identify absorption features at $z$ = 0, the redshift of the Milky Way. We then identify any ``proximate" absorption at the redshift of the QSO observed. Finally, we examine the spectra for Lyman series lines at redshifts $< z_{\rm QSO}$ to find serendipitous absorption systems. 
%this typically led to a search for various intervening absorbers. 
After identifying these features, we move to the redshift of the target galaxy to look for any absorption features within $\sim$300 km/s associated with \ion{C}{IV} ($\lambda\lambda$1548, 1550), similar to the COS-Halos Survey \citep{werk_2013}. For this paper we specifically focus on \ion{C}{IV} identifications and analysis even though other ions (e.g, \ion{Si}{IV} $\lambda\lambda$1393, 1402) were observable; in future work we plan on analysing other ion absorption features present. We obtain preliminary line profile fits including the following parameters: central velocity $v$, column density $N$, and Doppler parameter, $b$; we then input these user specified parameters into a Voigt profile fitting program. 

\subsubsection{Voigt Profile Fitting}\label{sec: voigt}

\begin{figure}
    \centering
    \includegraphics[width = \columnwidth]{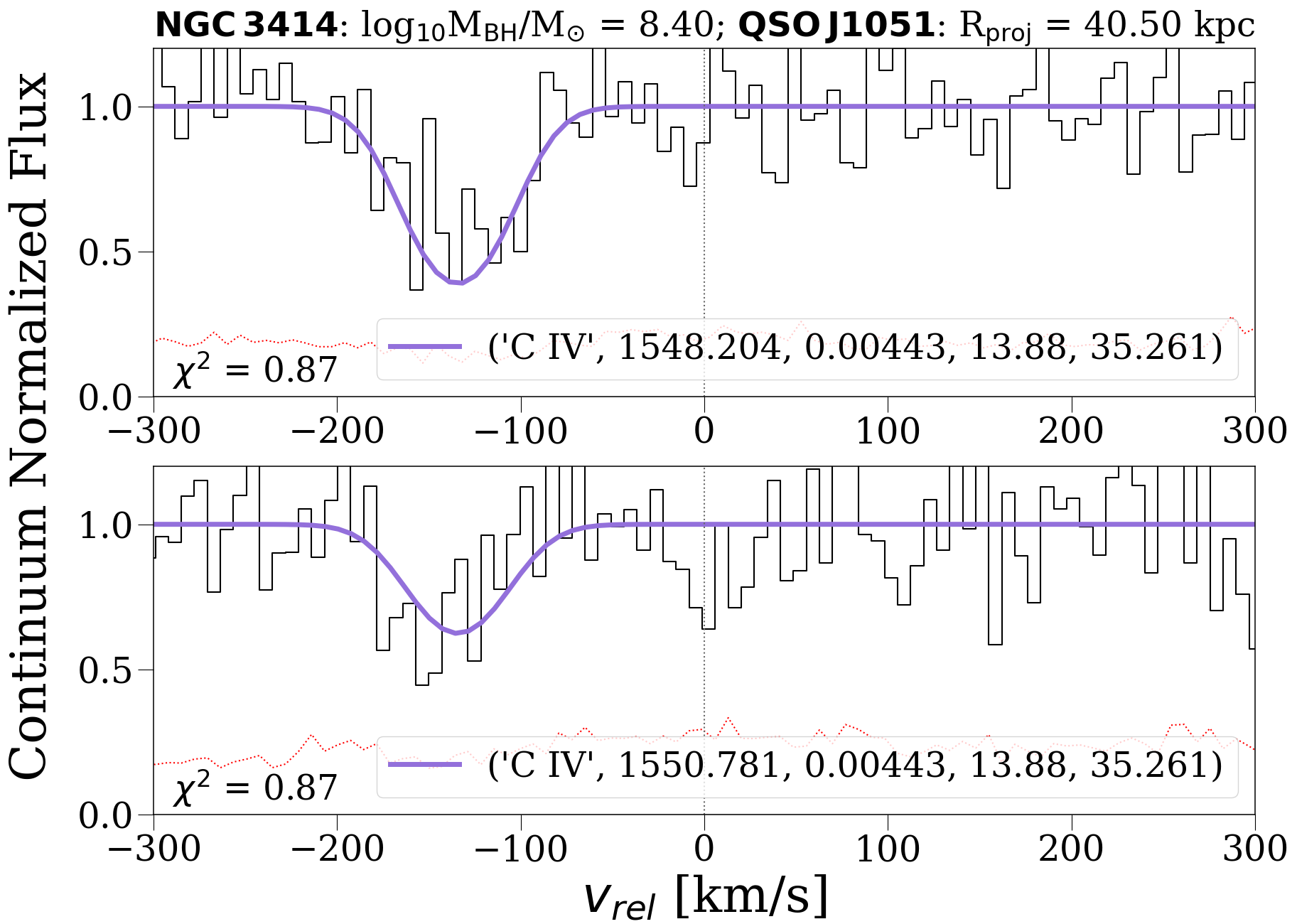}
    \caption{Representative \ion{C}{IV} absorption feature of a QSO-galaxy pair (SDSSJ1051-NGC 3414) set in the rest frame of the galaxy. The colored line represents the Voigt profile fit due to the $\lambda\lambda$1548 1550 lines in the top and bottom panels respectively. The values in the labels correspond to the following: ionic species, wavelength, absorption feature redshift, column density (log$_{10}$N$_\mathrm{C IV}$) [cm$^{-2}$], and doppler parameter [km s$^{-1}$]. In the bottom left hand corner of the figure is the reduced chi squared for the fit made to each absorption feature. The spectra for the rest of the COS-Holes sample are presented in Appendix \ref{sec: apendix A}.}
    \label{fig: civ_lineprofie_example}
\end{figure}

Based on the identifications from {\tt\string PyIGM IGMGuesses} GUI, we measure \ion{C}{IV} column densities, Doppler parameter, and the relative velocity of the absorption components using Voigt profile fitting with the package {\tt\string veeper}\footnote{https://zenodo.org/doi/10.5281/zenodo.10993983}, which uses {\tt\string scipy.optimize.least\_squares}\footnote{https://doi.org/10.1038/s41592-019-0686-2} 
to perfrom a least squares minimization. In five of our QSO-galaxy line-of-sight pairs we detect \ion{C}{IV} while the other four were non-detections and we report them as upper limits. In the spectral regions with no detected metal absorption, we calculate a 2$\sigma$ upper limit on the column density as estimated by the apparent optical depth method (AODM) with the {\tt\string linetools XSpectrum1D} package\footnote{https://doi.org/10.5281/zenodo.1036773} over a 100 km s$^{-1}$ velocity span centered on the galaxy redshift. By default, we use the stronger line at 1548\AA~ to estimate $2\sigma$ equivalent width upper limits, similar to the AOD method, using {\tt\string linetools XSpectrum1D}, but in cases where there is blending or contamination, we use the 1550\AA~ line. When multiple absorption components are found in a galaxy's search window, their column densities are summed, and then this total column density is associated with the galaxy. Figure \ref{fig: civ_lineprofie_example} displays the line profile for the \ion{C}{IV} ($\lambda\lambda$1548, 1550) absorption doublet for NGC 3414 as a representative Voigt profile for the entire COS-Holes sample. The spectra showing \ion{C}{IV} (or upper limits) for the rest of the survey are presented in Appendix \ref{sec: apendix A}.

\begin{table*}
\centering
\caption{COS-Holes Measurements}
\begin{tabular}{ccccccc}
\hline
Galaxy   & QSO ID & $z_\mathrm{abs}$     & log$_{10}$N$_\mathrm{C IV}$    & $b$            & EW      & $|v_\mathrm{rel}|$          \\
         &            &          & (cm$^{-2}$) & (km s$^{-1}$) & (m\AA) & (km s$^{-1}$) \\
(1)      & (2)    & (3)        & (4)      & (5)        & (6)          & (7)         \\ \hline
NGC 1097 & UVQSJ0246  & 0.00426  & 14.14$\pm$0.05      & 55.60$\pm$7.58         & 323.84$\pm$33.38       & 2.01       \\
NGC 1097 & UVQSJ0246  & 0.00471  & 13.71$\pm$0.10      & 15.36$\pm$6.09         & 109.54$\pm$27.27       & 145.38       \\
NGC 3414 & SDSSJ1051  & 0.00443  & 13.88$\pm$0.05      & 35.26$\pm$6.06        & 235.30$\pm$26.86        & 148.32      \\
NGC 3489 & SDSSJ1101  & 0.00229  & 13.44$\pm$0.10                & 22.95$\pm$8.90         & 106.52$\pm$19.55       & 30.88       \\
NGC 3627 & SDSSJ1123  & 0.00287  & 13.89$\pm$0.07      & 85.00$\pm$18.33          & 230.70$\pm$41.64       & 152.24      \\
NGC 4026 & SDSSJ1159  & 0.0033   & $<$13.24                &        & $<$73.10       & 0       \\
NGC 4258 & SDSSJ1220  & 0.001494         & $<$13.47                            &              & $<$60.32             & 0              \\
NGC 4258 & UVQSJ1222  & 0.001494 & $<$13.39 &                          & $<$50.20              & 0              \\
NGC 4564 & LBQS-1235  & 0.0038   & $<$13.40 &            & $<$58.21              & 0                    \\ 
NGC 4736 & SDSSJ1249  & 0.00054  & 13.75$\pm$0.05                & 32.39$\pm$6.30         & 186.38$\pm$17.07         & 148.55      \\
NGC 4736 & SDSSJ1249  & 0.00083  & 13.48$\pm$0.10                & 11.53$\pm$6.33       & 95.97$\pm$12.53        & 47.32       \\
NGC 4736 & SDSSJ1249  & 0.00111  & 13.86$\pm$0.05      & 37.78$\pm$5.91           & 217.64$\pm$19.25        & 43.94       \\
\hline
\end{tabular}
\label{tab: cos_holes_analysis}
\tablecomments{Comments on columns: (1) galaxy name; (2) QSO identification that is shortened from full name; (3) redshift of the absorption coefficient; (4) \ion{C}{IV} column density; (5) Doppler parameter; (6) equivalent width; (7) absolute value relative velocity of absorption component projected along the line of sight in the galaxy's frame.}
\end{table*}

\section{Archival Observations \& BH Mass Estimations}\label{sec: additional_lit_data}

%New papragraph: August 7th 2023 
We increase our sample size with CGM \ion{C}{IV} measurements using published HST/COS data from \citet{Borthakur_2013} (starbursts), \cite{werk_2013} (COS-Halos), \citet{Bordoloi_2014} (COS-Dwarfs), and \citet{Lehner_2020} (Project AMIGA). M31 has a measured SMBH mass of log$_{10}$M$_\mathrm{BH}$/M$_{\odot}$ = 8.15 $\pm$ 0.24 \citep{davis_2017} and a stellar mass of log$_{10}$M$_{\star}$/M$_{\odot}$ = 10.9 $\pm$ 0.22  \citep{Williams_2017}, both of which are within range of the COS-Holes galaxy properties. Through the use of several QSO sightlines, it has a well studied CGM (Project AMIGA; \cite{Lehner_2020}). To match the COS-Holes sample we only include sightlines from Project AMIGA if their corresponding impact parameter was $\leq$ 150 kpc and did not have any contamination from the Magellanic Stream; for a more detailed explanation of how this contamination was removed, see \cite{Lehner_2020}. If a sightline contained \ion{C}{IV} and had multiple absorption features, we sum the measured N$_\mathrm{C IV}$ to present a total column density, similar to COS-Halos. We note that having a plethora of QSO sightlines provides the opportunity to compare single QSO-galaxy sightline derived CGM properties to a galaxy with multiple sightlines. However, with this reduced sample of Project AMIGA observations, we choose to take an average of the column densities to represent a singular mean N$_\mathrm{C IV}$ for the Project AMIGA observations. This allows us to have a consistent Literature sample and not bias any results towards features seen in M31. 

Similarly to M31, we only include N$_\mathrm{C IV}$ measurements from galaxies if they had a stellar mass that fell within the range of our COS-Holes observations (log$_{10}$M$_{\star}$/M$_{\odot}$ = $10^{10} - 10^{11}$); by making these cuts we add four galaxies from \citet{Borthakur_2013}, five from \citet{Bordoloi_2014}, and two from \citet{werk_2013} (COS-Halos) to the Literature sample. Since galaxies from these surveys do not have dynamically measured BH masses, we estimate the SMBH mass for each galaxy using the following approximation from Eq. 7 in  \citet{Piotrowska_2022, saglia_2016}:
\begin{equation}\label{eq:mbh}
    \rm log_{10}\rm M_\mathrm{BH} = 5.246 \times \rm log_{10} \sigma_{c} - 3.77
\end{equation}
where $\sigma_{c}$ is the central stellar velocity dispersion of the galaxy, or the  random line-of-sight motion of stars due to the galaxy's gravitational potential well. 
%What are the typical uncertainties here? The main strength of the COS-Holes survey is that we have BH masses measured through a highly precise method, so some discussion on this more easily obtained measurement is warranted.  I fear that if we don't address this, the referee may question the motivation for our survey in the first place.

We obtain central stellar velocity dispersion measurements from the SDSS DR7 value-added catalog\footnote{gal\_info\_dr7\_v5\_2} \citep{Abdurro'uf_2022} for our selected sample of galaxies from \citet{Borthakur_2013} and \citet{Bordoloi_2014}. These SDSS $\sigma_{c}$ values are the superposition of many individual stellar spectra that were Doppler shifted due to the star's motion within each galaxy and their measurements were made by analyzing the integrated spectrum of the whole galaxy. We acknowledge that estimating measurements for $\sigma_{c}$ can be complex due to several components that can dominate the integrated spectra, either from different stellar populations and/or kinematics in the bulge and the disk. However, these complexities were taken into account in the SDSS catalog where velocity dispersion estimates were only measured for spheroidal systems whose spectra satisfied certain specifications (e.g. galaxy type, $z <$ 0.4, etc). In addition, it is recommended to only use SDSS velocity dispersion measurements $>$ 70 km s$^{-1}$ (due to the SDSS instrumental resolution) for spectra with a median per-pixel S/N $>$ 10; for more information about how these velocity dispersions were measured and how their biases were corrected see \citet{Bernardi_2007}. 

All the galaxies from \citet{Borthakur_2013} and \citet{Bordoloi_2014} (nine total) have median per-pixel S/N $>$ 10 and the average $\sigma_{\rm c}$ value for the galaxies with velocity dispersion measurements $>$ 70 km s$^{-1}$ is 111.59$\pm$17.18 km s$^{-1}$. Four galaxies from \citet{Bordoloi_2014} have stellar velocity dispersion measurements $<$ 70 km s$^{-1}$. We report these as upper limits and use 70 km s$^{-1}$ in our log$_{10}$M$_{\rm BH}$/M$_{\odot}$ estimates which correspond to a value of $<$ 5.91. As this is close to the lower bound of the BH mass range for the COS-Holes survey, and the SDSS fiber spectra is not sensitive to BH estimates lower than this value, we do not believe that adding these BH mass estimations bias the new combined sample. For those galaxies drawn from the COS-Halos sample \citep{werk_2013}, where SDSS fiber spectra of the galaxies were not available, we use the python package {\tt\string pPXF}\footnote{https://pypi.org/project/ppxf/} to analyze the Keck LRIS spectra \citep[COS-Halos;][]{Werk_2012}. This package calculates a central velocity dispersion from the optical LRIS spectrum; for more clarification on techniques see \cite{Koss_2022}. The uncertainties on the BH masses for the literature sample are roughly a factor of five larger than those from the COS-Holes sample with dynamically-measured BH-masses.

We use the package {\tt KaplanMeierFitter} from the python package {\tt LIFELINES}\footnote{https://doi.org/10.5281/zenodo.10456828} which implements Greenwood's uncertainty estimate, to determine the average log$_{10}$M$_{\rm BH}$/M$_{\odot}$ for both the literature and the COS-Holes sample. Kaplan-Meier is a non-parametric technique of estimating the survival probability of a set of data and is useful since it assumes that censored observations (upper limits) have the same survival prospects as observations that continue to be followed. For the literature sample, with 95$\%$ confidence intervals, we find the average  log$_{10}$M$_{\rm BH}$/M$_{\odot}$ to be 7.40(6.10, 7.71); this is comparable to the COS-Holes average log$_{10}$M$_{\rm BH}$/M$_{\odot}$ of 7.58(6.77, 8.255). 
This sample of 12 additional galaxies adds a wider range of galaxies black hole masses to the sample and statistical power to our analysis, especially in \S\ref{sec: cos+lit data and stats}. The collective information for the additional Literature sample can be seen in Appendix \ref{sec: apendix B} in Tables \ref{tab: lit_gal_info} and \ref{tab: lit_qso_info} respectively.

\section{Observational Results} \label{sec:Results}

\begin{figure*}%
    \centering
    \includegraphics[width=\textwidth]{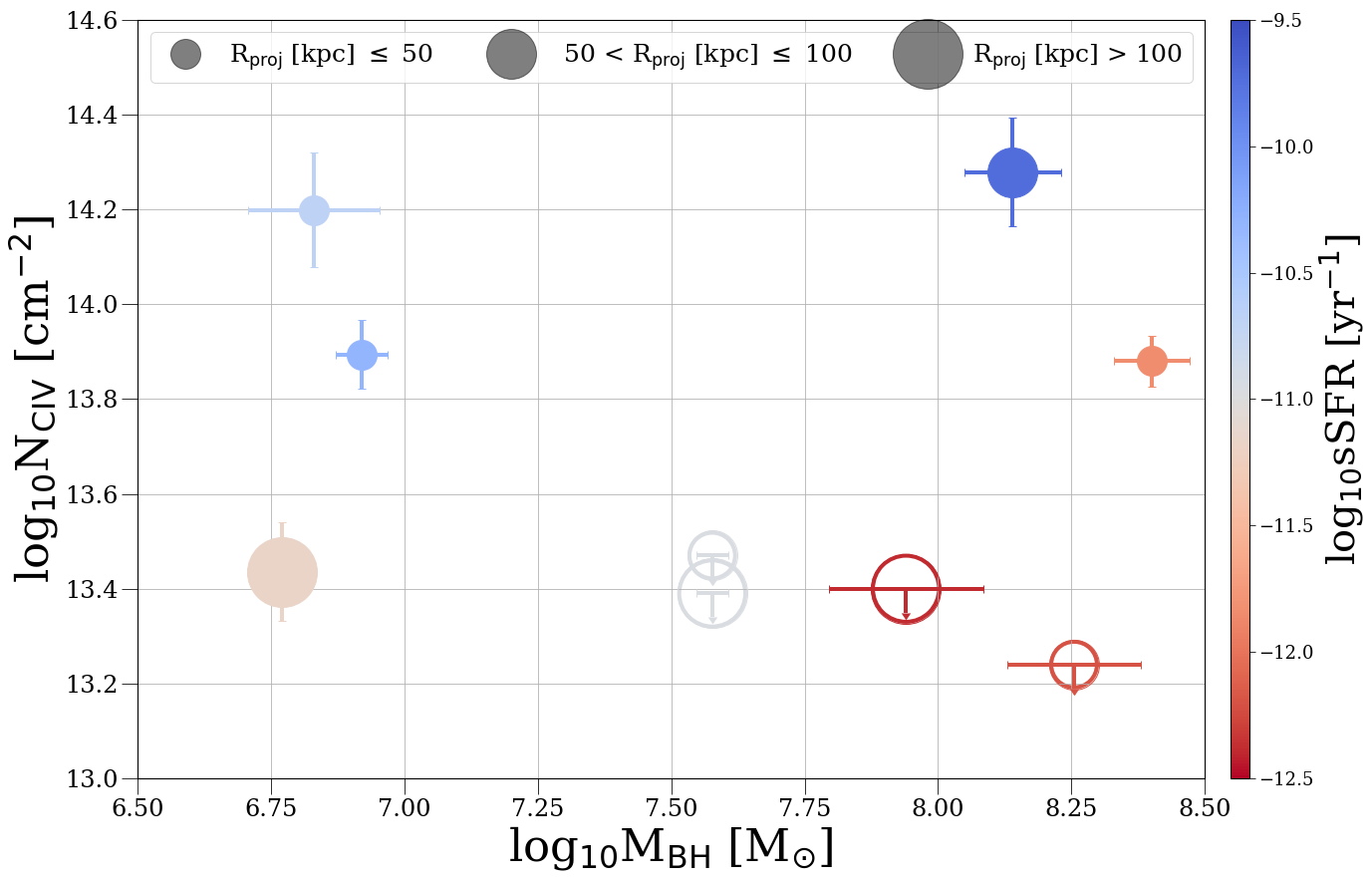}%
    \caption{Measured \ion{C}{IV} column densities versus log$_{10}$M$_\mathrm{BH}$. Each data point is colored with by the specific star formation rate (sSFR) and each marker size corresponds to the respective impact parameter. Unfilled circles represent 2$\sigma$ upper limits. There is a wide spread in the \ion{C}{IV} column densities as black hole mass increases; interestingly, there is a slight trend with sSFR and column density. Galaxies with low observed column density (log$_{10}$N$_\mathrm{C IV} \leq$ 13.5) tend to be less star forming (sSFR $ \lesssim$ -11.0) than galaxies with higher observed column density. }%
    \label{fig: gal_properties}
\end{figure*}

\begin{figure*}
    \centering
    \includegraphics[width = \textwidth]{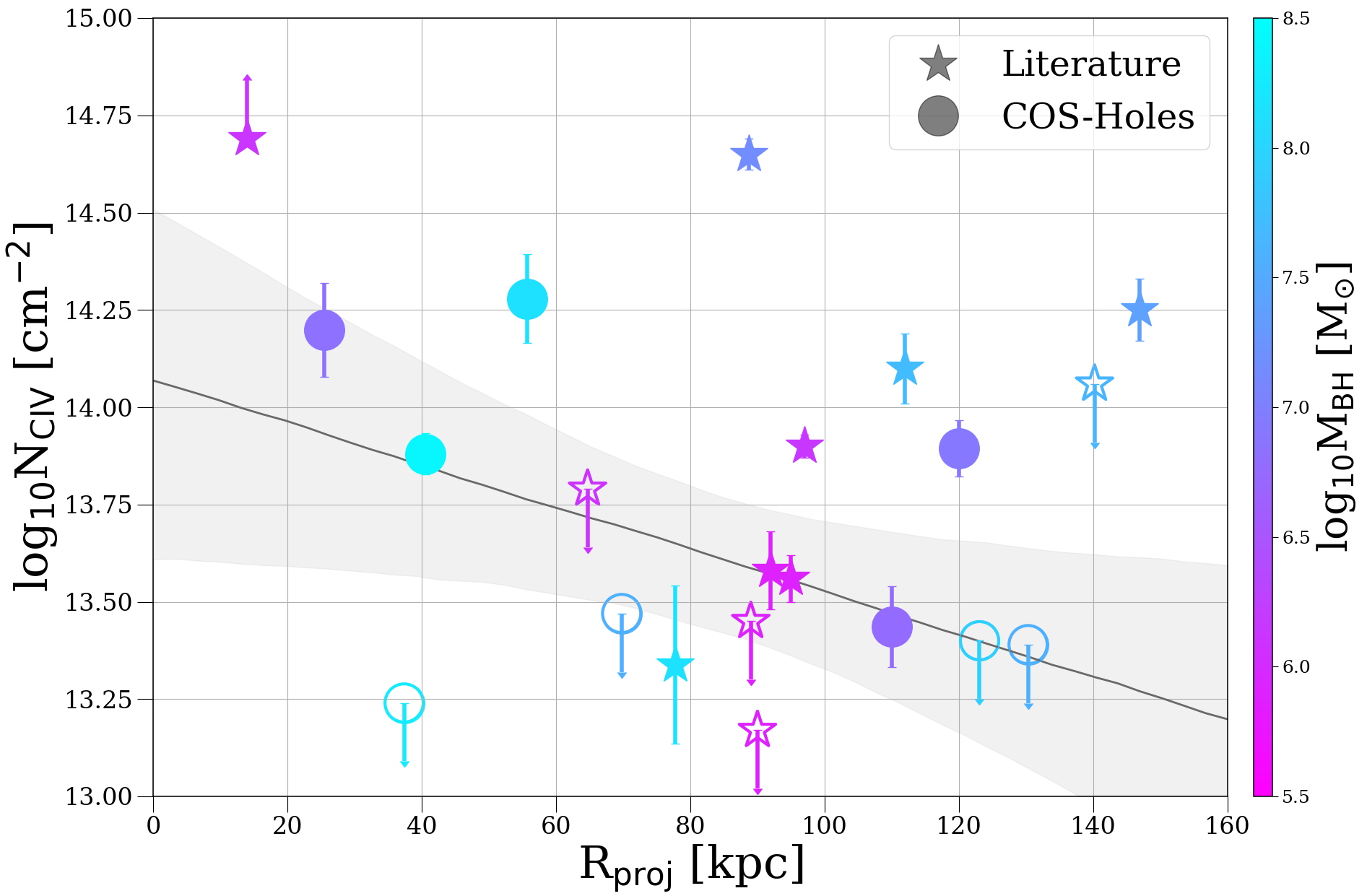}
    \caption{\ion{C}{IV} column densities assembled from previous QSO absorption line surveys probing the CGM of low-z, log$_{10}$M$_{\star}$/M$_{\odot}$ = $10^{10} - 10^{11}$ galaxies, including \cite{Borthakur_2013}, \cite{Bordoloi_2014}, \cite{werk_2013}, and \cite{Lehner_2020} along side our COS-Holes detections. Each observation is colored by its corresponding SMBH mass, whether that be their dynamically measured BH mass or estimated using Equation \ref{eq:mbh}. The dark grey line is a linear regression fit for the combined COS-Holes+Literature sample and is characterized by Equation \ref{eq: lr_nciv_rproj} with the shaded grey region representing a 95$\%$ confidence interval. We find there is a wide scatter in the COS-Holes+Literature radial profiles as impact parameter increases. From this relation alone, we see very little observational evidence that feedback from a SMBH heavily impacts the ionization state of its CGM.}
    \label{fig: coslit_sample}
\end{figure*} 

\begin{figure*}
    \centering
    \includegraphics[width = \textwidth]{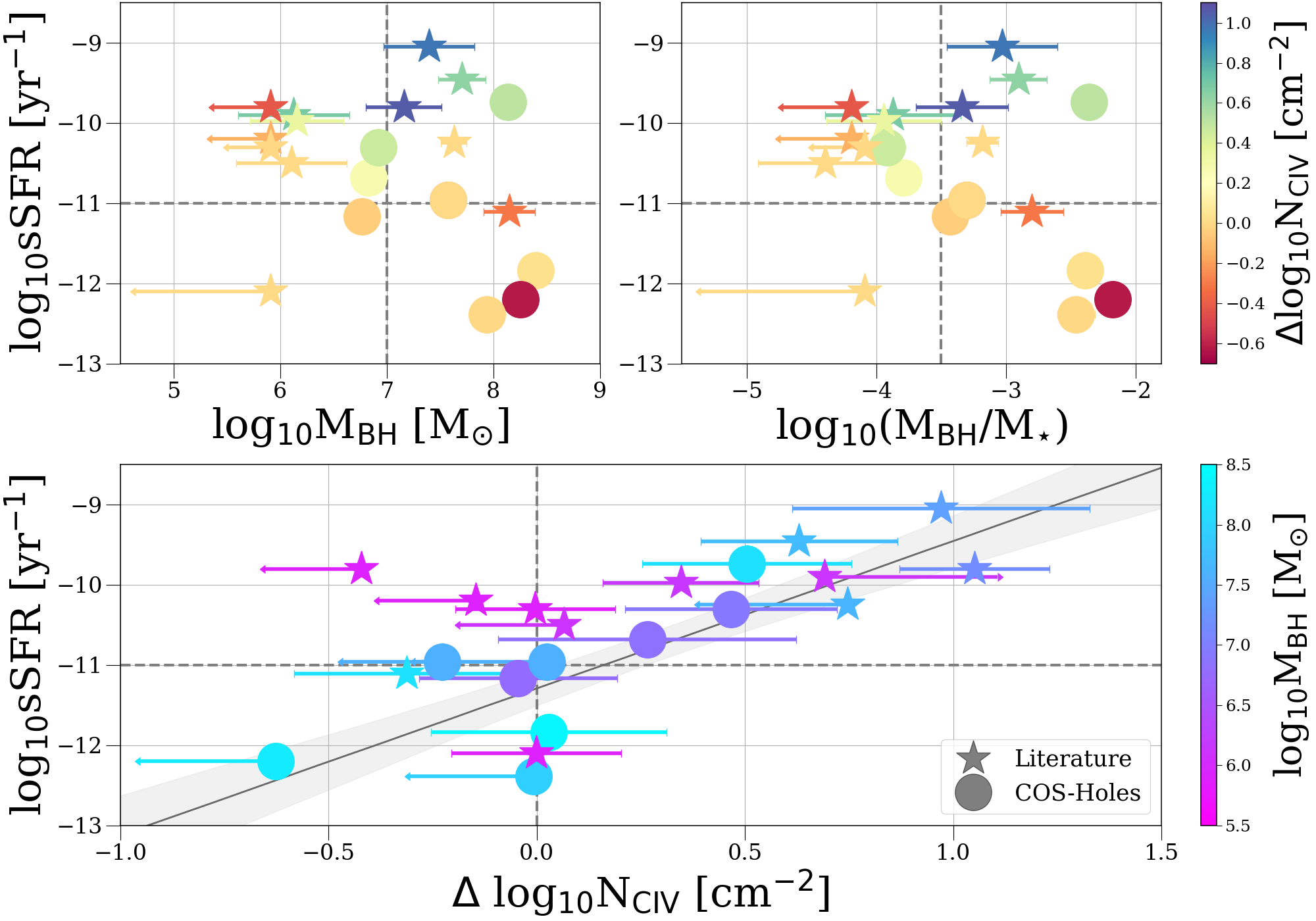}
    \caption{Investigating the role sSFR plays in driving the ionization content of the CGM across the combined sample's range of SMBH masses. Top panels: log$_{10}$sSFR as a function of log$_{10}$M$_{\rm BH}$ (left panel) and log$_{10}$M$_{\rm BH}$ normalized by the stellar mass (right panel) colored by $\Delta$log$_{10}$N$_{\rm CIV}$ (\ion{C}{IV} column density corrected for impact parameter). We present our upper limit observations that have higher column densities than those predicted by our model (Equation \ref{eq: lr_sSFR_deltaNCIV}) as having a $\Delta$log$_{10}$N$_{\rm CIV}$ of zero (yellow coloring). Stars with arrows pointing to the left represent upper limits on the log$_{10}$M$_{\rm BH}$ estimations; for a more detailed description of how these BH masses were estimated, see \S \ref{sec: additional_lit_data}. Bottom panel: log$_{10}$sSFR as a function of $\Delta$log$_{10}$N$_{\rm CIV}$ colored by log$_{10}$M$_{\rm BH}$. The dark grey line is a linear regression, similar to the fit for Figure \ref{fig: coslit_sample}, and characterized by Equation \ref{eq: lr_sSFR_deltaNCIV} with 95$\%$ confidence intervals (the shaded grey region). This strong correlation between sSFR and \ion{C}{IV} suggests that sSFR is more closely tied to the ionization state of the CGM than the BH mass.}
    \label{fig: multi_panel_fig}
\end{figure*} 

\begin{figure*}
    \centering
    \includegraphics[width = \textwidth]{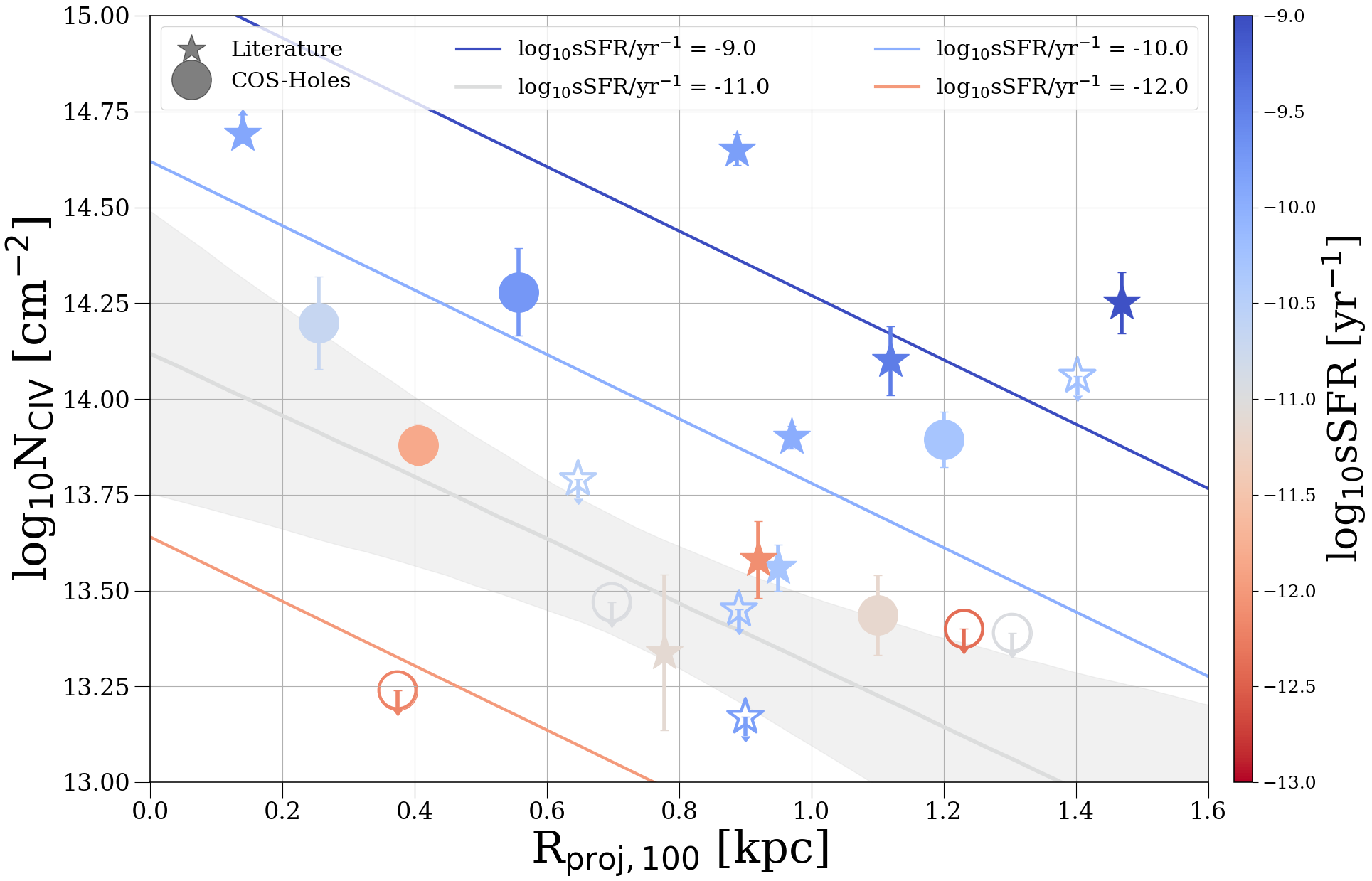}
    \caption{\ion{C}{IV} column density as a function of R$_\mathrm{proj,100}$ colored by log$_{10}$sSFR for the COS-Holes+Literature sample. The grey line is characterized by Equation \ref{eq: lr_nciv_rproj100_sSFR} and has 95$\%$ confidence intervals depicted as the grey shaded regions. The other solid lines represent an evaluation of a single fit for column density as a function of R$_\mathrm{proj,100}$ at different values of log$_{10}$sSFR. The value of log$_{10}$sSFR is denoted by their color and label. As log$_{10}$ sSFR increases from -9.0 to -12.0, the regression intercept increases and closely follows the gradient of the observations. This is further evidence showing how dominant sSFR is within the combined sample.}
    \label{fig: multivariate_fits}
\end{figure*}

%\begin{figure*}
%    \centering
%    \includegraphics[width = \textwidth]{km_plot_newLR_delta_nciv.png}%
%    \caption{Kaplan-Meier plots comparing the likelihood a particular sample (either the high or low SMBH sample) will have a particular column density. Blue represents the high mass black hole sub-sample (log$_{10}$M$_\mathrm{BH} >$ 7.0 M$_{\odot}$) while pink refers to the low mass black hole sub-sample (log$_{10}$M$_\mathrm{BH}$ $\leq$ 7.0 M$_{\odot}$). The left panel shows the Kaplan-Meier estimate for the observed column density (p-value = 0.63); the right panel shows the Kaplan-Meier estimate for the observed column density corrected by impact parameter (p-value = 0.72) using estimates from the multivariate linear regression (Equation \ref{eq: lr_nciv_rproj100_sSFR}).}
%    \label{fig: km-plot}
%\end{figure*}

In this section we examine the effects of SMBH feedback on the state of the CGM in $\sim$L$^{\star}$ galaxies by examining the observational data. The section proceeds as follows: we investigate the relationship, if any, between N$_\mathrm{C IV}$ and M$_{\rm BH}$ for the COS-Holes sample \S\ref{sec: gen_trends}; \S\ref{sec: cos+lit data and stats} describes analysis for the COS-Holes survey with the addition of a subset of published literature observations; and \S\ref{sec: carbon mass} shows our estimate of the minimum mass of carbon in the CGM of our sample. 

\subsection{COS-Holes General Trends}\label{sec: gen_trends}

In Figure \ref{fig: gal_properties}, we show our measured N$_\mathrm{C IV}$ for each line of sight versus M$_{\rm BH}$; each point is colored by its specific star formation rate (sSFR) and the size scales by impact parameter. We note that the two sightlines that intersect the halo of NGC 4258, colored in grey, are upper limits; even though the upper-limit observations are consistent with each other, studying the variations within this single halo is not possible with these two sightlines. For galaxies with a log$_{10}$M$_{\rm BH} <$ 7.0 we find a 100$\%$ covering fraction and a 33$\%$ covering fraction for galaxies with a log$_{10}$M$_{\rm BH} >$ 7.0. However, as black hole mass increases we show a large scatter of $>$1 dex in \ion{C}{IV} column density for this range of M$_{\rm BH}$. Due to this wide scatter and how small the sample size is, we suggest that there is no strong identifiable relationship between these two particular properties seen in the COS-Holes observations. 

Interestingly, there is a different correlation with another galaxy property; across the M$_\mathrm{BH}$ range of our sample, galaxies that have low observed \ion{C}{IV} column density (log$_{10}$N$_\mathrm{CIV} \lesssim$ 13.5 cm$^{-2}$) are less star forming (log$_{10}$sSFR/M$_{\star}$ $<$ -11) than galaxies with higher observed column density. Its difficult to determine if this trend is due to sample selection and if its causally or significantly correlated, since several factors could be influencing the \ion{C}{IV} content of the CGM. Even so, it raises the question, how much is the SMBH feedback really impacting observed N$_\mathrm{C IV}$ in the CGM and do other galaxy parameters, like sSFR, play a larger role in setting the ionization state? 

%Finally, the nearby galaxy NGC 4258 which has a highly accurate BH-mass measurement from megamaser kinematics \citep{miyoshi_1995}, is serendipitously intersected by two inner-CGM QSO sightlines at 70 and 130 kpc. We include both QSO targets in our final sample because it offers a rare opportunity to study subtle variations (e.g. column density, kinematics, etc.) within a single halo.
 
\subsection{COS-Holes \& Archival Data} \label{sec: cos+lit data and stats}

To examine if a relationship dependent on BH mass is observable in the radial profile, we show log$_{10}$N$_{\rm C IV}$ versus impact parameter for COS-Holes and the 12 additional literature observations (\S\ref{sec: additional_lit_data}) in Figure \ref{fig: coslit_sample}. Similar to Figure \ref{fig: gal_properties} the COS-Holes observations are represented by the circles and the literature observations are depicted as stars; all the observations are colored by their SMBH masses. 
%To continue the search for observable trends set by M$_\mathrm{BH}$, we present the relation between log$_{10}$N$_{\rm C IV}$ and impact parameter for the COS-Holes observations and the additional literature sample colored by their SMBH masses (Figure \ref{fig: coslit_sample}).
Log$_{10}$N$_\mathrm{C IV}$ weakly declines with the impact parameter within 150 kpc, a trend that has been discussed in a number of previous works \citep[e.g.][]{Bordoloi_2014}.
Although there is significant scatter ($>$ 1 dex) in the combined samples, we find that the average \ion{C}{IV} column density of the literature sample detections (average log$_{10}$N$_\mathrm{C IV, Lit}$ = 13.98$\pm$0.08 cm$^{-2}$) is comparable to the average \ion{C}{IV} column density of the COS-Holes sample (average log$_{10}$N$_\mathrm{C IV, COS-Holes}$ = 13.94$\pm$0.09 cm$^{-2}$). We find a 52$\%$ covering fraction for galaxies in the combined sample with \ion{C}{IV} absorption above log$_{10}$N$_\mathrm{C IV}$ = 13.5 cm$^{-2}$. 

To characterize the radial profile for the combined samples and get a quantifiable constraint on the observed scatter mentioned above, we fit the relation between impact parameter and column density with a linear model (the dark grey line with a dark grey line where the shaded grey region represents the 95\% confidence intervals). 
%%%% This is where I need to describe the model more then refer to this section for the basics and add more description later when we want to add more variables %%%% 
%Why I think we have the linear regression in Figure 4: To describe the radial profile? We fit for N(R) also so we could get the impact correected delta values which we talk about in the next section
%DESCRIPTION OF 1D LINEAR MODEL, CAN GO HERE OR MAYBE AFTER AN EXPLANATION OF WHY WE'RE FITTING FOR N(R)
%We fit the relation between impact parameter and column density with a linear model.
The mean column density at some $R_{\rm proj}$ is:
\begin{equation}\label{eq: lr_nciv_rproj}
    \log_{10} {\rm N_{C\,IV} / cm^{-2}} = \alpha {\rm R_{proj}/kpc} + \beta.
\end{equation}
From examining the distribution of column density measurements in narrow $R_{proj}$ ranges in \autoref{fig: coslit_sample}, it is clear that there is column density scatter beyond what can be explained by the observational uncertainties.
We model this additional scatter about the $\log_{10}$ mean column density trend as a Gaussian distribution with mean zero and standard deviation $\sigma$.
The prior probability distributions over these parameters are:
\begin{align}
\alpha &\sim \text{Normal}(0, 1^2)\\
\beta &\sim \text{Uniform}(10, 16)\\
\sigma &\sim \text{Gamma}(2, 4),
\end{align}
where the gamma distribution parameters are the shape and rate, respectively.
The priors over $\alpha$ and $\beta$ are broad but not infinite.
The prior over $\sigma$ is moderately informative: it has a mean of $1/2$ and a standard deviation of $1/\sqrt{2}\approx 0.7$.

The dataset includes three kinds of measurements which require different likelihood functions: detections, upper limits, and lower limits.
The likelihood for a detection is assumed to be a normal distribution with known mean and standard deviation. 
The result of convolution with the scatter term is also a normal distribution.
The likelihood for an upper limit is an improper uniform distribution between negative infinity and the upper limit value.
The convolution with the scatter term is the cumulative distribution function of a normal distribution with the mean column density trend and standard deviation $\sigma$. 
The likelihood and convolution with the scatter term for lower limits are similar to those of upper limits, but done in the opposite direction.

We implement this model using the {\tt\string NumPyro}\footnote{https://github.com/pyro-ppl/numpyro} probabilitistic programming library, which relies on {\tt\string JAX}\footnote{https://jax.readthedocs.io/en/latest/}, and {\tt\string ArviZ}\footnote{https://doi.org/10.5281/zenodo.10436212}.
To infer values of $\alpha$, $\beta$, and $\sigma$, we run MCMC using the No-U-Turn Sampler (NUTS) and collect samples from the posterior probability distribution.
%END DESCRIPTION
%\textcolor{red}{Using {\tt\string NumPyro}\footnote{\textcolor{red}{https://github.com/pyro-ppl/numpyro}}, which relies on {\tt\string JAX}\footnote{\textcolor{red}{https://jax.readthedocs.io/en/latest/}}, and {\tt\string ArviZ}\footnote{\textcolor{red}{https://doi.org/10.5281/zenodo.10436212}}, we use a Bayesian linear model to fit a regression to the COS-Holes+Literature sample which is described by the following equation: 
%\begin{equation}\label{eq: lr_nciv_rproj}
%    \log_{10} {\rm N_{C\,IV} / cm^{-2}} = \alpha {\rm R_{proj}/kpc} + \beta.
%\end{equation}
%Our model includes the possibility of scattering around a Gaussian model prediction; we include the width of this scatter in the fit. The model also calculates the likelihood distributions for the three types of data we observe (detections, upper limits, and lower limits). This creates a more dynamic linear regression model that is adaptable to the addition of multiple variables (e.g. analysis in \S\ref{sec:bayesian_model}). To infer values of unknown parameters in our model, we run MCMC using the No-U-Turn Sampler (NUTS) and collect samples from the target (posterior) distribution. } 
The best-fit coefficients with 95$\%$ confidence intervals are $\alpha$ = -0.0057 (-0.016, 0.0042) and $\beta$ = 14.08 (13.08, 14.93) respectively. Using this linear model we place constraints on the scatter in the combined COS-Holes+Literature sample and find that the slope of this relation is consistent with zero within error bars. %This leads us to suggest that there are no identifiable trends with M$_\mathrm{BH}$ as impact parameter increases.  %Yeah this sentence doesn;t make sense, its more about the impact parameter, maybe it means shows that you can't jsut characterize on impact parameter and need more variables? 

\subsubsection{Is sSFR Directly Linked to the \ion{C}{IV} Content of the CGM?}\label{sec: multivariate}

To investigate the possible trend suggested in Figure \ref{fig: gal_properties} between column density and other galaxy properties, we present the combined COS-Holes+Literature sample in three different ways as shown in Figure \ref{fig: multi_panel_fig}. In the top left panel we investigate sSFR as a function of black hole mass colored by $\Delta$log$_{10}$N$_\mathrm{C IV}$. These $\Delta$ log$_{10}$N$_\mathrm{C IV}$ values, which marginalize the large scatter in the radial profile (\S\ref{sec: cos+lit data and stats}, were calculated by subtracting the observed column densities by values estimated from the best fit line depicted in Figure \ref{fig: coslit_sample} (Equation \ref{eq: lr_nciv_rproj}) and is characterized by the following equation: 
\begin{equation}
    \Delta\rm log_{10}\rm N_\mathrm{C IV} = \rm log_{10} N_{\rm C IV, obs} - \rm log_{10}N_\mathrm{C IV, Eq 3}.
\end{equation}\label{delta_nciv}
We choose to color the data using these corrected column densities to normalize the observations with respect to impact parameter for the combined sample so we can focus on only four parameters: sSFR, M$_{\rm BH}$, M$_{\star}$ and N$_\mathrm{C IV}$. A similar relation is shown in the top right panel, where we present sSFR as a function of black hole mass normalized by stellar mass colored by $\Delta$ log$_{10}$N$_\mathrm{C IV}$. In both of the top panels there is a clear branching occurring at log$_{10}$M$_{\rm BH}$ $>$ 7.0 (log$_{10}$(M$_{\rm BH}$/M$_{\star}$) $\gtrsim$ -3.5); galaxies that have a log$_{10}$sSFR greater than -11.0 appear to have an excess of \ion{C}{IV} column density, while galaxies that have a log$_{10}$sSFR less than -11.0 seem to have much lower \ion{C}{IV} column densities. 

However, when we show sSFR as a function of $\Delta$log$_{10}$N$_\mathrm{C IV}$ colored by black hole mass in the bottom panel, we see that this branching falls away to reveal a correlation between sSFR and column density. We fit a linear regression to this relation, using the same method and packages as described for Equation \ref{eq: lr_nciv_rproj}, and is characterized by the following equation: 
\begin{equation}\label{eq: lr_sSFR_deltaNCIV}
    \rm log_{10} \rm sSFR / \rm yr^{-1} = \alpha \rm \Delta log_{10} \rm N_{\rm C IV} / \rm cm^{-2} + \beta.
\end{equation}
The best-fit coefficients with 95$\%$ confidence intervals are $\alpha$ = 1.8 (1.1, 2.5) and $\beta$ = -11.30 (-11.74, -10.88) respectively. Within this relation, we do not see any trends with respect to black hole mass, suggesting that the CGM properties are only loosely tied to black hole growth,if at all. In the CGM of our combined sample, the sSFR is more closely coupled with conditions in galactic atmospheres. Since CGM properties vary as a function of galaxy properties in various and complex ways, quantifying which of these is the primary driver of the ionization state is challenging. We present two methods of analysis in which we attempt to quantify the correlations seen within the combined sample, so we can further build our understanding of how CGM properties scale with galaxy properties. 

\subsubsection{Bayesian Analysis}\label{sec:bayesian_model}

\begin{table}
\centering
\caption{Multivariate Linear Regression Coefficients}
\begin{tabular}{cccc}
\hline
coeff       & mean    & $\sigma$     & 95$\%$ CI       \\
(1)         & (2)     & (3)    & (4)             \\ \hline
\multicolumn{4}{c}{Equation \ref{eq: lr_nciv_rproj100_multivariate}}             \\ \hline
$\alpha$ (slope)    & -0.87   & 0.49   & (-1.85, 0.09)   \\
$\gamma$ (log$_{10}$sSFR coeff)    & 0.57    & 0.21   & (0.16, 0.98)    \\
$\delta$ (log$_{10}$M$_{\star}$ coeff)    & -0.57   & 0.88   & (-3.20, 1.25)   \\
$\epsilon$ (log$_{10}$M$_{\rm BH}$ coeff)   & 0.26    & 0.33   & (-0.43, 0.76)   \\
$\beta$ (intercept)     & 14.04   & 0.44   & (13.14, 14.88)  \\ \hline
\multicolumn{4}{c}{Equation \ref{eq: lr_nciv_rproj100_sSFR}} \\ \hline
$\alpha$ (slope)    & -0.84   & 0.45   & (-1.73, 0.05)   \\
$\gamma$ (log$_{10}$sSFR coeff)    & 0.49    & 0.17   & (0.17, 0.84)    \\
$\beta$ (intercept)     & 14.13   & 0.38   & (13.37, 14.89)  \\ \hline
\end{tabular}
\tablecomments{Comments on Columns: (1) Coefficient; (2) mean coefficient value; (3) standard deviation; (4) 95$\%$ confidence intervals.}
\label{tab: coeffs}
\end{table}

To examine the effect of R$_\mathrm{proj}$, sSFR, M$_{\star}$, and M$_\mathrm{BH}$ on the \ion{C}{IV} column density at increasing impact parameter, we perform several multivariate linear regression analyses. Building upon the Bayseian linear regression model discussed in \S\ref{sec: cos+lit data and stats}, we include the galaxy properties mentioned above. 
We center log$_{10}$sSFR/yr$^{-1}$, log$_{10}$M$_{\star}$/M$_{\odot}$, and log$_{10}$M$_\mathrm{BH}$/M$_{\odot}$ at -11.0 (typical dividing point between star-forming and quenched galaxies), 10.5 (middle of the range for the combined sample), and 7.0 (middle of the range for our combined sample and point at which branching is seen in top panels of Figure \ref{fig: multi_panel_fig}) respectively.
This operation makes the intercept $\beta$ more interpretable but has essentially no effect on the linear relation slopes.
We also divide R$_\mathrm{proj}$ by 100 (R$_\mathrm{proj,100}$) so that all properties used in the regression would have a similar dynamic range. In addition, we acknowledge that there are some upper limit M$_\mathrm{BH}$ estimations for a few of our galaxies in the combined sample, however our multivariate linear regression treats these as detections; due to the error bars for these upper limits, this should not affect the best fit in a substantial way. The equation for our multivariate regression is described by the following:
\begin{equation}\label{eq: lr_nciv_rproj100_multivariate}
\begin{split}
    \log_{10} {\rm N_{C\,IV} / cm^{-2}} = \alpha {\rm R_{proj,100}/kpc} \\ 
    \ + \ \gamma {\rm \log_{10}(sSFR - (-11.0))/ \rm yr^{-1}} \\
    + \delta {\ log_{10} (\rm M_{\star} - 10.5)/ \rm M_{\odot}} \\
    + \ \epsilon {\log_{10} (\rm M_{\rm BH} - 7.0)/ \rm M_{\odot}} + \beta.
\end{split}
\end{equation}

Looking at the best fit mean coefficients and their standard deviations, given in Table \ref{tab: coeffs}, we can immediately rule out strong correlations with log$_{10}$M$_{\star}$ and log$_{10}$M$_\mathrm{BH}$. Both of these parameters' mean coefficients have significance less than 1$\sigma$ and are unlikely to be driving the regression or impacting the ionization. The most dominant galaxy parameter, which is greater than zero with a significance of nearly 3$\sigma$, is log$_{10}$sSFR and is likely the main driving component in this relation. To test this assertion, we run a similar multivariate linear regression, but only including R$_\mathrm{proj}$ and sSFR, where the coefficients and statistics are shown in the bottom half of Table \ref{tab: coeffs} and is characterized by the following equation: 
\begin{equation}\label{eq: lr_nciv_rproj100_sSFR}
\begin{split}
    \log_{10} {\rm N_{C\,IV} / cm^{-2}} = \alpha {\rm R_{proj,100}/kpc} \\ 
    \ + \ \gamma {\rm \log_{10}(sSFR - (-11.0))/ \rm yr^{-1}} + \beta.
\end{split}
\end{equation}

This relation is shown in Figure \ref{fig: multivariate_fits} as the grey line with its 95$\%$ confidence intervals depicted as the grey shaded regions. The coefficients and standard deviations for the new regression remain essentially the same with log$_{10}$M$_{\star} $ and log$_{10}$M$_\mathrm{BH}$ removed, confirming that they are subdominant in setting the ionization content of the CGM of our combined sample. This is further demonstrated by the other multivariate linear regressions included in Figure \ref{fig: multivariate_fits} where log$_{10}$N$_\mathrm{C IV}$ is plotted as a function of R$_\mathrm{proj,100}$ colored by log$_{10}$sSFR; each line evaluates a single fit for column density as a function of R$_\mathrm{proj,100}$ at different values of log$_{10}$sSFR. As the value of sSFR increases, the regression intercept increases and changes the relation substantially, and follows the gradient of the observations, showing how dominant sSFR is within the combined sample. Based on the strong correlation seen in Figure \ref{fig: multi_panel_fig}, we suggest that sSFR of a galaxy is directly linked to the \ion{C}{IV} content of the CGM.

\subsubsection{Frequentist Analysis}\label{sec: freq_analysis}
% Need to talk more about what Kendall's tau is and why we use it, and then discuss what the values of t and p mean in the context here. 
We also investigate the relationship between log$_{10}$N$_{\rm C IV}$, log$_{10}$M$_\mathrm{BH}$, and log$_{10}$sSFR using frequentist non-parameteric tests.
We first use Kendall's rank correlation test (also known as a $\tau$ test) to check for a dependence between column density and black hole mass.
Specifically, we use the {\tt cenken} function in the {\tt NADA} R package \citep{lee_2020}, which can handle censoring (i.e., non-detections).
The test $p$-value for a correlation between log$_{10}$N$_{\rm C IV}$ and log$_{10}$M$_\mathrm{BH}$ is greater than 0.05, indicating no evidence for a dependence.
The test $p$-value for log$_{10}$N$_{\rm C IV}$ and log$_{10}$sSFR is 0.017, which would correspond to about $2.3\sigma$ for a normal distribution: a somewhat significant correlation.

%by computing the Kendall's $\tau$ using the {\tt cenken}\footnote{\textcolor{red}{https://www.rdocumentation.org/packages/NADA/versions/1.6-1.1/topics/cenken}} function in the . This function computes the Akritas-Theil-Sen non parametric line, with the Turnbull estimate for the intercept. {\tt cenken} computes the slope, intercept, $\tau$ correlation coefficient, and p-value for singly or doubly censored data. The Tau correlation coefficient, $\tau$, returns a value from 0 to 1 where 0 corresponds to no relationship and 1 corresponds to a perfect relationship. Kendall's rank correlation is very similar to Spearman's rank correlation test, however it is the preferred method to test this data set since the combined COS-Holes+Literature sample is on the smaller side. The corresponding $\tau$ and p-value for log$_{10}$N$_{\rm C IV}$ and log$_{10}$M$_\mathrm{BH}$ is 0 and 1 respectively. Under the null hypothesis of independence of X and Y, log$_{10}$N$_{\rm C IV}$ and M$_\mathrm{BH}$ are distinct. We also perform a Kendall's $\tau$ test for log$_{10}$N$_{\rm C IV}$ and log$_{10}$sSFR and find a $\tau$ of 0.371 and a p-value of 0.017. For this properties, we see a stronger correlation, but with high significance.}

We repeat these tests on an impact-parameter-trend-corrected column density, $\Delta$log$_{10}$N$_{\rm C IV}$.
The {\tt cenken} function in {\tt NADA} provides the Akritas-Theil-Sen estimator for the slope and the Turnbull estimator for the intercept of the linear relation between two variables. 
We use this functionality to determine the linear relation between log$_{10}$N$_{\rm C IV}$ and R$_{\rm proj}$, use that linear relation to get a predicted log$_{10}$N$_{\rm C IV}$ for each observation, and subtract that from the observed value to get $\Delta$log$_{10}$N$_{\rm C IV}$.
We then run Kendall's rank correlation tests from the previous paragraph replacing log$_{10}$N$_{\rm C IV}$ with $\Delta$log$_{10}$N$_{\rm C IV}$.

For $\Delta$log$_{10}$N$_{\rm C IV}$ and log$_{10}$M$_\mathrm{BH}$, we find a $\tau$ of 0.33 and a p-value of 0.85; for $\Delta$log$_{10}$N$_{\rm C IV}$ and log$_{10}$sSFR we find a $\tau$ and p-value of 0.40 and 0.0094, respectively. The results for N$_{\rm C IV}$ and $\Delta$log$_{10}$N$_{\rm C IV}$ are comparable and consistent with results from \S\ref{sec:bayesian_model}, where we see no correlation between \ion{C}{IV} column density and black hole mass, but there is a possible correlation with sSFR. This further supports our earlier conclusion that the SMBH does not have as significant effect on the state of the CGM as predicted, and that sSFR, with the stronger correlation, is more directly linked with the \ion{C}{IV} content of the CGM.

%\textcolor{red}{To calculate a similar $\Delta$log$_{10}$N$_{\rm C IV}$ as that found in \S\ref{sec: multivariate}, we compute the Kendall's $\tau$ for log$_{10}$N$_{\rm C IV}$ and R$_{\rm proj}$; we plugged the slope ($\alpha$) and intercept ($\beta$) values, -0.0039 and 13.94 respectfully, into Equation \ref{eq: lr_nciv_rproj} to get predicted log$_{10}$N$_{\rm C IV}$. We then plug these predicted values (log$_{10}$N$_{\rm CIV}$) into Equation \ref{delta_nciv}. It is these new Kendall's $\tau$ calculated $\Delta$log$_{10}$N$_{\rm C IV}$ values that we use find the Kendall's $\tau$ correlation coefficient and p value for log$_{10}$M$_\mathrm{BH}$ and log$_{10}$sSFR. For $\Delta$log$_{10}$N$_{\rm C IV}$ and log$_{10}$M$_\mathrm{BH}$, we find a $\tau$ of 0.33 and a p-value of 0.85; for $\Delta$log$_{10}$N$_{\rm C IV}$ and log$_{10}$sSFR we find a $\tau$ and p-value of 0.40 and 0.0094 respectively. The results for N$_{\rm C IV}$ and $\Delta$log$_{10}$N$_{\rm C IV}$ are comparable and consistent with results from \S\ref{sec:bayesian_model}, where we see no correlation between \ion{C}{IV} column density and black hole mass, but there is a possible correlation with sSFR. This further supports our earlier conclusion that the SMBH does not have as significant effect on the state of the CGM as predicted, and that sSFR, with the stronger correlation, may be the more dominant quantity.}

\subsection{Minimum Mass of Carbon in the CGM} \label{sec: carbon mass}

Following the methods used in \cite{Bordoloi_2014}, we estimate the carbon mass in the CGM around our sample of $\sim$L$^{\star}$ galaxies. \cite{Bordoloi_2014} obtained their upper limit on carbon mass (M$_{\rm carbon}$) by applying a conservative ionization correction (assuming $f_{\rm C IV}$ = 0.3) to their values of \ion{C}{IV} mass (M$_{\rm C IV}$); these estimates were made by assuming ionization equilibrium and including collisional- and photo-ionization using the CLOUDY photoionization code \citep{ferland_1998, Chatzikos_2023}. 
The minimum carbon mass can be written as: 
\begin{equation}
\begin{split}
    M_{ \rm carbon} \gtrsim 1.12 \times 10^6 \rm \: M_{\odot} \left(\frac{N_{\rm C IV,\:mean}}{10^{14} \: \rm cm^{-2}}\right) \\
    \times \left(\frac{ \rm R_{\rm proj}}{110\: \rm kpc}\right){^2} \times \left(\frac{0.3}{f_{\rm CIV}}\right)\times \rm C_{f}. 
\end{split}
\end{equation}
This calculation assumes that these galaxies conform to global stellar metallicity relations and the gas-phase mass-metallicity relation. 

Inserting typical values for the COS-Holes sample, R$_{\rm proj} =$ 140 kpc, covering fraction C$_{\rm f}$\footnote{All C$_{\rm f}$ values are determined above log$_{10}$N$_{\rm CIV}$ = 13.5 cm$^{-2}$} 44$\%$, and mean column density of our detections N$_{\rm C IV, mean} =$ $10^{13.94}$ cm$^{-2}$, we get a minimum mass of M$_\mathrm{carbon}$/M$_{\odot} =$  7.41 $\times$ $10^{5}$. This is about a factor of 2.5 lower than the M$_{\rm carbon}$ value presented in \cite{Bordoloi_2014} found for both star-forming and non star-forming dwarf galaxies using Voigt profile fitted N$_{\rm C IV}$ (1.9$\times 10^{6}$ M$_{\odot}$). Comparing this value to the total carbon mass in the ISM of L$^{\star}$ galaxies, we find that our minimum carbon mass is approximately a factor of three lower \citep[e.g.][]{Peeples_2014}. 

We repeated this calculation for the whole COS-Holes+Literature sample (R$_{\rm proj} =$ 150 kpc, covering fraction C$_{\rm f} =$ 52$\%$, and mean column density of our detections N$_{\rm C IV, mean} =$ $10^{13.98}$ cm$^{-2}$) and just the star-forming galaxies (log$_{10}$sSFR/yr$^{-1}$ $\geq$ -11) in the COS-Holes+Literature sample (R$_{\rm proj} =$ 150 kpc, covering fraction C$_{\rm f} =$ 60$\%$, and mean column density of our detections N$_{\rm C IV, mean} =$ $10^{14.17}$ cm$^{-2}$) to find a minimum mass of carbon for both samples to be 1.11 $\times$ $10^{6}$ M$_{\odot}$ and 1.98 $\times$ $10^{6}$ M$_{\odot}$ respectively. These values are comparable to those reported in \citet{Bordoloi_2014}; the combined COS-Holes + Literature sample has a carbon mass 1.7 times lower than the value reported for both star-forming and non-star forming galaxies (1.9$\times 10^{6}$ M$_{\odot}$), while the the star-forming-galaxy-only combined sample has a carbon mass only a factor of 1.3 lower (2.6$\times 10^{6}$ M$_{\odot}$). 

\section{Simulation Results}\label{sec: sim_results}
In this section we compare the observational results of the combined COS-Holes+Literature sample to results from simulations. The section proceeds as follows: we describe the three simulations in \S\ref{sec: EAGLE},\ref{sec: R25}, and \ref{sec: TNG} and discuss their various \ion{C}{IV} column density predictions in \S\ref{sec: theory_pred}; we compare the combined COS-Holes+Literature sample to simulated values from our three simulations in \S\ref{sec: comp_to_sim}; and in \S\ref{sec: smohalos} we mimic the COS-Holes survey across all three simulations used in \S\ref{sec: comp_to_sim}.  

\subsection{Simulation Descriptions}\label{sec: sim_descriptions}

Here, we briefly describe the three simulations to which we will compare our results. For more details on these well-known and widely used simulations, we refer the reader to the citations referenced throughout these sections. 

\subsubsection{Evolution and Assembly of GaLaxies and their Environments (EAGLE)} \label{sec: EAGLE}

We compare our observations to a sample galaxies from the EAGLE  main `Reference' simulation volume (Ref-L100N1504), originally published in \citep{schaye_2015, crain_2015}. This (100 comoving Mpc)$^{3}$, $1504^{3}$ dark matter and smooth particle hydrodynamic (SPH) particle run uses a heavily modified version of the \textit{N}-body code {\sc GADGET} \citep{Springel_2005}. EAGLE applies the pressure-entropy SPH formulation from \citep{Hopkins_2013}, extra parameters referred to as {\sc ANARCHY} \citep{schaye_2015}, and assumes cosmology from the  \citet{PCXI_2013} ($\Omega_{m}$ = 0.307, $\Omega_{\Lambda}$ = 0.693, $H_{0}$ = 67.77 km $\rm s^{-1}~Mpc^{-1}$). The initial dark matter and SPH particle masses are 9.7$\times 10^{6} M_{\odot}$ and 1.8$\times 10^{6} M_{\odot}$ respectively. 

EAGLE implements the following subgrid physics modules: radiative cooling \citep{Wiesrsma_Schaye_Smith_2009a}, star formation \citep{Schaye_dalla_vecchia_2008}, stellar evolution and metal enrichment \citep{Wiersma_2009b}, stellar feedback \citep{dalla_vecchia_and_schaye_2012}, BH formation accretion, feedback \citep{booth_and_schaye_2009, rosas-guevara_2015}. In regards to the black holes, EAGLE follows BHs from seed black hole particles with mass $10^{5} \textit{h}^{-1} M_{\odot}$ (where \textit{h}=0.6777) placed at the center of every halo that exceeds a mass of $10^{10} \textit{h}^{-1} M_{\odot}$. The BH particles grow via \cite{Bondi_Hoyle_1944} gas accretion as well as as mergers with other BHs using the prescription derived by \citet{booth_and_schaye_2009}. Stellar and BH feedback operate via thermal prescriptions that heat surrounding gas to $10^{7.5}\;$K and $10^{8.5}\;$K, respectively. Further information on these processes and their calibrations are described in \citet{crain_2015}.   
 
The BH energy feedback rate is calculated by tracking the accretion rate onto BHs using the efficiency
\begin{equation}
    \dot{E}_{BH} = \frac{\epsilon_f \epsilon_r}{1 - \epsilon_r} \dot{M}_{BH} c^2
\end{equation}\label{eqn:EdotBH}
where $\epsilon_{r}=0.1$ is the radiative efficiency of the accretion disk and $\epsilon_{f}=0.15$ is the thermal feedback efficiency. The combined efficiency prefactors result in a total BH efficiency of 1.67\% of the rest mass energy accreted onto the BH. The BH feedback operates via a single-mode thermal prescription that heats surround gas particles to $10^{8.5}\;$K. Energy is stored until a gas particle or particles can be heated to this temperature to ensure that the feedback is numerically efficient.  

\subsubsection{Romulus25}\label{sec: R25}

We also compare our observations to galaxies from the cosmological volume {\sc Romulus25} \citep[hereafter R25;][]{tremmel_2017}. The {\sc Romulus25} (25 Mpc)$^3$ volume was run with a $\Lambda$CDM cosmology from \cite{Plank2016} with $\Omega_0$ = 0.3086, $\Lambda$ = 0.6914, $h$ = 0.67, $\sigma_{8}$ = 0.77. R25 is run using the smooth particle hydrodynamics code, Charm N-body GrAvity Solver \citep[ChaNGa][]{menon_2015}. ChaNGa adopts the same models as {\sc GASOLINE} \citep{wadsley04,wadsley17}, including the following physical prescriptions: cosmic UV background \citep{haardt&madau2012}, star formation (using a \citealp{kroupa2001} IMF), and blastwave supernova feedback \citep{ostriker&mckee1988,stinson2012}, which includes both SNIa and SNII \citep{thielemann1986,woosley&weaver1995}. R25 has a Plummer softening length of 250 pc and a mass resolution of $3.4 \times 10^{5} M_{\odot}$ and $2.1 \times 10^{5} M_{\odot}$ for dark matter and gas, respectively.

%BEN!! look at line 982 on page 16, and tell us if this is accurate
R25 includes independent subgrid physics modules for the black hole formation, accretion, feedback, and dynamical friction as introduced in \citet{tremmel_2017}.  Unlike other simulations that use a threshold halo mass to initiate a BH seeding, BH seed particles with initial mass $10^6 M_{\odot}$ are required to form in dense, extremely low metallicity gas to better model SMBH populations across galaxy mass scales as described in as described in \citet{tremmel_2017} \S5.1. The BH accretion utilizes a modified Bondi-Hoyle prescription that considers angular momentum support from nearby gas, resulting in a different physical growth model which uses fewer free parameters. Thermal feedback energy from the BH is imparted onto the 32 nearest gas particles every time step in the form
\begin{equation}
E_{BH} = \epsilon_r \epsilon_f \dot{M}_{BH} c^2 dt,
\end{equation}
where the radiative and feedback efficiencies are $\epsilon_r$ = 0.1 and $\epsilon_f$ = 0.02, resulting in a total rest mass energy efficiency of 0.2\%.  The energy is released every timestep $dt$ in contrast to the EAGLE prescription that stores energy until a surrounding gas particle can be heated to its threshold energy.  
%$dt$ indicates the timestep over which the accretion is assumed to remain constant. 
%For additional details on the BH prescriptions, see \cite{tremmel_2017}. 

To calculate the ion column densities from the Romulus25 galaxy suite, we use the public analysis software, {\tt\string Pynbody}\footnote{https://pynbody.github.io/pynbody/index.html} \citep{pynbody}. Oxygen and metal enrichment from SN and winds is traced throughout the integration of the simulation. Then, ionization states are post-processed, assuming optically thin conditions, collicaiton ionization equilibrium, and a \citet{haardt_manau_2012} UV radiation field. Finally, we create models using the CLOUDY software package \citep{Stinson_2012, Ferland_2013} for varying redshift, temperature, and density to calculate individual ion fractions for each gas particle in every simulated galaxy. %Added parapgraph from Nicole 

\subsubsection{IllustrisTNG} \label{sec: TNG}

The last cosmological simulation that we compare our galaxies to is the IllustrisTNG simulation, hereafter TNG. The TNG simulations were run with the moving mesh code { \sc AREPO} \citep{springel10}, including a magnetic hydrodynamic (MHD) solver that is seeded with the cosmologically motivated initial conditions and then follows the magnetic field self-consistently \citep{pakmor13}. TNG utilizes values consistent with the \cite{Plank2016} results ($\Omega_{m}$ = 0.3089, $\Omega_{\Lambda}$ = 0.6911, $h$ = 0.6774).
%which uses Newtonian self-gravity solved within an expanding Universe.  

The (100 Mpc)$^3$ TNG simulation, also known as TNG100, is the middle of the three TNG volume series, providing a balance of volume and resolution, particularly for intermediate mass halos. The simulation implements several subgrid processes as part of the TNG model including primordial/metal-line radiative cooling on microphysical scales, star formation based on a two-phase subgrid ISM model, evolution of stellar populations and the expected chemical enrichment/mass loss, galactic-scale outflows from energy-driven, kinetic winds from stellar feedback, and the seeding, growth and feedback from BHs \citep{pillepich_2018}.  Black holes seeds with mass $8\times 10^{5} \textit{h}^{-1} M_{\odot}$ are initially seeded in halos of $5\times 10^{10} \textit{h}^{-1} M_{\odot}$.  

The black hole prescriptions are introduced in \citet{weinberger17}.  Their dual-model AGN model, incorporates a `thermal' mode injects thermal energy at high Eddington accretion rates and `kinetic' mode injects kinetic energy at low Edditington accretion rates.  The feedback efficiency for the thermal mode uses $\epsilon_{r}=0.1$ $\epsilon_{f}= 0.2$ from Equation \ref{eqn:EdotBH} for a total rest-mass accretion efficiency of $\sim 2$\%, which is distributed thermally over surrounding gas cells. The kinetic mode injects kinetic ``pulses'' at a total efficiency that can achieve $\sim 20$\% of the accreted rest-mass energy (via the physical mechanism of \citet{blandfordznajek1977}). Randomly oriented, jet-like pulsed feedback events apply energy directionally, imparting significant momentum stored across multiple timesteps to avoid dependence on the simulation timestep. This low accretion rate, kinetic mode generally dominates for late growing SMBHs above a threshold mass of $M_{\rm BH}\approx 10^{8.1}\, M_{\odot}$ \citep{Davies_2020, terrazas20}.   %including a steep transition of galaxy and gaseous halo properties
%For additional IllustrisTNG descriptions and details, see \citet{pillepich_2018} and \cite{Nelson_2018, Nelson_2019b}. %and \cite{Oppenheimer_2021}.  

\subsection{Simulation Predictions} \label{sec: theory_pred}

Using EAGLE, \cite{Oppenheimer_2020} asserts that the efficiency with which a SMBH feedback energy is coupled to the CGM is critical for understanding the process of secular galaxy formation. This can be thought about in a three step pathway: (1) the formation of the halo, (2) the rapid growth of the BH, and (3) the lifting by AGN feedback of the baryonic halo reducing the supply of fuel for star formation. The last point is evidenced by a decrease in heavy metals in the CGM (such as \ion{C}{IV}). These results were heavily based on the work by \cite{Davies_2019} that published an inverse correlation between M$_\mathrm{BH}$ and $\textit{f}_\mathrm{CGM}$ in EAGLE, suggesting a link between the BH and the removal of a significant portion of gas from the halo, essentially reducing CGM accretion and galactic star formation. \citet{Oppenheimer_2020} find that the galaxy BH mass is generally a good indicator of its past feedback history at masses above log$_{10}$M$_{\rm BH}$/M$_{\odot}\sim $ 7.0. They used high-cadence snapshot outputs from the EAGLE simulation to determine that significant AGN episodes directly lift the CGM and significantly reduce (in some cases quenching) star formation on a $<100$ Myr timescale. Ion tracers including \ion{C}{IV} take longer ($0.5-2.5$ Gyr) to respond, but this sequence generally happens at $z>1$ for L$^{\star}$  galaxies meaning that this ion is an indicator of CGM gas content by $z=0$. 

TNG shows a dramatic decline in the covering factor of \ion{O}{VI} from star-forming to quenched galaxies, as presented by \citet{Nelson_2018_oxygen}.  \citet{Davies_2020} explored both TNG and EAGLE to determine how CGM mass depends on BH mass, finding that the BH feedback energy released during the low-Eddington kinetic mode in TNG is most strongly anti-correlated with the gas content of halos in the mass range corresponding to our samples explored here.  Hence, the BH mass itself is not directly deterministic for baryon lifting in TNG, but the energy released during the kinetic mode \citep{Davies_2020,voit2023}.
%This paper and \citet{voit2023} demonstrated that it is not as much the BH mass that is deterministic for baryon lifting in TNG, but the energy released during the kinetic mode that is most deterministic for the CGM.  
This manifests itself in a strong anti-correlation between BH and CGM mass, but for BH masses that are above $10^{8}\;M_{\odot}$ where kinetic mode (and therefore baryon lifting) operate in TNG.  Taking these results together, BH feedback drives the results of star-forming (less massive BHs) galaxies having higher \ion{O}{VI} column density compared to quiescent (more massive BHs) galaxies \citep{Nelson_2018_oxygen}.  Hence, the driving force depleting ionized oxygen (and mostly likely \ion{C}{IV} as well) is the ejection of mass from the CGM due to black hole feedback. 
%For instance, as the total energy injected in the kinetic wind by a black hole increases (or move from the star-forming to quenched regime) the halo gas becomes hotter, less dense, and less metal enriched.

In contrast, \cite{sanchez_2019} uses R25 to examine the effects of SMBH feedback and star formation history (SFH) on the column densities of \ion{O}{VI} in the CGM of galaxies. They determine that the host galaxy's SMBH transports metal-rich gas out of the galaxy disk where metals are formed and propagates it into the CGM. From these results they posit that galaxies with lower mass BHs (which have experienced less accretion and therefore less feedback) are likely to have a lower metallicity CGM and vice versa for galaxies with higher mass BHs leading them to have more metal-enriched material in their host CGM. Therefore, SMBH feedback impacts the total metal mass in the CGM (but not the total gas mass) and may play a critical role in galaxy quenching \citep{sanchez_2021}. In a follow up paper, \citet{Sanchez_2023} measure the SMBH masses and CGM metal content from a sample of galaxies from R25. They find higher CGM metal fractions in galaxies with more massive black holes (compared to their host's stellar dispersion). In further contrast to EAGLE and TNG, \citet{Sanchez_2023} find no correlation between $M_\mathrm{BH}$ and $\textit{f}_{CGM}$ indicating that in their galaxies, the SMBH's influence is more local, impacting the galaxy's disk and enriching the CGM without evacuating gas from the halo.

All of these simulations support the idea that SMBHs transport gas and metals into the CGM of the host galaxy. However, they differ on their predictions of how the mass of the SMBH regulates the amount of \ion{C}{IV} present in the CGM. The COS-Holes survey provides the opportunity to constrain these feedback processes by comparing EAGLE, R25, and TNG predictions. \S \ref{sec: comp_to_sim} compares these theoretical SMBH feedback prescriptions with our observations. 

\subsection{Comparison to Simulations}\label{sec: comp_to_sim}

\begin{figure}%
    \centering
    \includegraphics[width=0.99\columnwidth]{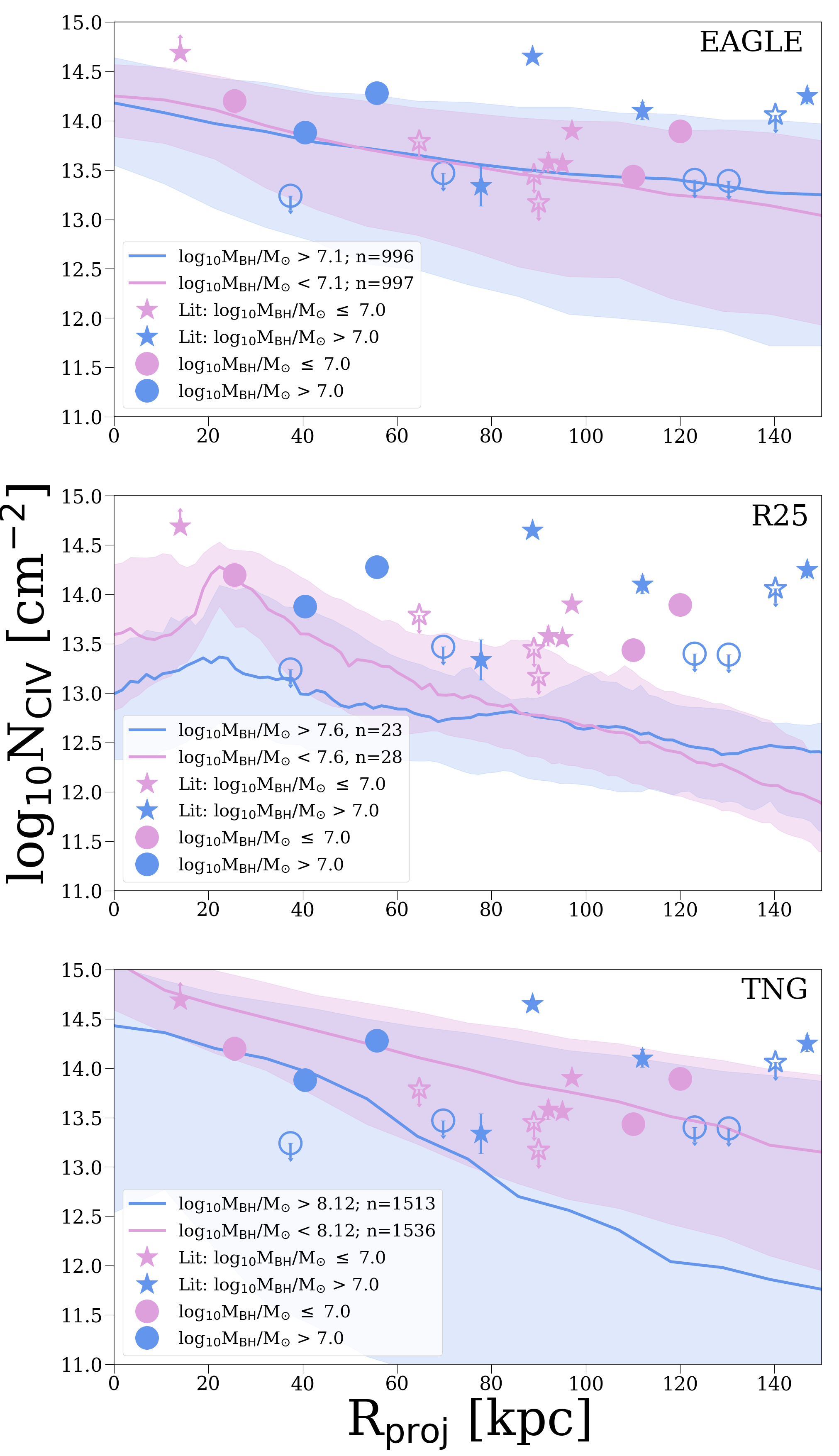}%
    \caption{Column densities of the combined COS-Holes+Literature sample versus impact parameter compared to predictions from the EAGLE (top panel), R25 (middle panel) and TNG (bottom panel) simulations. Blue represents the ``high" mass black hole sample (log$_{10}$M$_\mathrm{BH} >$ 7.0 $M_{\odot}$) while pink refers to the ``low" mass black hole sample (log$_{10}$M$_\mathrm{BH}$ $\leq$ 7.0 M$_{\odot}$). The corresponding blue and pink lines are the median \ion{C}{IV} radial profile predictions from each simulation (also split based on black hole mass) each with 16-84$\%$ confidence spreads represented as the shaded region around each prediction. Like previous figures, unfilled markers represent an upper limit for that observation. Combined sample column densities agree reasonably with predictions from EAGLE and TNG and lie above the R25 predictions. }%
    \label{fig: onepanel_sim}
\end{figure}  

We begin our comparison to simulations by showing the entire observed sample (COS-Holes+Literature; log$_{10}$M$_\mathrm{BH}$ = 5.91 - 8.4 M$_{\odot}$) to the three simulations. In Figure \ref{fig: onepanel_sim}, we split the observations into two bins divided by log$_{10}$M$_\mathrm{BH}$ = 7.0, a ``low''-M$_\mathrm{BH}$ sample (10 observations, log$_{10}$M$_\mathrm{BH}$ $\leq$ 7.0) and a ``high''-M$_\mathrm{BH}$ sample (11 observations, log$_{10}$M$_\mathrm{BH}$ $>$ 7.0). This split is strategically made to take advantage of the range of BH masses that make up the combined COS-Holes+Literature sample, but to also investigate the question, ``do galaxies with similar stellar masses, but hosting differing SMBH masses show different CGM metal contents?"

We first make a broad-brush type comparison between data and simulations by choosing all galaxies with stellar masses between log$_{10}$M$_{\star}$/M$_{\odot}$ = $10^{10} - 10^{11}$ in the three simulations, and dividing the samples into two M$_\mathrm{BH}$ bins in Fig. \ref{fig: onepanel_sim}. This means that the split black hole mass is different amongst the simulations as they have different M$_\mathrm{BH}$ distributions as discussed in \S\ref{sec: sim_descriptions}. In EAGLE there are 1993 central galaxies at \textit{z} $=$ 0.00 with a dividing M$_\mathrm{BH}$=$10^{7.10}$\;M$_{\odot}$. In R25 there are 52 central galaxies at \textit{z} $=$ 0.05 with a dividing M$_\mathrm{BH}$=$10^{7.6}$\;M$_{\odot}$. In TNG there are 3049 central galaxies at \textit{z} $=$ 0.00 with a dividing M$_\mathrm{BH}$=$10^{8.12}$\;M$_{\odot}$. The distribution of black hole masses for TNG is narrower and more massive than the distributions for EAGLE and R25; see \S\ref{sec: smohalos} for more details. For EAGLE and TNG the \ion{C}{IV} column density radial profiles were calculated using a projection along the $z$ axis with a total depth of 2 Mpc, while for R25, they were averaged down to R$_\mathrm{200c}$. We choose to plot the simulations predicted column densities versus the true projected impact parameter,  R$_\mathrm{proj}$ [kpc], since it is a more direct measurement that does not rely on estimations from an indirectly observable property (like using R$_\mathrm{proj}$/R$_\mathrm{200c}$ which uses dark matter mass). 

The radial profiles shown in the top panel of Figure \ref{fig: onepanel_sim} show that the EAGLE simulation results are in reasonable agreement with the COS-Holes$+$Literature absorption observations. Interestingly that there is no discernible difference in the low and high BH mass samples (average log$_{10}$N$_\mathrm{C IV}$ = 13.6 cm$^{-2}$ for both samples). We note that \citet{Oppenheimer_2020} predicted an anti-correlation between \ion{C}{IV} and M$_\mathrm{BH}$, but we do not see such a correlation here. This likely is a result of the slightly larger stellar mass range probed by our COS-Holes galaxy sample compared to that of Oppenheimer et al. 2020 \citep[$10^{10-11}$ M$_{\odot}$ vs. $10^{10.2-10.7}$ M$_{\odot}$;][]{Oppenheimer_2020}. Additionally,  they found a measurable difference in the reduction of \ion{C}{IV} column densities for only the highest {\it{quartile}} of BH masses, whereas here, we have divided the sample into two, leading us to have similar radial profiles for the split samples. 

In the middle panel of Figure \ref{fig: onepanel_sim} where the observations are compared to N$_\mathrm{C IV}$ values from R25, we see that the mean predicted column density (log$_{10}$N$_\mathrm{C IV}$ = 12.8 cm$^{-2}$) is on average $\sim$1 dex below our combined sample; thus, there is little agreement between the combined sample and those predicted by R25. We see more of a difference between the low and high mass sample R25 predictions than EAGLE until $\sim$90 kpc where the two samples become more indiscernible. These results are comparable to averaged radial profiles presented in Fig 14 of \citet{Sanchez_2023} (log$_{10}$N$_\mathrm{C IV}$ vs log$_{10}$(R$_\mathrm{proj}$/R$_\mathrm{200c}$)) at the same BH mass split but are under predicting the \ion{C}{IV} column density observed.

For TNG, in the last panel of Figure \ref{fig: onepanel_sim}, we see the largest differentiation between the low and high samples (average log$_{10}$N$_\mathrm{C IV}$ = 14.0 and 13.1 cm$^{-2}$ respectively), which reasonably overlaps with the \ion{C}{IV} column densities for the combined sample. Interestingly, TNG has a much larger spread for their high-mass sample than any other simulation and sample.  While we do see a population of low \ion{C}{IV} values for high M$_\mathrm{BH}$, there is also a population with high \ion{C}{IV} for high M$_\mathrm{BH}$.  A fuller investigation determining why there is such a spread in \ion{C}{IV} for high black hole masses is beyond the scope of this paper.

We note these general trends between each of the simulations, but we cannot directly compare them to the COS-Holes survey without reproducing the observed survey, which we now do in \S\ref{sec: smohalos}.  

\subsubsection{Mocking up the COS-Holes Survey} \label{sec: smohalos}

We create a mock up of the COS-Holes Survey of the nine sight lines by using the SMOHALOS (Simulation Mocker Of Hubble Absorption Line Observational Surveys) used first in \citet{oppenheimer_2016} across the three simulations. In our implementation here, SMOHALOS matches the impact parameter, stellar mass, and SFR of observed galaxies using a selection of central galaxies taken from $z=0.00$ simulation outputs. We also attempt to match black hole mass, but only divide the sample into two using a black hole mass split, M$_{\rm BH, {\rm split}}$ that is defined differently for each simulation based on the distribution of simulated M$_\mathrm{BH}$ values. In SMOHALOS, the 1-$\sigma$ range of BH masses in EAGLE spans $10^{6.65}-10^{7.94} {\rm M}_{\odot}$, in R25 spans $10^{7.27}-10^{8.35} {\rm M}_{\odot}$, and in TNG spans $10^{7.94}-10^{8.43} {\rm M}_{\odot}$. Briefly, we choose a sight line by selecting a random pixel in a \ion{C}{IV} column density map at an impact parameter within 5 kpc of an observed sight line around a simulated galaxy that matches the observed galaxy. We use a projection along the $z$ axis with a total depth of 2 Mpc. SMOHALOS selects a matching simulated galaxy by taking the observed galaxy values and adding a random error assuming a Gaussian dispersion, then finding the simulated galaxy that best fits the observed galaxy's dispersed values.  We assume dispersions of 0.3 dex to M$_{\star}$ and 0.5 dex to SFR, therefore ensuring we are selecting galaxies that are similar to the COS-Holes sample but have a random scatter based on reasonable uncertainties in stellar masses and star formation rates.  

We run SMOHALOS for 100 realizations, reporting the results in Table \ref{tab:smohalos}.  We first report the median \ion{C}{IV} column density from our observations noting the large range on the high M$_{\rm BH}$ sample due to upper limits (cf. Fig. \ref{fig: gal_properties}) indicating the uncertainty in which the sample has more \ion{C}{IV}. For the simulations, we are not limited by upper limits, therefore we present the equivalent of noiseless column densities for the median and $1-\sigma$ spread in their distribution.  

\begin{table}
\centering
\caption{SMOHALOS Simulation \ion{C}{IV} Comparison}
\begin{tabular}{cccc}
\hline
Dataset   & M$_\mathrm{BH}$(split) & Low M$_\mathrm{BH}$     & High M$_\mathrm{BH}$ \\
          &                        & N$_\mathrm{CIV}$ [cm$^{-2}$]        & N$_\mathrm{CIV}$ [cm$^{-2}$] \\
(1)      & (2)    & (3)        & (4)     \\ \hline
Observed & 7.6 & $13.44-13.47$ & $<13.40-13.88$ \\
EAGLE    & 7.49 &  $13.59^{+0.63}_{-1.15}$  & $13.71^{+0.53}_{-1.06}$ \\
R25      & 8.10 & $12.75^{+1.05}_{-0.77}$   &  $12.78^{+0.64}_{-0.63}$ \\
TNG      & 8.23 & $13.85^{+0.69}_{-1.49}$ &    $13.79^{+0.78}_{-2.61}$ \\
 \hline
\end{tabular}
\label{tab:smohalos}
\tablecomments{Comments on columns: (1) Dataset- observed or simulation; (2) black hole mass used to divide sample; (3,4) median and 1-sigma split for N$_\mathrm{C IV}$ for low and high M$_\mathrm{BH}$ samples, respectively; for observations the best estimate for median given upper limits}
\end{table}

The first result to note is how each simulation compares with the observed dataset. EAGLE has values that are consistent with both samples, R25 has values that are significantly lower than observations, and TNG agrees best with the high M$_{\rm BH}$ sample but appears to over-predict the low M$_{\rm BH}$ sample.  In detail, TNG predicts the largest reduction in \ion{C}{IV} with M$_{\rm BH}$ while EAGLE predicts the largest increase.  This agrees with the trends in Fig. \ref{fig: onepanel_sim}, but we note the SMOHALOS sample as well as the split $M_{\rm BH}$ are different.  By selecting matched galaxies, we are sampling a distribution that has a much smaller difference than, for example TNG would predict for a typical galaxy in the bottom panel of that figure.

The second result to note is that the simulations all show $0.12$ dex or less differences in their log$_{10}$N$_{\rm CIV}$ medians indicating that a COS-Holes is not large enough to distinguish the different behaviours across the simulations.  Even if there exists different \ion{C}{IV} absorption patterns relating to M$_{\rm BH}$, our SMOHALOS exploration finds that COS-Holes is too insensitive due to its small sample size and heterogeneous sample of galaxies.  

Finally, we estimate the number of sightlines needed to distinguish between different \ion{C}{IV} distributions as a function of BH mass by replicating the results from R25. For a set sample size, we interpolate a \ion{C}{IV} column density from a random impact parameter (between 0 and 150 kpc) and assign it to either the low or high BH mass sample to create a uniform sample. We fit these random replications to a linear regression model for increasing sample sizes iteratively to create a distribution. We determine that at least 60 sightlines for each high and low mass sample would be needed to distinguish between the samples with a 2$\sigma$ confidence and over 100 sightlines in each sample to tell with a 3$\sigma$ confidence. 

\section{Discussion}\label{sec:Discuss}

\subsection{$\Delta$N$_{\rm CIV}$ Dependence on sSFR}\label{sec: sSFR dom}

We find a $>$2$\sigma$ correlation between the impact-parameter-corrected column density ($\Delta$N$_{\rm CIV}$) and sSFR, as shown in Figure \ref{fig: multi_panel_fig}. In the top panels of Figure \ref{fig: multi_panel_fig}, we see a distinct split in $\Delta$N$_{\rm CIV}$ between star-forming and non-starforming galaxies at log$_{10}$M$_\mathrm{BH} >$ 7.0. This dividing point occurs at log$_{10}$sSFR $\approx$ -11.0. This is consistent with star-forming (log$_{10}$sSFR$ > -11.0$) and passive (log$_{10}$sSFR$ < -11.0$) galaxies in the COS-Halos survey \citep{Tumlinson11}. COS-Halos found that star-forming galaxies exhibited an OVI covering fraction $>$80\%, and higher N$_\mathrm{O VI}$ than their passive galaxy counterparts (f$_{\rm C}$ $\approx$ 30\%). Building off of these results, subsequent studies \citep{Johnson_2015, Zahedy_2019, Tchernyshyov_2023} have established an evident dichotomy in the amount of \ion{O}{VI} present in star-forming and passive galaxies. Controlling for stellar/halo mass, \citet{Tchernyshyov_2023} demonstrated that this dichotomy persists at high statistical significance. For the first time, we tentatively confirm with $>2\sigma$ significance that this correlation exists in the \ion{C}{IV}-bearing gas phase as well, even when we control for other potential variables (see \S\ref{sec:Results} for a discussion of our multivariate analysis). 

There is little \ion{C}{IV} coverage in other surveys of L$^{\star}$ galaxies, and our current sample size is only 21 galaxies.  The CIViL$^{\star}$ survey \citep{berg_civil_2022} will fill this gap in previous COS absorption-galaxy studies by adding NUV data covering \ion{C}{IV} for many of the L$^{\star}$ galaxies of COS-Halos and other surveys which also have \ion{O}{VI} coverage. With the addition of data from this survey, we will be able to test whether \ion{C}{IV} acts more like \ion{O}{VI} than a tracer of the photo-ionized, 10$^{4}$K gas phase. Our current sample indicates that \ion{C}{IV} is more \ion{O}{VI}-like than ``low-ion-like," where low-ionization state gas traced by singly and doubly ionized species shows no correlation with galaxy star-forming properties \citep{werk_2013}.  

While we find a clear trend that exists between $\Delta$N$_\mathrm{C IV}$ and sSFR (\ref{sec: multivariate}), we note a possible second order connection between the sSFR and $M_{\rm BH}$ as they relate to N$_\mathrm{C IV}$ (Figure \ref{fig: multi_panel_fig}, top panels). As discussed above galaxies with higher sSFRs and M$_{\rm BH}$s show higher \ion{C}{IV} content in the CGM, while galaxies with similarly high black hole masses but low sSFR maintain lower \ion{C}{IV} column densities. This split trend could indicate that the relationship between sSFR and black hole mass could result in varying N$_\mathrm{C IV}$ possibly connected to overmassive or undermassive black hole characteristics; however, other evolutionary factors such as galaxy formation time may also play a role (e.g. \citet{Sharma_2020} connects overmassive black hole formation to earlier galaxy formation). 

Results using EAGLE and TNG, have shown that there is an inter-relationship between intrinsic galaxy halo properties and the properties of the central galaxy such as sSFR \citep{Davies_2019, Davies_2020}. In these simulations, galaxies with overmassive BHs are more likely to be quenched, and vice vera \citep[Fig 2]{Davies_2020}, and these quenched systems almost always have an evacuated CGM; due to the BH's influence on the CGM of the central galaxy through suppressing cooling the total sSFR is reduced. Therefore, sSFR, BH mass (and its subsequent growth), and the CGM are highly interconnected. However, for our current sample (including the additional literature values), is too small to directly test these interdependent relationships seen in simulations. Exploring whether this sSFR vs N$_{\rm C IV}$ trend appears cosmological simulations could shed light on the underlying physics driving this apparent connection.

\subsection{Do BHs evacuate their CGM?}\label{sec: bh_evacuate}

%Additionally, most SMBHs are generally thought to accumulate the bulk of their mass through luminous gas accretion \citep{soltan_1982}, which then in turn power active galactic nuclei (AGNs). The majority of galaxy formation theories (and the numerical simulations that explore them) now rely on violent episodes of AGN feedback to terminate star formation by expelling the interstellar medium from the galaxy's disk \citep[e.g.,][]{Bower_2017, Nelson_2019}. Theorists also typically invoke AGN feedback to ``maintain" passive galaxies via a relatively low-energy `radio-mode' \citep[e.g.,][]{croton_2006, Vogelsberger_2013} to keep the halo too hot for cold gas accretion. Both of these processes imply a link between black hole growth history (and hence its total mass) and the gas properties of the surrounding halos.

Cosmological hydrodynamical simulation suites are now able to self-consistently recreate an array of galaxy observables (e.g., EAGLE, \cite{schaye_2015}; Illustris-TNG (TNG), \cite{pillepich_2018}; Romulus25, \citet{tremmel_2017}), including not only the galaxy mass function but specific SMBH-related observables including the AGN luminosity function \cite[EAGLE,][]{rosas_guevara_2016} and the M$_\mathrm{BH}-\sigma$ relation (IllustrisTNG, \citet{sijacki_2015}; Romulus25, \citet{tremmel_2015}). From these simulations, numerical and analytical calculations predict that even a small percentage of the energy from SMBH assembly, when coupled to its surrounding halo, will unbind the CGM from the dark matter halo \citep{Davies_2019,Oppenheimer_2020}. 
%\nns{Add Romulus25 to first sentence (cite Tremmel+2015) and M-sigma sentence (cite Ricarte+2019).}
%Moreover, these calculations suggest that this results from the cumulative effects of energy imparted to the CGM from short-lived AGN growth phases. 
This evacuation of the CGM has a preventative effect such that the reduction in CGM gas density leads to long cooling times for the gas in the inner halo \citep{Davies_2020}; meanwhile, this lower global gas density (not short lived cavities or bubbles carved by AGN-mode feedback) causes galaxies to quench and stay quenched \citep{davies_2021}. 

Both EAGLE \citep{Oppenheimer_2020} and TNG \citep{Nelson_2019b} predict that the ionized gas in the CGM traced by \ion{C}{IV} will be far lower-density (and thus lower measured column density) in galaxies with over-massive black holes (relative to their stellar mass). In contrast, results from \cite{sanchez_2019} which used {\sc Romulus25} \citep{tremmel_2017}, suggest that galaxies with high mass BHs will have higher measured column densities (higher metallicity) in their CGM due to the BH ejecting material out into the diffuse parts of the halo. The COS-Holes observations directly tests these predictions to quantify the imprint of different implementations and efficiencies of BH feedback on the physical state of the CGM. 

%Specifically, to add physical constraints to the assertion that more massive SMBHs should be correlated with more heavy lifting of the CGM that corresponds to less overall bound gaseous content traced by the CGM \citep{sanchez_2019, Oppenheimer_2020}, an effect predicted to be especially dramatic in the observed column densities within 150 kpc. 

Comparing between the simulations (Figure \ref{fig: onepanel_sim}), EAGLE and TNG agree better with the combined observed sample than R25. EAGLE appears to perform the best in matching the column densities of the Low $M_{\rm BH}$ sample as well as reproducing the $M_{\rm BH}$ values themselves. This may not be the case for every ion as \citet{Nelson_2018_oxygen} finds TNG shows better agreement for \ion{O}{VI} around the COS-Halos galaxies than found in EAGLE or EAGLE zoom simulations \citep{oppenheimer_2016}.

%ADD REASONS 
%I've taken this out and added a couple sentences above to be relatively agnostic about which simulation is the best.  I don't know if there is much more to add, at least here.  I think what I should do is address your first point on why TNG shows such a large spread vs. EAGLE in Fig. 8, which is the last of 3 tasks you gave me for this week.  

Interestingly, R25 seems to be under predicting the observed \ion{C}{IV} column density. The AGN feedback in R25 has been shown to be more moderate in comparison to TNG and EAGLE \citep{Tremmel_2019, Chadayammuri_2021, Jung_2022}, possibly due to R25's lack of metal cooling. \citet{Sanchez_2023} shows that a result of this less powerful feedback is that the CGM of these galaxies are significantly less evacuated at these masses. However, it may be that the metal rich gas evacuated by the SMBH in these galaxies remains somewhat nearby to the galaxies, \textless 50 kpc, as in the MW-mass galaxies explored by \citet{sanchez_2019}, which may explain the predicted peak in N$_\mathrm{C IV}$ around 30-40 kpc and the subsequent decline at high impact parameter. 

Despite the combined sample reasonably aligning with EAGLE and TNG, there is no striking evidence in the COS-Holes survey that more massive SMBHs have lower observed column densities, which would indicate this ``cleared" CGM \citep{Oppenheimer_2020} or that more massive SMBHs have a more metal enriched CGM due to the BH ejecting material out into the halo \citep{sanchez_2019}. However, we note that from the SMOHALOS exploration (\ref{sec: smohalos}), we do not yet have a sample large enough to determine if SMBHs are impacting the content or the ionization state of the CGM.  

%While we discuss the clear trend that exists between $\Delta$N$_\mathrm{C IV}$ and sSFR (\ref{sec: sSFR dom}), we note a possible second order connection between the sSFR and $M_{\rm BH}$ as they relate to N$_\mathrm{C IV}$ (Figure \ref{fig: multi_panel_fig}, top panels). We note that galaxies with higher sSFRs and $M_{rm BH}$s show higher \ion{C}{4} content in the CGM, while galaxies with similarly high black hole masses but low sSFR maintain lower \ion{C}{4} column densities. This split trend could indicate that the relationship between sSFR and black hole mass could result in varying N$_\mathrm{C IV}$ possibly connected to overmassive or undermassive black hole characteristics; however, other evolutionary factors such as galaxy formation time may also play a role (\citet{Sharma_2020}, connects overmassive black hole formation to earlier galaxy formation). Exploring whether this trend appears in our and other cosmological simulations could shed light on the underlying physics driving this apparent connection.  

\section{Summary \& Conclusion}\label{sec:End}

%I think that I can tighten up this section and the abstract once I have a better idea of what we are including in the abstract. 

The COS-Holes survey, in combination with a detailed comparison to cosmological simulations, offers the first assessment of the role of BH growth in the regulation of the baryonic content of extended gaseous halos. Broadly, our observations, when combined with data from the literature, are in reasonable agreement with simulation predictions, but do not provide definitive evidence that SMBH feedback significantly impacts the state of the CGM, in either evacuation or through metal enrichment. While our results do not rule out that a galaxy's central SMBH plays an important role in setting the state of the CGM, we find that the sSFR is more correlated with properties of the CGM. Specifically, our key results are: 

\begin{enumerate}
    \item  There is no identifiable relationship between the \ion{C}{IV} content of the CGM and the mass of the assumed host galaxy's SMBH. We attribute this lack of a correlation to both the COS-Holes survey's small sample size and the  large scatter of $>$1 dex in \ion{C}{IV} column density as BH mass increases.  
    
    %Specifically, out of nine QSO lines of sight, we report five \ion{C}{IV} detections and four upper limits (non-detections). We find a 100$\%$ covering fraction for galaxies with a log$_{10}$M$_{\rm BH} <$ 7.0 and a 33$\%$ covering fraction for galaxies with a log$_{10}$M$_{\rm BH} >$ 7.0. 

    %I think the detailed references here are not necessary. Can send the reader to the section in the text where you are describing this. (When reading conclusion, I prefer them to be focused and point me to where I can find more details in the paper.)
    \item When we augment the COS-Holes sample of eight galaxies with 12 additional galaxies from the literature for which we can estimate SMBH masses from ground-based spectroscopy and which have \ion{C}{IV} coverage along paired QSO sightlines, we again find no significant trend between CGM \ion{C}{IV} column densities and SMBH mass with increasing impact parameter (Figure \ref{fig: coslit_sample}).

%We estimate the SMBH mass for galaxies from previously published surveys and only include values from the other surveys if they are of similar stellar mass to our galaxies (log$_{10}$M$_{\star}$/M$_{\odot}$ = $10^{10} - 10^{11}$). We find that COS-Holes observations are comparable to the literature sample where the average \ion{C}{IV} column densities of the COS-Holes and literature sample are log$_{10}$N$_\mathrm{C IV, CH}$ = 13.94 cm$^{-2}$ and log$_{10}$N$_\mathrm{C IV, Literature}$ = 13.91 cm$^{-2}$ respectively. 

    \item  We find that galaxy sSFR is correlated with the ionized content of the CGM as traced by \ion{C}{IV}; this is evidenced by a large spread in sSFR for log$_{10}$M$_\mathrm{BH} >$  7.0, where \ion{C}{IV} strength shows clear dependence on sSFR but not M$_\mathrm{BH}$. Our multivariate analysis tentatively confirms, with $>$2$\sigma$ significance, that a correlation between sSFR and CGM \ion{C}{IV} content exists, similar to that of CGM \ion{O}{VI} \citep{Tchernyshyov_2023}. Combined with items 1 and 2, above, our results suggest that the mass of the SMBH is a subdominant factor in the processes that contribute to the content or ionization state of the z$\sim$0 CGM and that it is the galaxy sSFR that that is more tightly tied to the \ion{C}{IV}-bearing gas phase of the CGM.

    \item We compare \ion{C}{IV} column densities to simulated column densities from the EAGLE, R25, and TNG simulations (Figure \ref{fig: onepanel_sim}). Upon splitting the combined sample into two SMBH mass bins, we find there are only small differences in median column densities between different simulations.
    %We split the combined sample into two SMBH mass bins where the ``low mass" black hole galaxies have a SMBH mass of log$_{10}$M$_\mathrm{BH}$ $\leq$ 7.0 M$_{\odot}$ and ``high mass" black hole galaxies have a SMBH mass of log$_{10}$M$_{\rm BH}$ $\geq$ 7.0 M$_{\odot}$. The simulations are also split into two mass bins, however the mass split is different for each simulation due to each simulation having their own M$_\mathrm{BH}$ distributions as discussed in \S\ref{sec: simulations}. 
    The combined COS-Holes + Literature sample measurements of N$_{CIV}$ are in reasonable agreement with predictions from EAGLE and TNG, but are higher than predictions from R25. 

    \item We create a mock up of the nine lines of sight from the COS-Holes Survey in all three simulations: EAGLE, R25, and TNG.
    %Using SMOHALOS (Simulation Mocker Of Hubble Absorption Line Observational Surveys), we mock the nine lines of sight from the COS-Holes survey across EAGLE, R25, and TNG. 
    We conclude that COS-Holes does not contain enough QSO-galaxy pairs to distinguish the different behaviours across all three simulations. To do so, the sample size would need to be increased from 9 lines of sight to 120 lines of sight. See \S\ref{sec: smohalos} and Table \ref{tab:smohalos} for further details. 
    
    %Our results lead us to infer that the mass of a galaxy's central SMBH (and its internal growth process) does not play a considerable role in setting the state of the CGM and instead, we suggest that sSFR is a more dominant process, and driving ionization relations.  
\end{enumerate}

Since we targeted nearby, spatially extended galaxies with ancillary data (e.g., resolved ALMA and VLA maps of the molecular and neutral ISM, rotation curves, stellar population ages, and metallicity gradients), our QSO spectroscopy will enable a variety of studies well beyond the scope of the goals of this paper. For example, our observed absorption line kinematics will aid in differentiating between material recently launched from the central galaxy via feedback and gas accreting from the larger-scale environment \citep{Bowen_2016, Ho_2017}. Closer to the purview of the paper, this COS spectroscopy will also be crucial for differentiating between other AGN feedback prescriptions invoked in models and simulations that are in the public domain in addition to EAGLE and R25 described and used throughout this paper; these include TNG with dramatic SMBH feedback \citep{pillepich_2018, Nelson_2019b} and the FIRE simulations suite \citep{angles_alcazar_2017, pandya_2021}. Given the significant investment in observational resources required to establish independent and well-constrained SMBH mass measurements and the rarity of UV-bright QSOs, it is unlikely that additional sightlines will become available until the next generation of ELTs and HWO are in active use. However, these future UV observatories will be pivotal for increasing the sample size and allowing us to test the effect of SMBH feedback on the state of the CGM. 

\begin{acknowledgments}
Based on observations with the NASA/ESA Hubble Space Telescope obtained at the Space Telescope Science Institute, which is operated by the Association of Universities for Research in Astronomy, Incorporated, under NASA contract NAS5- 26555. Support for Program number HST-GO-16650 was provided through a grant from the STScI under NASA contract NAS5- 26555. This research has made use of the NASA/IPAC Extragalactic Database (NED), which is operated by the Jet Propulsion Laboratory, California Institute of Technology, under contract with the National Aeronautics and Space Administration. 

We thank an anonymous referee for their insightful comments with helped improve the clarity of this paper. S.L.G recognizes the unceded traditional lands of the Duwamish and Puget Sound Salish Tribes, on which she is grateful to love and work. S.L.G. also thanks Trystyn Berg for their enlightening science discussions on this topic and their valuable input. NNS acknowledges support from NSF MPS-Ascend award ID 2212959 and NASA award SMD-21-75544133. MCB gratefully acknowledges support from the NSF through grant AST-2009230. KHRR acknowledges partial support from NSF grant AST-2009417. YF acknowledges support by NASA award 19-ATP19-0023 and NSF award AST-2007012.
\end{acknowledgments}

\vspace{5mm}
\facilities{\textit{Hubble Space Telescope/Cosmic Origins Spectrograph}}

Data Availability: {\it HST}/COS spectra can be found on in MAST: \dataset[doi:10.17909/njhs-4a40]{http://dx.doi.org/10.17909/njhs-4a40}.

\software{{\tt ArviZ} \citep{Martin_arviz_2023},
{\tt astropy} \citep{astropy:2013, astropy:2018, astropy:2022},
{\tt JAX} \citep{jax2018github},
{\tt LIFELINES} \citep{Davidson-Pilon_lifelines_2024},
{\tt linetools} \citep{prochaska_linetools_2017}, 
{\tt matplotlib} \citep{Hunter:2007},
{\tt numpy} \citep{harris2020array},
{\tt NumPyro} \citep{phan2019composable, bingham2019pyro},
{\tt pandas} \citep{pandas_2024},
{\tt pPXF} \citep{Cappellari2023},
{\tt PyIGM} \citep{prochaska_pyigm_2017},
{\tt scipy} \citep{2020SciPy-NMeth}, 
{\tt NADA} \citep{lee_2020},
{\tt veeper} \citep{burchett_veeper_2024}.} 

\appendix

\section{C IV Absorption Profiles}\label{sec: apendix A}

\begin{figure*}%
    \centering
    {{\includegraphics[width=0.43\textwidth]{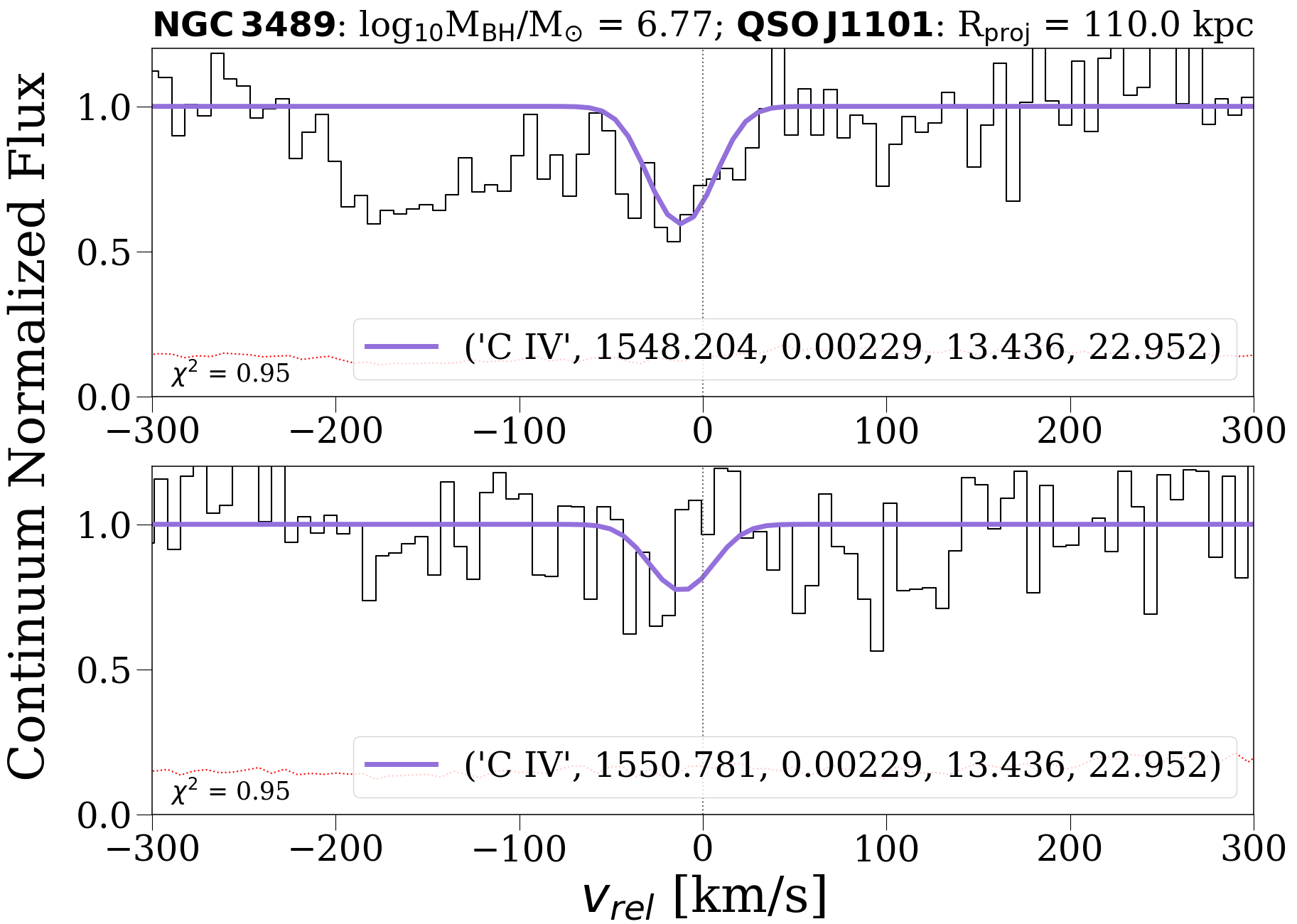}}}%
    \qquad
    {{\includegraphics[width=0.43\textwidth]{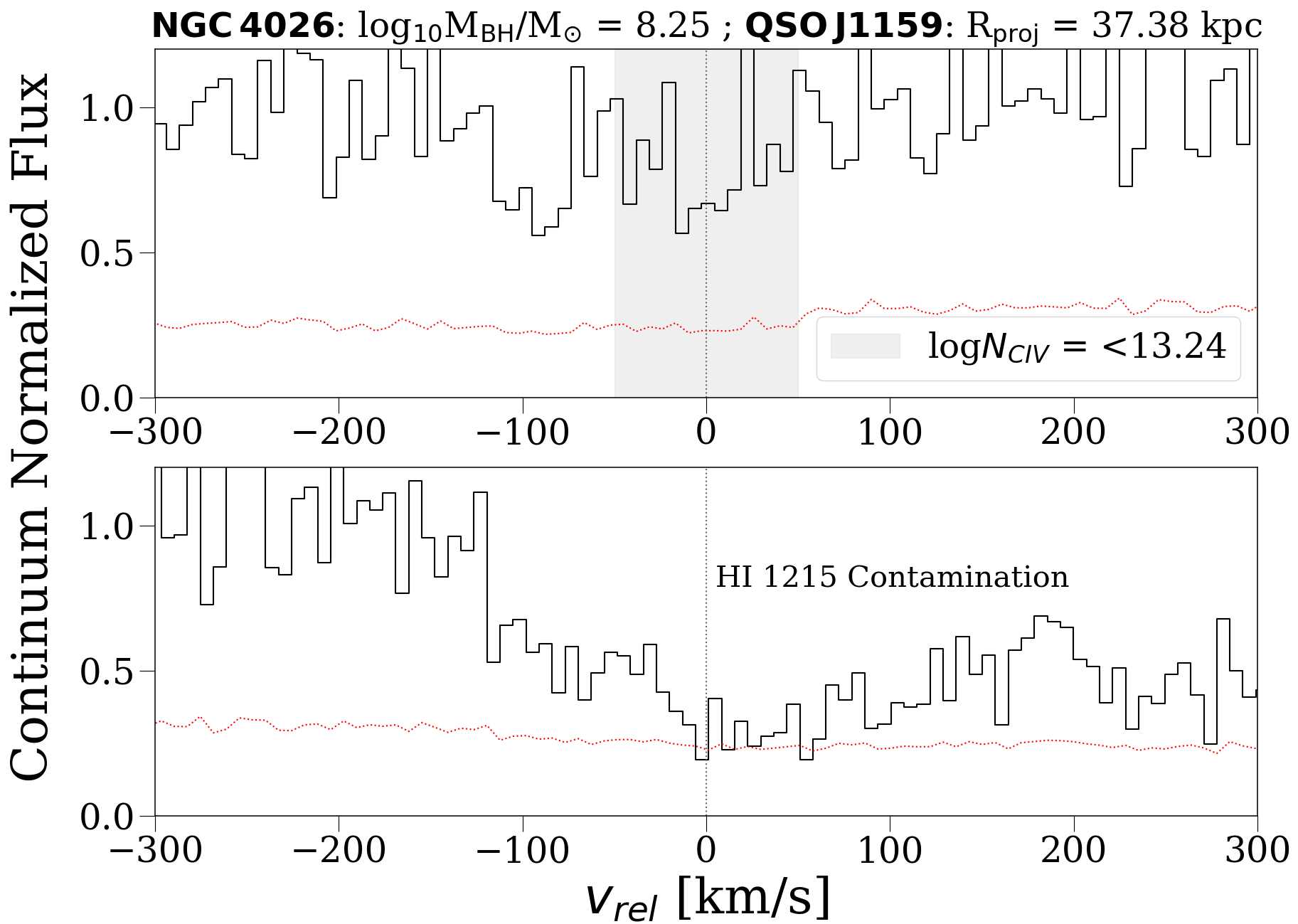}}}%
    \qquad
    {{\includegraphics[width=0.43\textwidth]{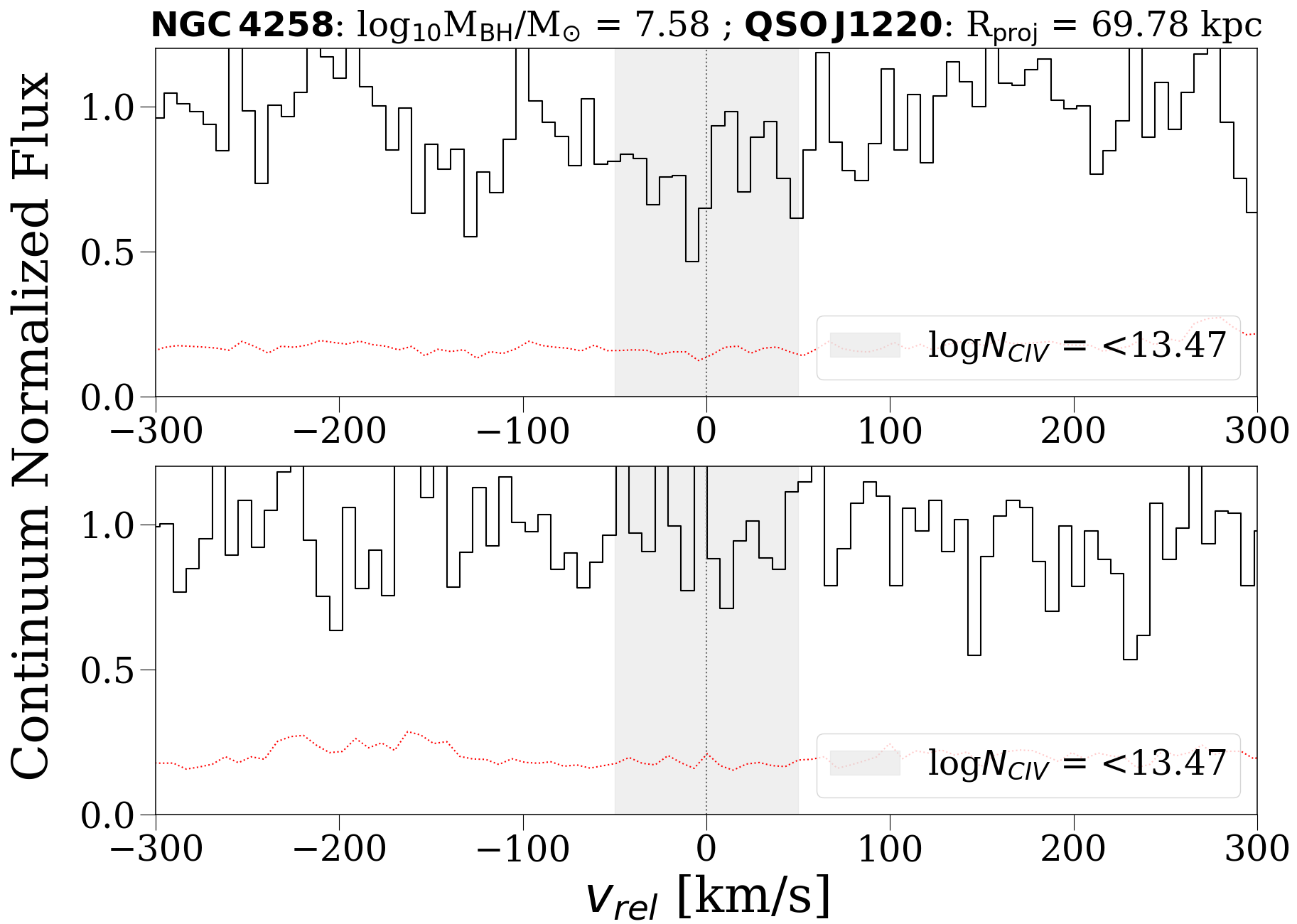}}}%
    \qquad
    {{\includegraphics[width=0.43\textwidth]{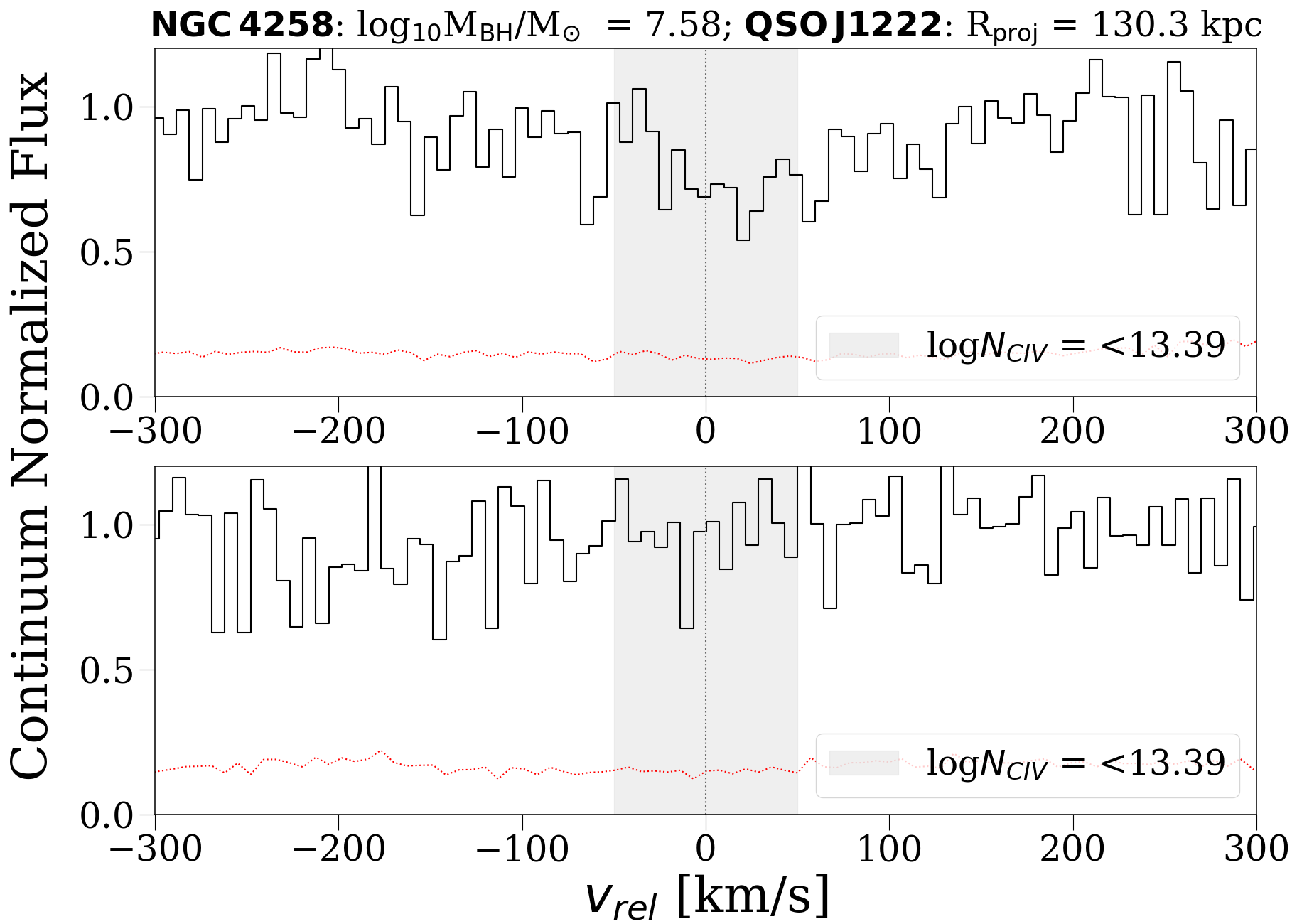}}}%
     \qquad
     {{\includegraphics[width=0.43\textwidth]{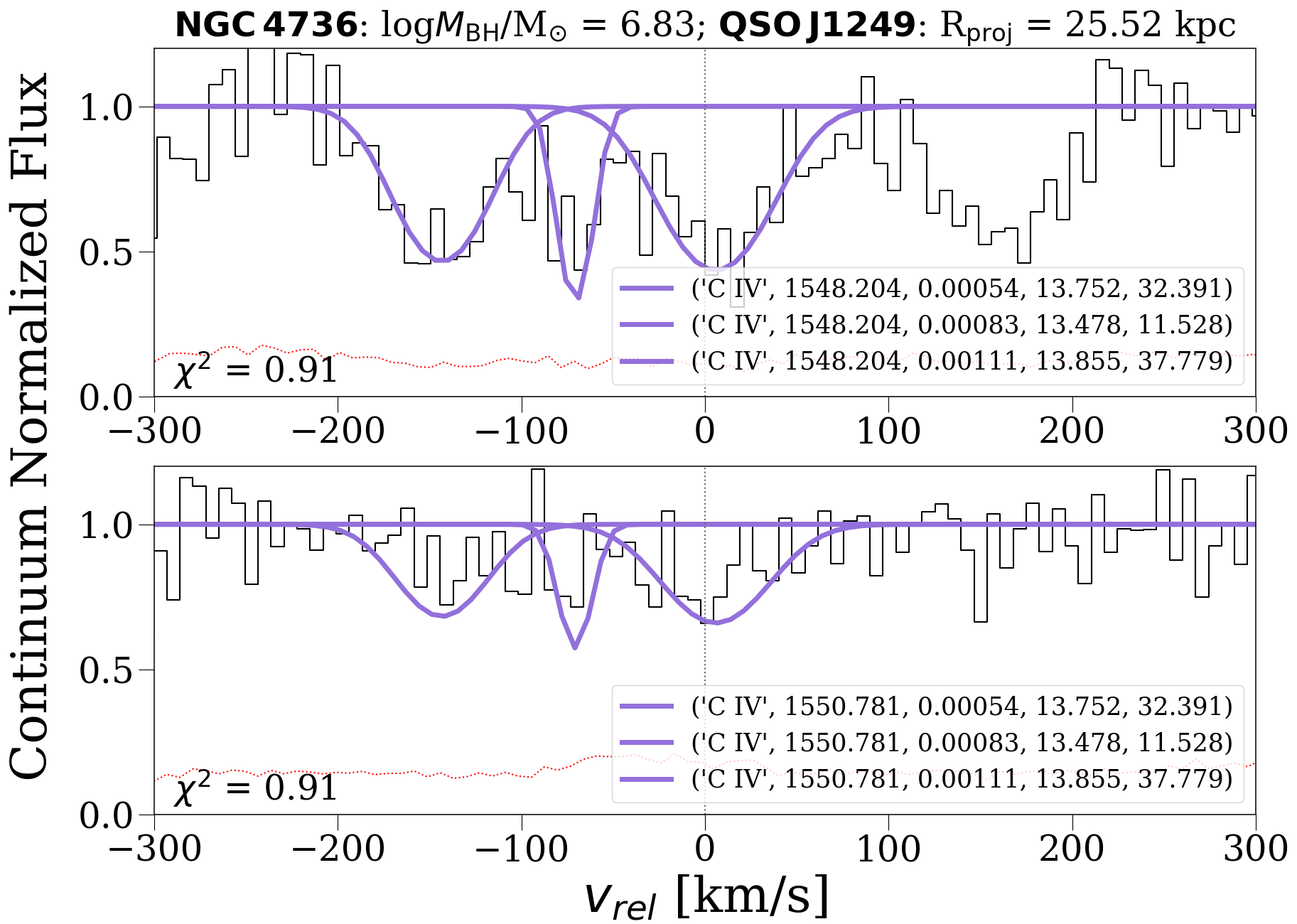} }}%
    \qquad
    {{\includegraphics[width=0.43\textwidth]{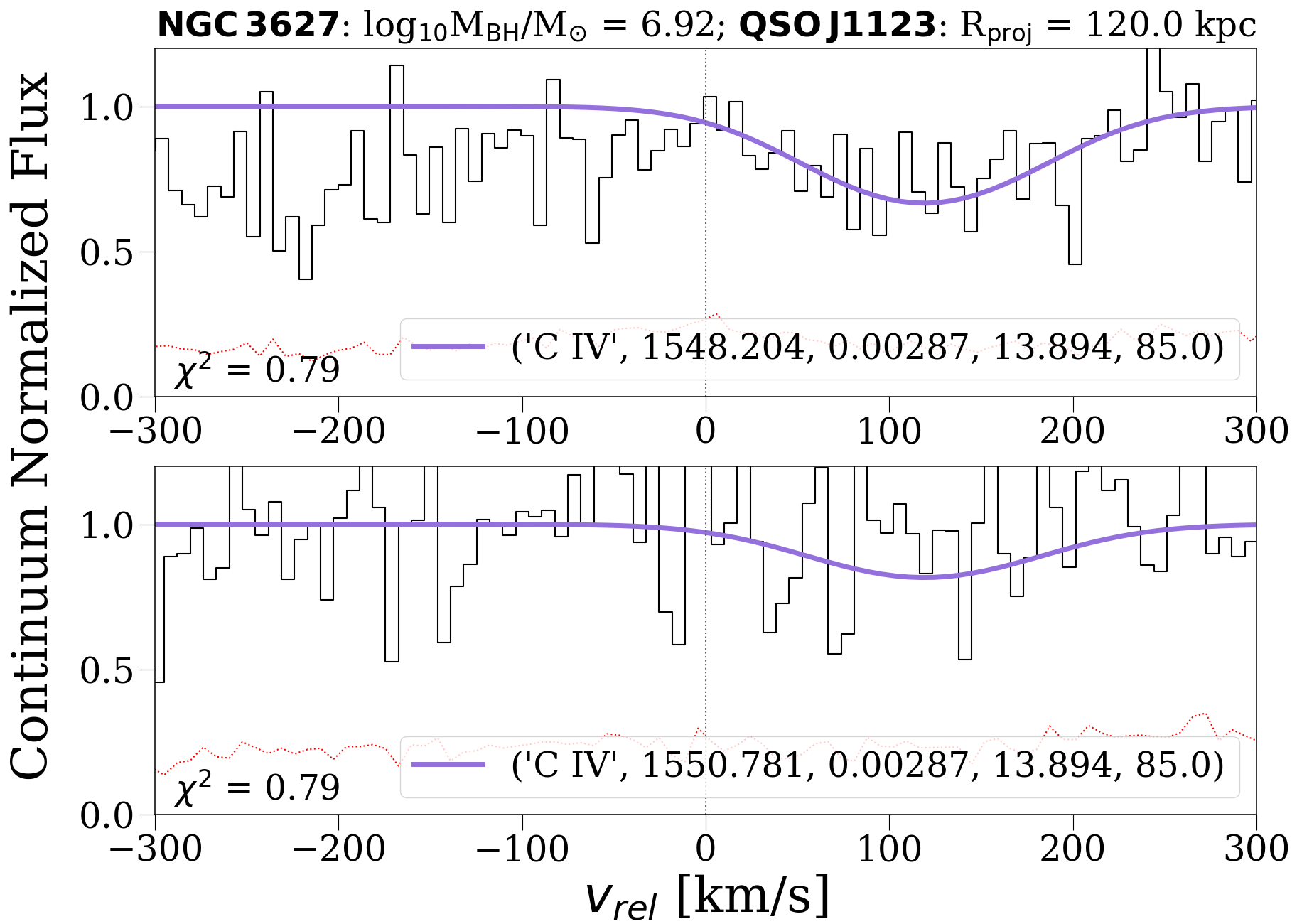} }}%
    \qquad
    {{\includegraphics[width=0.43\textwidth]{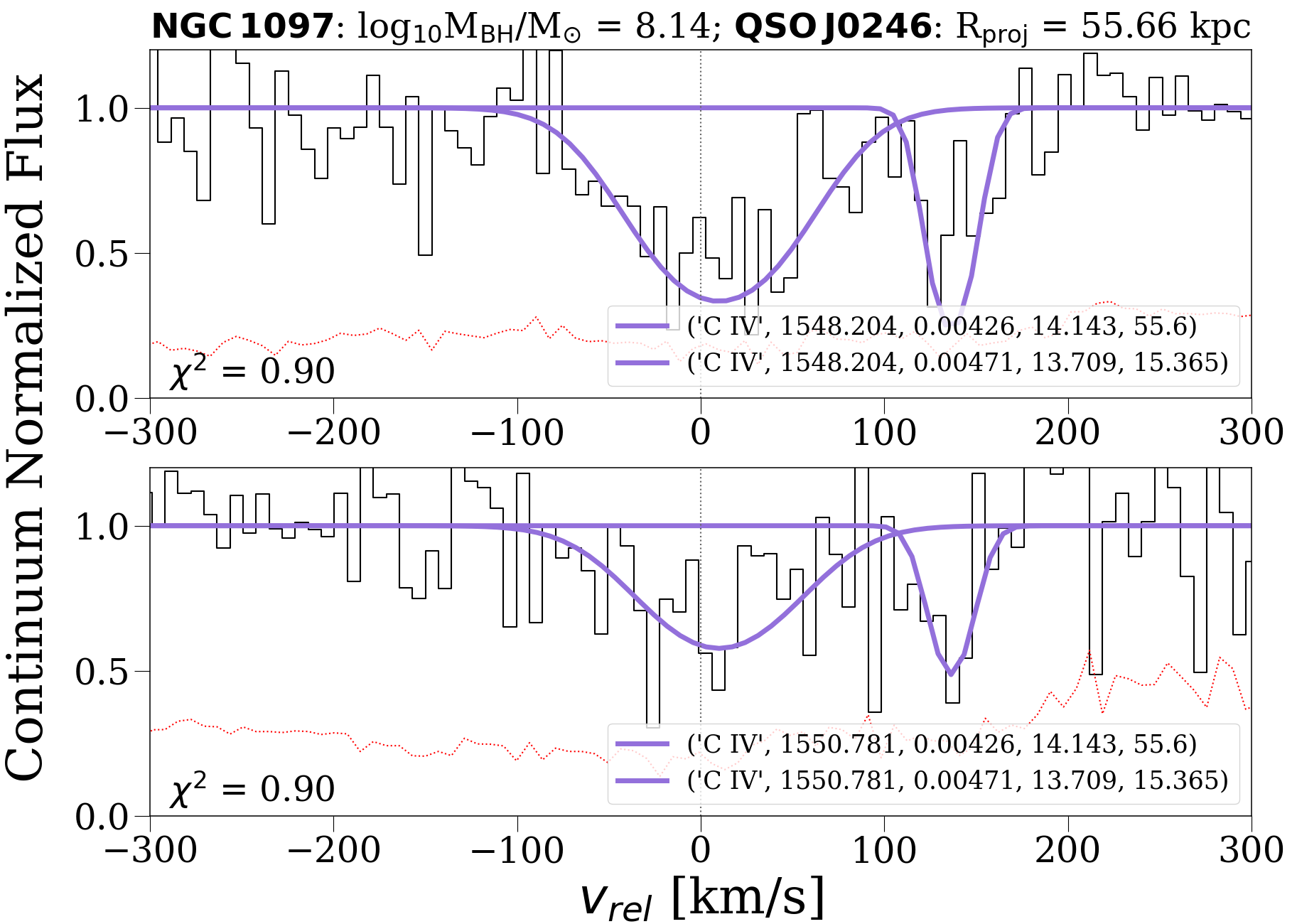} }}%
    \qquad
    {{\includegraphics[width=0.43\textwidth]{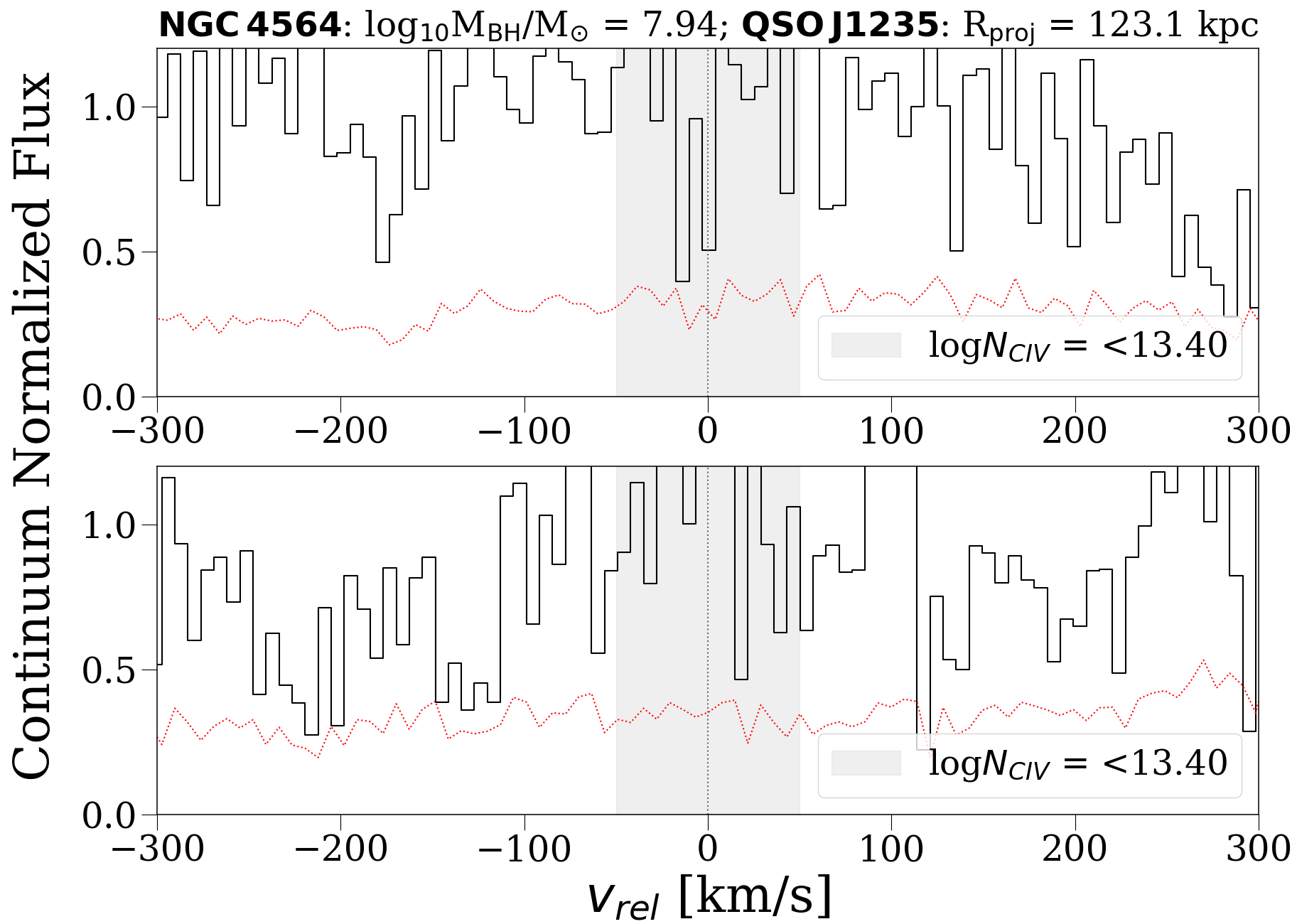} }}%
    \caption{The remaining \ion{C}{IV} absorption features for the COS-Holes survey. Same conventions are used as described in Figure \ref{fig: civ_lineprofie_example}.}%
    \label{fig: cont figure}
\end{figure*}

Figure \ref{fig: cont figure} presents absorption line profiles for the COS-Holes sample. Some items of note are: (1) for NGC 4026 (QSO: SDSSJ1159) there is a prominent blend in $\lambda$1550 which we identified corresponds to H I $\lambda$1215 at z = 0.28; due to this contamination and no features in $\lambda$1548 we report an upper limit for the \ion{C}{IV} column density; (2) for NGC 4258 (QSO: SDSSJ1220 and SDSSJ1222) and NGC 4564 (QSO: SDSSJ1235) we do not detect any \ion{C}{IV} absorption, and report only upper limits on \ion{C}{IV} column density.

\section{Literature Sample Galaxy and QSO Tables}\label{sec: apendix B}

\begin{center}
\begin{table*}
\centering
\caption{Literature Sample Galaxy Information}
\begin{tabular}{cccccccc}
\hline
Galaxy              & RA      & Dec    & z & sSFR       & M$_{*}$           & M$_\mathrm{BH}$              & Ref \\
                    & (deg)   & (deg)  &  & (log$_{10}$yr$^{-1}$)&
                    (log$_{10}$M$_{\odot}$) & (log$_{10}$M$_{\odot}$)      &     \\
(1)                   & (2)       & (3)      & (4)        & (5)           & (6)                & (7) & (8)  \\ \hline
J102846.43+391842.9 & 157.194 & 39.312 & 0.1135   & -9.8 & 10.5        & 7.16 $\pm$ 0.35  & a   \\
J132150.89+033034.1 & 200.462 & 3.509  & 0.0816   & -10.3 & 10.8       & 7.63 $\pm$ 0.12  & a   \\
J140502.20+470525.9 & 211.259 & 47.091 & 0.1452   & -9.0 & 10.4       & 7.4 $\pm$ 0.43   & a   \\
J154527.12+484642.2 & 236.363 & 48.778 & 0.0752   & -10.5 & 10.5        & 6.11 $\pm$  0.52 & a   \\
J0925+4535\_227\_334             & 141.379 & 45.533 & 0.014    & -10.3 & 10.0        & $<$5.91   & b   \\
J0959+0503\_318\_13             & 149.813 & 5.068  & 0.059    & -9.9 & 10.0        & 6.13 $\pm$ 0.52  & b   \\
J1121+0325\_73\_198              & 170.362 & 3.445  & 0.023    & -10.2 & 10.1        & $<$5.91  & b   \\
J1211+3657\_312\_196            & 182.761 & 36.998 & 0.023    & -9.8 & 10.1        & $<$5.91  & b   \\
PG1202+281\_165\_95             & 181.183 & 27.878 & 0.051    & -12.1 & 10.0        & $<$5.91  & b   \\
J0910+1014\_34\_46  & 137.626 & 10.24  & 0.1427   & -9.5 & 10.61       & 7.71 $\pm$ 0.22  & c   \\
J1619+3342\_113\_40 & 244.831 & 33.706 & 0.1414   & -9.9 & 10.1        & 6.16 $\pm$ 0.44  & c   \\
M31                 & 10.685  & 41.269 & -0.00099 & -11.1 & 10.9        & 8.15 $\pm$ 0.24  & d   \\ \hline
\end{tabular}
\label{tab: lit_gal_info}
\tablecomments{Comments on columns: (1) galaxy name; (2-3) RA and Dec; (4) redshift of the galaxy; (5) specific star formation; (6) stellar mass; (7) black hole mass; (8) reference sample where a, b, c, and d are \cite{Borthakur_2013}, \cite{Bordoloi_2014}, \cite{werk_2013}, and \cite{Lehner_2020} respectively.}
\end{table*}
\end{center}

\begin{table*}
\centering
\caption{Literature Sample QSO Information}
\begin{tabular}{cccccccc}
\hline
QSO                 & RA      & Dec    & z     & R$_\mathrm{proj}$ & $R_\mathrm{proj}$/R$_{200c}$ & log$_{10}$N$_\mathrm{CIV}$   & Ref \\
                    & (deg)   & (deg)  &       & (kpc)      &                    & (cm$^{-2}$)         &     \\
(1)                 & (2)     & (3)    & (4)   & (5)        & (6)                & (7)                 & (8) \\ \hline

J102847.00+391800.4 & 157.2   & 39.3   & 0.473 & 88.7       & 0.52              & 14.65 $\pm$ 0.04     & a   \\
J132144.97+033055.7 & 200.44  & 3.52   & 0.269 & 140.2      & 0.52              & \textless{}14.06    & a   \\
J140505.77+470441.1 & 211.27  & 47.08  & 1.24  & 146.9      & 0.90              & 14.25 $\pm$ 0.08     & a   \\
J154530.23+484608.9 & 236.38  & 48.77  & 0.399 & 64.7       & 0.37              & \textless{}13.79    & a   \\
J09525+4535         & 141.478 & 45.596 & 0.329 & 95.0         & 0.76              & 13.56 $\pm$ 0.06     & b   \\
J0959+0503          & 149.815 & 5.065  & 0.162 & 14.0         & 0.11              & \textgreater{}14.69 & b   \\
PG1202+281          & 181.175 & 27.903 & 0.165 & 92.0         & 0.73              & 13.58 $\pm$ 0.10     & b   \\
J1121+0325          & 170.309 & 3.43   & 0.152 & 89.0         & 0.68              & \textless{}13.45    & b   \\
J1211+3657          & 182.811 & 36.961 & 0.171 & 90.0         & 0.68              & \textless{}13.17    & b   \\
J0910+1014          & 137.624 & 10.237 & 0.462 & 112.0        & 0.58              & 14.1 $\pm$ 0.09      & c   \\
J1619+3342          & 244.819 & 33.711 & 0.47  & 97.0         & 0.72               & 13.9 $\pm$ 0.03      & c   \\
HS0033+4300         & 9.096   & 43.278 & 0.12  & 30.5       & 0.133              & 14.1 $\pm$ 0.05      & d   \\
HS0058+4213         & 15.38   & 42.493 & 0.19  & 48.6       & 0.211              & 13.33 $\pm$ 0.18     & d   \\
RX\_J0043.6+3725    & 10.927  & 37.422 & 0.08  & 50.5       & 0.22               & 13.85 $\pm$ 0.03     & d   \\
Zw535.012           & 9.087   & 45.665 & 0.048 & 59.7       & 0.26               & 12.99 $\pm$ 0.30     & d   \\
RX\_J0050.8+3536    & 12.711  & 35.612 & 0.058 & 77.1       & 0.335              & 13.45 $\pm$ 0.07     & d   \\
IRAS\_F00040+4325   & 1.652   & 43.708 & 0.163 & 93         & 0.404              & 13.23 $\pm$ 0.11     & d   \\
MRK352              & 14.972  & 31.827 & 0.015 & 131.7      & 0.573              & 13.5 $\pm$ 0.15      & d   \\
RX\_J0043.6+3725    & 10.927  & 37.422 & 0.08  & 50.5       & 0.22               & \textless{}12.92    & d   \\
RXS\_J0118.8+3836   & 19.706  & 38.606 & 0.216 & 97.2       & 0.423              & \textless{}12.9     & d   \\
RX\_J0028.1+3103    & 7.045   & 31.063 & 0.5   & 139.1      & 0.605              & \textless{}13.11    & d   \\
Project AMIGA Avg   &         &        &       & 77.79      & 0.33               & 13.34 $\pm$ 0.20     & d   \\ \hline
\end{tabular}
\label{tab: lit_qso_info}
\tablecomments{Comments on columns: (1) QSO ID; (2-3) RA and Dec; (4) redshift of the galaxy; (5) impact parameter; (6) impact parameter normalized by the virial radius; (7) CIV column density;(8) reference sample where a, b, c, and d are \cite{Borthakur_2013}, \cite{Bordoloi_2014}, \cite{werk_2013}, and \cite{Lehner_2020} respectively.}
\end{table*}

We increase our COS-Holes sample size by adding values from  \cite{Lehner_2020} (Project AMIGA), \cite{werk_2013} (COS-Halos), \cite{Borthakur_2013} (starbursts), and \cite{Bordoloi_2014} (COS-Dwarfs). Project AMIGA (M31; log$_{10}$M$_{\star} = $9$\pm$2 $\times$ 10$^{10}$ M$_{\odot}$, \cite{Williams_2017}; log$_{10}$M$_{\rm BH}$ = 8.15 $\pm$ 0.24, \cite{davis_2017}; total SFR = 0.7 M$_{\odot}$ yr$^{-1}$, \cite{Lewis_2015}) was specifically designed to span M31's project major and minor axis and intermediate orientations, thus it provides the unique opportunity to probe one high mass SMBH galaxy at increasing impact parameter. It is important to mention that there were data values that had possible contamination from the Magellanic Stream, however we only plot values they denote as uncontaminated; for more a more detailed explanation of how this contamination was removed see \cite{Lehner_2020}. Using this smaller sample of Project AMIGA observations, we take an average of the detections to report mean Project AMIGA column density. We use this single data point to represent the Project AMIGA observations. 

We also include \cite{Borthakur_2013} which found highly ionized gas traced by \ion{C}{IV} in 80$\%$ of their starburst galaxies. They assert that it is extremely unlikely that these absorbers were photoionized from either the metagalactic background or the stellar radiation from the starburst; using CLOUDY models, they suggest that this observed \ion{C}{IV} would arise from shock-ionization and be accelerated by the ram pressure of the wind and thus enriching the CGM. Similar high detections of \ion{C}{IV} were seen in \cite{Bordoloi_2014}, where they detected the ion out to 100 kpc in their sample of sub-L$^{\star}$ galaxies. They find that strong \ion{C}{IV} absorption observations were detected around star-forming galaxies and they are kinematically consistent with being bound to the dark matter halos of their hosts. In conclusion, they assert that the metallic content of the CGM around their galaxy sample is best explained by the addition of strong outflows in addition to tidal debris and ram pressure stripping. Taken together, both of these archival studies support the idea that energy-driven feedback is needed to explain the presence of highly ionized ions, such as \ion{C}{IV} in the CGM. Thus, they are interesting samples to compare and expand upon the COS-Holes observations.

In addition, we also include \cite{werk_2013} which presented column density measurements of the CGM from QSO-galaxy pairs (low-z, L $\approx$ L$^{\star}$) drawn from the COS-Halos survey. One of their main results is finding that column densities derived for intermediate ionization state metal lines decrease with increasing impact parameter; they interpret this trend to mean there is a decline in the metal surface density profle of the CGM within its inner 160kpc. They also see that the gas kinematics derived from Voigt profile fits to their observations suggest that the CGM is mostly bound to its host galaxy's dark matter halo similar to results seen in \cite{Bordoloi_2014}. The collective information for \cite{werk_2013} and the rest of our additional literature sample can be seen in Table \ref{tab: lit_gal_info} and \ref{tab: lit_qso_info} respectively.

\bibliography{references}{}

\begin{thebibliography}{}
\expandafter\ifx\csname natexlab\endcsname\relax\def\natexlab#1{#1}\fi
\providecommand{\url}[1]{\href{#1}{#1}}
\providecommand{\dodoi}[1]{doi:~\href{http://doi.org/#1}{\nolinkurl{#1}}}
\providecommand{\doeprint}[1]{\href{http://ascl.net/#1}{\nolinkurl{http://ascl.net/#1}}}
\providecommand{\doarXiv}[1]{\href{https://arxiv.org/abs/#1}{\nolinkurl{https://arxiv.org/abs/#1}}}

\bibitem[{{Abdurro'uf} {et~al.}(2022){Abdurro'uf}, {Accetta}, {Aerts}, {Silva Aguirre}, {Ahumada}, {Ajgaonkar}, {Filiz Ak}, {Alam}, {Allende Prieto}, {Almeida}, {Anders}, {Anderson}, {Andrews}, {Anguiano}, {Aquino-Ort{\'\i}z}, {Arag{\'o}n-Salamanca}, {Argudo-Fern{\'a}ndez}, {Ata}, {Aubert}, {Avila-Reese}, {Badenes}, {Barb{\'a}}, {Barger}, {Barrera-Ballesteros}, {Beaton}, {Beers}, {Belfiore}, {Bender}, {Bernardi}, {Bershady}, {Beutler}, {Bidin}, {Bird}, {Bizyaev}, {Blanc}, {Blanton}, {Boardman}, {Bolton}, {Boquien}, {Borissova}, {Bovy}, {Brandt}, {Brown}, {Brownstein}, {Brusa}, {Buchner}, {Bundy}, {Burchett}, {Bureau}, {Burgasser}, {Cabang}, {Campbell}, {Cappellari}, {Carlberg}, {Wanderley}, {Carrera}, {Cash}, {Chen}, {Chen}, {Cherinka}, {Chiappini}, {Choi}, {Chojnowski}, {Chung}, {Clerc}, {Cohen}, {Comerford}, {Comparat}, {da Costa}, {Covey}, {Crane}, {Cruz-Gonzalez}, {Culhane}, {Cunha}, {Dai}, {Damke}, {Darling}, {Davidson}, {Davies}, {Dawson}, {De Lee}, {Diamond-Stanic}, {Cano-D{\'\i}az}, {S{\'a}nchez},
  {Donor}, {Duckworth}, {Dwelly}, {Eisenstein}, {Elsworth}, {Emsellem}, {Eracleous}, {Escoffier}, {Fan}, {Farr}, {Feng}, {Fern{\'a}ndez-Trincado}, {Feuillet}, {Filipp}, {Fillingham}, {Frinchaboy}, {Fromenteau}, {Galbany}, {Garc{\'\i}a}, {Garc{\'\i}a-Hern{\'a}ndez}, {Ge}, {Geisler}, {Gelfand}, {G{\'e}ron}, {Gibson}, {Goddy}, {Godoy-Rivera}, {Grabowski}, {Green}, {Greener}, {Grier}, {Griffith}, {Guo}, {Guy}, {Hadjara}, {Harding}, {Hasselquist}, {Hayes}, {Hearty}, {Hern{\'a}ndez}, {Hill}, {Hogg}, {Holtzman}, {Horta}, {Hsieh}, {Hsu}, {Hsu}, {Huber}, {Huertas-Company}, {Hutchinson}, {Hwang}, {Ibarra-Medel}, {Chitham}, {Ilha}, {Imig}, {Jaekle}, {Jayasinghe}, {Ji}, {Johnson}, {Jones}, {J{\"o}nsson}, {Katkov}, {Khalatyan}, {Kinemuchi}, {Kisku}, {Knapen}, {Kneib}, {Kollmeier}, {Kong}, {Kounkel}, {Kreckel}, {Krishnarao}, {Lacerna}, {Lane}, {Langgin}, {Lavender}, {Law}, {Lazarz}, {Leung}, {Leung}, {Lewis}, {Li}, {Li}, {Lian}, {Liang}, {Lin}, {Lin}, {Lin}, {Lintott}, {Long}, {Longa-Pe{\~n}a}, {L{\'o}pez-Cob{\'a}}, {Lu},
  {Lundgren}, {Luo}, {Mackereth}, {de la Macorra}, {Mahadevan}, {Majewski}, {Manchado}, {Mandeville}, {Maraston}, {Margalef-Bentabol}, {Masseron}, {Masters}, {Mathur}, {McDermid}, {Mckay}, {Merloni}, {Merrifield}, {Meszaros}, {Miglio}, {Di Mille}, {Minniti}, {Minsley}, {Monachesi}, {Moon}, {Mosser}, {Mulchaey}, {Muna}, {Mu{\~n}oz}, {Myers}, {Myers}, {Nadathur}, {Nair}, {Nandra}, {Neumann}, {Newman}, {Nidever}, {Nikakhtar}, {Nitschelm}, {O'Connell}, {Garma-Oehmichen}, {Luan Souza de Oliveira}, {Olney}, {Oravetz}, {Ortigoza-Urdaneta}, {Osorio}, {Otter}, {Pace}, {Padilla}, {Pan}, {Pan}, {Parikh}, {Parker}, {Peirani}, {Pe{\~n}a Ram{\'\i}rez}, {Penny}, {Percival}, {Perez-Fournon}, {Pinsonneault}, {Poidevin}, {Poovelil}, {Price-Whelan}, {B{\'a}rbara de Andrade Queiroz}, {Raddick}, {Ray}, {Rembold}, {Riddle}, {Riffel}, {Riffel}, {Rix}, {Robin}, {Rodr{\'\i}guez-Puebla}, {Roman-Lopes}, {Rom{\'a}n-Z{\'u}{\~n}iga}, {Rose}, {Ross}, {Rossi}, {Rubin}, {Salvato}, {S{\'a}nchez}, {S{\'a}nchez-Gallego}, {Sanderson}, {Santana
  Rojas}, {Sarceno}, {Sarmiento}, {Sayres}, {Sazonova}, {Schaefer}, {Schiavon}, {Schlegel}, {Schneider}, {Schultheis}, {Schwope}, {Serenelli}, {Serna}, {Shao}, {Shapiro}, {Sharma}, {Shen}, {Shetrone}, {Shu}, {Simon}, {Skrutskie}, {Smethurst}, {Smith}, {Sobeck}, {Spoo}, {Sprague}, {Stark}, {Stassun}, {Steinmetz}, {Stello}, {Stone-Martinez}, {Storchi-Bergmann}, {Stringfellow}, {Stutz}, {Su}, {Taghizadeh-Popp}, {Talbot}, {Tayar}, {Telles}, {Teske}, {Thakar}, {Theissen}, {Tkachenko}, {Thomas}, {Tojeiro}, {Hernandez Toledo}, {Troup}, {Trump}, {Trussler}, {Turner}, {Tuttle}, {Unda-Sanzana}, {V{\'a}zquez-Mata}, {Valentini}, {Valenzuela}, {Vargas-Gonz{\'a}lez}, {Vargas-Maga{\~n}a}, {Alfaro}, {Villanova}, {Vincenzo}, {Wake}, {Warfield}, {Washington}, {Weaver}, {Weijmans}, {Weinberg}, {Weiss}, {Westfall}, {Wild}, {Wilde}, {Wilson}, {Wilson}, {Wilson}, {Wolf}, {Wood-Vasey}, {Yan}, {Zamora}, {Zasowski}, {Zhang}, {Zhao}, {Zheng}, {Zheng}, \& {Zhu}}]{Abdurro'uf_2022}
{Abdurro'uf}, {Accetta}, K., {Aerts}, C., {et~al.} 2022, \apjs, 259, 35, \dodoi{10.3847/1538-4365/ac4414}

\bibitem[{{Angl{\'e}s-Alc{\'a}zar} {et~al.}(2017){Angl{\'e}s-Alc{\'a}zar}, {Faucher-Gigu{\`e}re}, {Quataert}, {Hopkins}, {Feldmann}, {Torrey}, {Wetzel}, \& {Kere{\v{s}}}}]{angles_alcazar_2017}
{Angl{\'e}s-Alc{\'a}zar}, D., {Faucher-Gigu{\`e}re}, C.-A., {Quataert}, E., {et~al.} 2017, \mnras, 472, L109, \dodoi{10.1093/mnrasl/slx161}

\bibitem[{{Asmus} {et~al.}(2020){Asmus}, {Greenwell}, {Gandhi}, {Boorman}, {Aird}, {Alexander}, {Assef}, {Baldi}, {Davies}, {H{\"o}nig}, {Ricci}, {Rosario}, {Salvato}, {Shankar}, \& {Stern}}]{Asmus_2020}
{Asmus}, D., {Greenwell}, C.~L., {Gandhi}, P., {et~al.} 2020, \mnras, 494, 1784, \dodoi{10.1093/mnras/staa766}

\bibitem[{{Assef} {et~al.}(2018){Assef}, {Stern}, {Noirot}, {Jun}, {Cutri}, \& {Eisenhardt}}]{Assef_2018}
{Assef}, R.~J., {Stern}, D., {Noirot}, G., {et~al.} 2018, \apjs, 234, 23, \dodoi{10.3847/1538-4365/aaa00a}

\bibitem[{{Astropy Collaboration} {et~al.}(2013){Astropy Collaboration}, {Robitaille}, {Tollerud}, {Greenfield}, {Droettboom}, {Bray}, {Aldcroft}, {Davis}, {Ginsburg}, {Price-Whelan}, {Kerzendorf}, {Conley}, {Crighton}, {Barbary}, {Muna}, {Ferguson}, {Grollier}, {Parikh}, {Nair}, {Unther}, {Deil}, {Woillez}, {Conseil}, {Kramer}, {Turner}, {Singer}, {Fox}, {Weaver}, {Zabalza}, {Edwards}, {Azalee Bostroem}, {Burke}, {Casey}, {Crawford}, {Dencheva}, {Ely}, {Jenness}, {Labrie}, {Lim}, {Pierfederici}, {Pontzen}, {Ptak}, {Refsdal}, {Servillat}, \& {Streicher}}]{astropy:2013}
{Astropy Collaboration}, {Robitaille}, T.~P., {Tollerud}, E.~J., {et~al.} 2013, \aap, 558, A33, \dodoi{10.1051/0004-6361/201322068}

\bibitem[{{Astropy Collaboration} {et~al.}(2018){Astropy Collaboration}, {Price-Whelan}, {Sip{\H{o}}cz}, {G{\"u}nther}, {Lim}, {Crawford}, {Conseil}, {Shupe}, {Craig}, {Dencheva}, {Ginsburg}, {Vand erPlas}, {Bradley}, {P{\'e}rez-Su{\'a}rez}, {de Val-Borro}, {Aldcroft}, {Cruz}, {Robitaille}, {Tollerud}, {Ardelean}, {Babej}, {Bach}, {Bachetti}, {Bakanov}, {Bamford}, {Barentsen}, {Barmby}, {Baumbach}, {Berry}, {Biscani}, {Boquien}, {Bostroem}, {Bouma}, {Brammer}, {Bray}, {Breytenbach}, {Buddelmeijer}, {Burke}, {Calderone}, {Cano Rodr{\'\i}guez}, {Cara}, {Cardoso}, {Cheedella}, {Copin}, {Corrales}, {Crichton}, {D'Avella}, {Deil}, {Depagne}, {Dietrich}, {Donath}, {Droettboom}, {Earl}, {Erben}, {Fabbro}, {Ferreira}, {Finethy}, {Fox}, {Garrison}, {Gibbons}, {Goldstein}, {Gommers}, {Greco}, {Greenfield}, {Groener}, {Grollier}, {Hagen}, {Hirst}, {Homeier}, {Horton}, {Hosseinzadeh}, {Hu}, {Hunkeler}, {Ivezi{\'c}}, {Jain}, {Jenness}, {Kanarek}, {Kendrew}, {Kern}, {Kerzendorf}, {Khvalko}, {King}, {Kirkby}, {Kulkarni},
  {Kumar}, {Lee}, {Lenz}, {Littlefair}, {Ma}, {Macleod}, {Mastropietro}, {McCully}, {Montagnac}, {Morris}, {Mueller}, {Mumford}, {Muna}, {Murphy}, {Nelson}, {Nguyen}, {Ninan}, {N{\"o}the}, {Ogaz}, {Oh}, {Parejko}, {Parley}, {Pascual}, {Patil}, {Patil}, {Plunkett}, {Prochaska}, {Rastogi}, {Reddy Janga}, {Sabater}, {Sakurikar}, {Seifert}, {Sherbert}, {Sherwood-Taylor}, {Shih}, {Sick}, {Silbiger}, {Singanamalla}, {Singer}, {Sladen}, {Sooley}, {Sornarajah}, {Streicher}, {Teuben}, {Thomas}, {Tremblay}, {Turner}, {Terr{\'o}n}, {van Kerkwijk}, {de la Vega}, {Watkins}, {Weaver}, {Whitmore}, {Woillez}, {Zabalza}, \& {Astropy Contributors}}]{astropy:2018}
{Astropy Collaboration}, {Price-Whelan}, A.~M., {Sip{\H{o}}cz}, B.~M., {et~al.} 2018, \aj, 156, 123, \dodoi{10.3847/1538-3881/aabc4f}

\bibitem[{{Astropy Collaboration} {et~al.}(2022){Astropy Collaboration}, {Price-Whelan}, {Lim}, {Earl}, {Starkman}, {Bradley}, {Shupe}, {Patil}, {Corrales}, {Brasseur}, {N{"o}the}, {Donath}, {Tollerud}, {Morris}, {Ginsburg}, {Vaher}, {Weaver}, {Tocknell}, {Jamieson}, {van Kerkwijk}, {Robitaille}, {Merry}, {Bachetti}, {G{"u}nther}, {Aldcroft}, {Alvarado-Montes}, {Archibald}, {B{'o}di}, {Bapat}, {Barentsen}, {Baz{'a}n}, {Biswas}, {Boquien}, {Burke}, {Cara}, {Cara}, {Conroy}, {Conseil}, {Craig}, {Cross}, {Cruz}, {D'Eugenio}, {Dencheva}, {Devillepoix}, {Dietrich}, {Eigenbrot}, {Erben}, {Ferreira}, {Foreman-Mackey}, {Fox}, {Freij}, {Garg}, {Geda}, {Glattly}, {Gondhalekar}, {Gordon}, {Grant}, {Greenfield}, {Groener}, {Guest}, {Gurovich}, {Handberg}, {Hart}, {Hatfield-Dodds}, {Homeier}, {Hosseinzadeh}, {Jenness}, {Jones}, {Joseph}, {Kalmbach}, {Karamehmetoglu}, {Ka{l}uszy{'n}ski}, {Kelley}, {Kern}, {Kerzendorf}, {Koch}, {Kulumani}, {Lee}, {Ly}, {Ma}, {MacBride}, {Maljaars}, {Muna}, {Murphy}, {Norman}, {O'Steen},
  {Oman}, {Pacifici}, {Pascual}, {Pascual-Granado}, {Patil}, {Perren}, {Pickering}, {Rastogi}, {Roulston}, {Ryan}, {Rykoff}, {Sabater}, {Sakurikar}, {Salgado}, {Sanghi}, {Saunders}, {Savchenko}, {Schwardt}, {Seifert-Eckert}, {Shih}, {Jain}, {Shukla}, {Sick}, {Simpson}, {Singanamalla}, {Singer}, {Singhal}, {Sinha}, {Sip{H{o}}cz}, {Spitler}, {Stansby}, {Streicher}, {{{S}}umak}, {Swinbank}, {Taranu}, {Tewary}, {Tremblay}, {Val-Borro}, {Van Kooten}, {Vasovi{'c}}, {Verma}, {de Miranda Cardoso}, {Williams}, {Wilson}, {Winkel}, {Wood-Vasey}, {Xue}, {Yoachim}, {Zhang}, {Zonca}, \& {Astropy Project Contributors}}]{astropy:2022}
{Astropy Collaboration}, {Price-Whelan}, A.~M., {Lim}, P.~L., {et~al.} 2022, \apj, 935, 167, \dodoi{10.3847/1538-4357/ac7c74}

\bibitem[{{Bahcall} \& {Spitzer}(1969)}]{Bahcall_Spitzer_1969}
{Bahcall}, J.~N., \& {Spitzer}, Lyman, J. 1969, \apjl, 156, L63, \dodoi{10.1086/180350}

\bibitem[{{Behroozi} {et~al.}(2019){Behroozi}, {Wechsler}, {Hearin}, \& {Conroy}}]{Behrooz_2019}
{Behroozi}, P., {Wechsler}, R.~H., {Hearin}, A.~P., \& {Conroy}, C. 2019, \mnras, 488, 3143, \dodoi{10.1093/mnras/stz1182}

\bibitem[{{Bell}(2003)}]{bell_2003}
{Bell}, E.~F. 2003, \apj, 586, 794, \dodoi{10.1086/367829}

\bibitem[{{Bentz} \& {Katz}(2015)}]{bentz_and_katz_2015}
{Bentz}, M.~C., \& {Katz}, S. 2015, \pasp, 127, 67, \dodoi{10.1086/679601}

\bibitem[{{Bentz} {et~al.}(2023){Bentz}, {Onken}, {Street}, \& {Valluri}}]{Bentz_2023}
{Bentz}, M.~C., {Onken}, C.~A., {Street}, R., \& {Valluri}, M. 2023, \apj, 944, 29, \dodoi{10.3847/1538-4357/acab62}

\bibitem[{{Berg} {et~al.}(2022){Berg}, {Bordoloi}, {Ellison}, {Oppenheimer}, \& {Werk}}]{berg_civil_2022}
{Berg}, T., {Bordoloi}, R., {Ellison}, S.~L., {Oppenheimer}, B.~D., \& {Werk}, J.~K. 2022, {The C IV in L* galaxies (CIViL*) survey - Pinpointing the physical conditions and evolutionary stages of gaseous halos}, HST Proposal. Cycle 30, ID. \#17076

\bibitem[{{Berg} {et~al.}(2018){Berg}, {Ellison}, {Tumlinson}, {Oppenheimer}, {Horton}, {Bordoloi}, \& {Schaye}}]{berg_2018}
{Berg}, T. A.~M., {Ellison}, S.~L., {Tumlinson}, J., {et~al.} 2018, \mnras, 478, 3890, \dodoi{10.1093/mnras/sty962}

\bibitem[{{Bergeron}(1986)}]{bergeron1986}
{Bergeron}, J. 1986, \aap, 155, L8

\bibitem[{{Bernardi}(2007)}]{Bernardi_2007}
{Bernardi}, M. 2007, \aj, 133, 1954, \dodoi{10.1086/512611}

\bibitem[{{Best} \& {Heckman}(2012)}]{best_2012}
{Best}, P.~N., \& {Heckman}, T.~M. 2012, \mnras, 421, 1569, \dodoi{10.1111/j.1365-2966.2012.20414.x}

\bibitem[{Bingham {et~al.}(2019)Bingham, Chen, Jankowiak, Obermeyer, Pradhan, Karaletsos, Singh, Szerlip, Horsfall, \& Goodman}]{bingham2019pyro}
Bingham, E., Chen, J.~P., Jankowiak, M., {et~al.} 2019, J. Mach. Learn. Res., 20, 28:1.
\newblock \url{http://jmlr.org/papers/v20/18-403.html}

\bibitem[{{Blakeslee} {et~al.}(2010){Blakeslee}, {Cantiello}, {Mei}, {C{\^o}t{\'e}}, {Barber DeGraaff}, {Ferrarese}, {Jord{\'a}n}, {Peng}, {Tonry}, \& {Worthey}}]{Blakeslee_2010}
{Blakeslee}, J.~P., {Cantiello}, M., {Mei}, S., {et~al.} 2010, \apj, 724, 657, \dodoi{10.1088/0004-637X/724/1/657}

\bibitem[{{Blandford} \& {Znajek}(1977)}]{blandfordznajek1977}
{Blandford}, R.~D., \& {Znajek}, R.~L. 1977, \mnras, 179, 433, \dodoi{10.1093/mnras/179.3.433}

\bibitem[{{Blanton} \& {Roweis}(2007)}]{Blanton_2007}
{Blanton}, M.~R., \& {Roweis}, S. 2007, \aj, 133, 734, \dodoi{10.1086/510127}

\bibitem[{{Bondi} \& {Hoyle}(1944)}]{Bondi_Hoyle_1944}
{Bondi}, H., \& {Hoyle}, F. 1944, \mnras, 104, 273, \dodoi{10.1093/mnras/104.5.273}

\bibitem[{{Booth} \& {Schaye}(2009)}]{booth_and_schaye_2009}
{Booth}, C.~M., \& {Schaye}, J. 2009, \mnras, 398, 53, \dodoi{10.1111/j.1365-2966.2009.15043.x}

\bibitem[{{Bordoloi} {et~al.}(2011){Bordoloi}, {Lilly}, {Knobel}, {Bolzonella}, {Kampczyk}, {Carollo}, {Iovino}, {Zucca}, {Contini}, {Kneib}, {Le Fevre}, {Mainieri}, {Renzini}, {Scodeggio}, {Zamorani}, {Balestra}, {Bardelli}, {Bongiorno}, {Caputi}, {Cucciati}, {de la Torre}, {de Ravel}, {Garilli}, {Kova{\v{c}}}, {Lamareille}, {Le Borgne}, {Le Brun}, {Maier}, {Mignoli}, {Pello}, {Peng}, {Perez Montero}, {Presotto}, {Scarlata}, {Silverman}, {Tanaka}, {Tasca}, {Tresse}, {Vergani}, {Barnes}, {Cappi}, {Cimatti}, {Coppa}, {Diener}, {Franzetti}, {Koekemoer}, {L{\'o}pez-Sanjuan}, {McCracken}, {Moresco}, {Nair}, {Oesch}, {Pozzetti}, \& {Welikala}}]{Bordoloi_2011}
{Bordoloi}, R., {Lilly}, S.~J., {Knobel}, C., {et~al.} 2011, \apj, 743, 10, \dodoi{10.1088/0004-637X/743/1/10}

\bibitem[{{Bordoloi} {et~al.}(2014){Bordoloi}, {Tumlinson}, {Werk}, {Oppenheimer}, {Peeples}, {Prochaska}, {Tripp}, {Katz}, {Dav{\'e}}, {Fox}, {Thom}, {Ford}, {Weinberg}, {Burchett}, \& {Kollmeier}}]{Bordoloi_2014}
{Bordoloi}, R., {Tumlinson}, J., {Werk}, J.~K., {et~al.} 2014, \apj, 796, 136, \dodoi{10.1088/0004-637X/796/2/136}

\bibitem[{{Borthakur} {et~al.}(2013){Borthakur}, {Heckman}, {Strickland}, {Wild}, \& {Schiminovich}}]{Borthakur_2013}
{Borthakur}, S., {Heckman}, T., {Strickland}, D., {Wild}, V., \& {Schiminovich}, D. 2013, \apj, 768, 18, \dodoi{10.1088/0004-637X/768/1/18}

\bibitem[{{Bowen} {et~al.}(2016){Bowen}, {Chelouche}, {Jenkins}, {Tripp}, {Pettini}, {York}, \& {Frye}}]{Bowen_2016}
{Bowen}, D.~V., {Chelouche}, D., {Jenkins}, E.~B., {et~al.} 2016, \apj, 826, 50, \dodoi{10.3847/0004-637X/826/1/50}

\bibitem[{Bradbury {et~al.}(2018)Bradbury, Frostig, Hawkins, Johnson, Leary, Maclaurin, Necula, Paszke, Vander{P}las, Wanderman-{M}ilne, \& Zhang}]{jax2018github}
Bradbury, J., Frostig, R., Hawkins, P., {et~al.} 2018, {JAX}: composable transformations of {P}ython+{N}um{P}y programs, 0.3.13.
\newblock \url{http://github.com/google/jax}

\bibitem[{Burchett(2024)}]{burchett_veeper_2024}
Burchett, J. 2024, {the Veeper}, v1.0,  Zenodo, \dodoi{10.5281/zenodo.10993984}

\bibitem[{{Burchett} {et~al.}(2018){Burchett}, {Tripp}, {Wang}, {Willmer}, {Bowen}, \& {Jenkins}}]{Burchett_2018}
{Burchett}, J.~N., {Tripp}, T.~M., {Wang}, Q.~D., {et~al.} 2018, \mnras, 475, 2067, \dodoi{10.1093/mnras/stx3170}

\bibitem[{{Burchett} {et~al.}(2016){Burchett}, {Tripp}, {Bordoloi}, {Werk}, {Prochaska}, {Tumlinson}, {Willmer}, {O'Meara}, \& {Katz}}]{burchett_2016}
{Burchett}, J.~N., {Tripp}, T.~M., {Bordoloi}, R., {et~al.} 2016, \apj, 832, 124, \dodoi{10.3847/0004-637X/832/2/124}

\bibitem[{{Cappellari}(2023)}]{Cappellari2023}
{Cappellari}, M. 2023, MNRAS, 526, 3273, \dodoi{10.1093/mnras/stad2597}

\bibitem[{{Chadayammuri} {et~al.}(2021){Chadayammuri}, {Tremmel}, {Nagai}, {Babul}, \& {Quinn}}]{Chadayammuri_2021}
{Chadayammuri}, U., {Tremmel}, M., {Nagai}, D., {Babul}, A., \& {Quinn}, T. 2021, \mnras, 504, 3922, \dodoi{10.1093/mnras/stab1010}

\bibitem[{{Chatzikos} {et~al.}(2023){Chatzikos}, {Bianchi}, {Camilloni}, {Chakraborty}, {Gunasekera}, {Guzm{\'a}n}, {Milby}, {Sarkar}, {Shaw}, {van Hoof}, \& {Ferland}}]{Chatzikos_2023}
{Chatzikos}, M., {Bianchi}, S., {Camilloni}, F., {et~al.} 2023, \rmxaa, 59, 327, \dodoi{10.22201/ia.01851101p.2023.59.02.12}

\bibitem[{{Crain} {et~al.}(2015){Crain}, {Schaye}, {Bower}, {Furlong}, {Schaller}, {Theuns}, {Dalla Vecchia}, {Frenk}, {McCarthy}, {Helly}, {Jenkins}, {Rosas-Guevara}, {White}, \& {Trayford}}]{crain_2015}
{Crain}, R.~A., {Schaye}, J., {Bower}, R.~G., {et~al.} 2015, \mnras, 450, 1937, \dodoi{10.1093/mnras/stv725}

\bibitem[{{Dalla Vecchia} \& {Schaye}(2012)}]{dalla_vecchia_and_schaye_2012}
{Dalla Vecchia}, C., \& {Schaye}, J. 2012, \mnras, 426, 140, \dodoi{10.1111/j.1365-2966.2012.21704.x}

\bibitem[{{Danforth} {et~al.}(2016){Danforth}, {Keeney}, {Tilton}, {Shull}, {Stocke}, {Stevans}, {Pieri}, {Savage}, {France}, {Syphers}, {Smith}, {Green}, {Froning}, {Penton}, \& {Osterman}}]{danforth_2016}
{Danforth}, C.~W., {Keeney}, B.~A., {Tilton}, E.~M., {et~al.} 2016, \apj, 817, 111, \dodoi{10.3847/0004-637X/817/2/111}

\bibitem[{Davidson-Pilon(2024)}]{Davidson-Pilon_lifelines_2024}
Davidson-Pilon, C. 2024, {lifelines, survival analysis in Python}, v0.28.0,  Zenodo, \dodoi{10.5281/zenodo.10456828}

\bibitem[{{Davies} {et~al.}(2019){Davies}, {Crain}, {McCarthy}, {Oppenheimer}, {Schaye}, {Schaller}, \& {McAlpine}}]{Davies_2019}
{Davies}, J.~J., {Crain}, R.~A., {McCarthy}, I.~G., {et~al.} 2019, \mnras, 485, 3783, \dodoi{10.1093/mnras/stz635}

\bibitem[{{Davies} {et~al.}(2020){Davies}, {Crain}, {Oppenheimer}, \& {Schaye}}]{Davies_2020}
{Davies}, J.~J., {Crain}, R.~A., {Oppenheimer}, B.~D., \& {Schaye}, J. 2020, \mnras, 491, 4462, \dodoi{10.1093/mnras/stz3201}

\bibitem[{{Davies} {et~al.}(2021){Davies}, {Crain}, \& {Pontzen}}]{davies_2021}
{Davies}, J.~J., {Crain}, R.~A., \& {Pontzen}, A. 2021, \mnras, 501, 236, \dodoi{10.1093/mnras/staa3643}

\bibitem[{{Davies} {et~al.}(2022){Davies}, {Pontzen}, \& {Crain}}]{davies_2022}
{Davies}, J.~J., {Pontzen}, A., \& {Crain}, R.~A. 2022, \mnras, 515, 1430, \dodoi{10.1093/mnras/stac1742}

\bibitem[{{Davies} {et~al.}(2024){Davies}, {Pontzen}, \& {Crain}}]{Davies_2024}
---. 2024, \mnras, 527, 4705, \dodoi{10.1093/mnras/stad3456}

\bibitem[{{Davis} {et~al.}(2017){Davis}, {Graham}, \& {Seigar}}]{davis_2017}
{Davis}, B.~L., {Graham}, A.~W., \& {Seigar}, M.~S. 2017, \mnras, 471, 2187, \dodoi{10.1093/mnras/stx1794}

\bibitem[{{Faerman} {et~al.}(2022){Faerman}, {Pandya}, {Somerville}, \& {Sternberg}}]{Faerman_2022}
{Faerman}, Y., {Pandya}, V., {Somerville}, R.~S., \& {Sternberg}, A. 2022, \apj, 928, 37, \dodoi{10.3847/1538-4357/ac4ca6}

\bibitem[{{Faerman} {et~al.}(2020){Faerman}, {Sternberg}, \& {McKee}}]{Faerman_2020}
{Faerman}, Y., {Sternberg}, A., \& {McKee}, C.~F. 2020, \apj, 893, 82, \dodoi{10.3847/1538-4357/ab7ffc}

\bibitem[{{Ferland} {et~al.}(1998){Ferland}, {Korista}, {Verner}, {Ferguson}, {Kingdon}, \& {Verner}}]{ferland_1998}
{Ferland}, G.~J., {Korista}, K.~T., {Verner}, D.~A., {et~al.} 1998, \pasp, 110, 761, \dodoi{10.1086/316190}

\bibitem[{{Ferland} {et~al.}(2013){Ferland}, {Porter}, {van Hoof}, {Williams}, {Abel}, {Lykins}, {Shaw}, {Henney}, \& {Stancil}}]{Ferland_2013}
{Ferland}, G.~J., {Porter}, R.~L., {van Hoof}, P.~A.~M., {et~al.} 2013, \rmxaa, 49, 137, \dodoi{10.48550/arXiv.1302.4485}

\bibitem[{{Ferrarese} \& {Merritt}(2000)}]{Ferrarese_Merrit_2000}
{Ferrarese}, L., \& {Merritt}, D. 2000, \apjl, 539, L9, \dodoi{10.1086/312838}

\bibitem[{{Froning} \& {Green}(2009)}]{Froning_and_Green_2009}
{Froning}, C.~S., \& {Green}, J.~C. 2009, \apss, 320, 181, \dodoi{10.1007/s10509-008-9758-y}

\bibitem[{{Gebhardt} {et~al.}(2000){Gebhardt}, {Bender}, {Bower}, {Dressler}, {Faber}, {Filippenko}, {Green}, {Grillmair}, {Ho}, {Kormendy}, {Lauer}, {Magorrian}, {Pinkney}, {Richstone}, \& {Tremaine}}]{Gebhardt_2000}
{Gebhardt}, K., {Bender}, R., {Bower}, G., {et~al.} 2000, \apjl, 539, L13, \dodoi{10.1086/312840}

\bibitem[{{Gnat} \& {Sternberg}(2007)}]{Gnat_2007}
{Gnat}, O., \& {Sternberg}, A. 2007, \apjs, 168, 213, \dodoi{10.1086/509786}

\bibitem[{{Green} {et~al.}(2012){Green}, {Froning}, {Osterman}, {Ebbets}, {Heap}, {Leitherer}, {Linsky}, {Savage}, {Sembach}, {Shull}, {Siegmund}, {Snow}, {Spencer}, {Stern}, {Stocke}, {Welsh}, {B{\'e}land}, {Burgh}, {Danforth}, {France}, {Keeney}, {McPhate}, {Penton}, {Andrews}, {Brownsberger}, {Morse}, \& {Wilkinson}}]{Green_2012}
{Green}, J.~C., {Froning}, C.~S., {Osterman}, S., {et~al.} 2012, \apj, 744, 60, \dodoi{10.1088/0004-637X/744/1/6010.1086/141956}

\bibitem[{{Haardt} \& {Madau}(2012{\natexlab{a}})}]{haardt_manau_2012}
{Haardt}, F., \& {Madau}, P. 2012{\natexlab{a}}, \apj, 746, 125, \dodoi{10.1088/0004-637X/746/2/125}

\bibitem[{{Haardt} \& {Madau}(2012{\natexlab{b}})}]{haardt&madau2012}
---. 2012{\natexlab{b}}, \apj, 746, 125, \dodoi{10.1088/0004-637X/746/2/125}

\bibitem[{{Haehnelt} {et~al.}(1998){Haehnelt}, {Natarajan}, \& {Rees}}]{Haehnelt_Natarajan_Rees1998}
{Haehnelt}, M.~G., {Natarajan}, P., \& {Rees}, M.~J. 1998, \mnras, 300, 817, \dodoi{10.1046/j.1365-8711.1998.01951.x}

\bibitem[{{H{\"a}ring} \& {Rix}(2004)}]{Haring_Rix_2004}
{H{\"a}ring}, N., \& {Rix}, H.-W. 2004, \apjl, 604, L89, \dodoi{10.1086/383567}

\bibitem[{Harris {et~al.}(2020)Harris, Millman, van~der Walt, Gommers, Virtanen, Cournapeau, Wieser, Taylor, Berg, Smith, Kern, Picus, Hoyer, van Kerkwijk, Brett, Haldane, del R{\'{i}}o, Wiebe, Peterson, G{\'{e}}rard-Marchant, Sheppard, Reddy, Weckesser, Abbasi, Gohlke, \& Oliphant}]{harris2020array}
Harris, C.~R., Millman, K.~J., van~der Walt, S.~J., {et~al.} 2020, Nature, 585, 357, \dodoi{10.1038/s41586-020-2649-2}

\bibitem[{{Ho} {et~al.}(1997){Ho}, {Filippenko}, \& {Sargent}}]{Ho_1997}
{Ho}, L.~C., {Filippenko}, A.~V., \& {Sargent}, W. L.~W. 1997, \apj, 487, 568, \dodoi{10.1086/304638}

\bibitem[{{Ho} {et~al.}(2017){Ho}, {Martin}, {Kacprzak}, \& {Churchill}}]{Ho_2017}
{Ho}, S.~H., {Martin}, C.~L., {Kacprzak}, G.~G., \& {Churchill}, C.~W. 2017, \apj, 835, 267, \dodoi{10.3847/1538-4357/835/2/267}

\bibitem[{{Hopkins}(2013)}]{Hopkins_2013}
{Hopkins}, P.~F. 2013, \mnras, 428, 2840, \dodoi{10.1093/mnras/sts210}

\bibitem[{{Hu} \& {Kravtsov}(2003)}]{Hu_2003}
{Hu}, W., \& {Kravtsov}, A.~V. 2003, \apj, 584, 702, \dodoi{10.1086/345846}

\bibitem[{Hunter(2007)}]{Hunter:2007}
Hunter, J.~D. 2007, Computing in Science \& Engineering, 9, 90, \dodoi{10.1109/MCSE.2007.55}

\bibitem[{{Johnson} {et~al.}(2015){Johnson}, {Chen}, \& {Mulchaey}}]{Johnson_2015}
{Johnson}, S.~D., {Chen}, H.-W., \& {Mulchaey}, J.~S. 2015, \mnras, 449, 3263, \dodoi{10.1093/mnras/stv553}

\bibitem[{{Jung} {et~al.}(2022){Jung}, {Rennehan}, {Saeedzadeh}, {Babul}, {Tremmel}, {Quinn}, {Loubser}, {O'Sullivan}, \& {Yi}}]{Jung_2022}
{Jung}, S.~L., {Rennehan}, D., {Saeedzadeh}, V., {et~al.} 2022, \mnras, 515, 22, \dodoi{10.1093/mnras/stac1622}

\bibitem[{{Kacprzak} {et~al.}(2019){Kacprzak}, {Vander Vliet}, {Nielsen}, {Muzahid}, {Pointon}, {Churchill}, {Ceverino}, {Arraki}, {Klypin}, {Charlton}, \& {Lewis}}]{Kacprazak_2019}
{Kacprzak}, G.~G., {Vander Vliet}, J.~R., {Nielsen}, N.~M., {et~al.} 2019, \apj, 870, 137, \dodoi{10.3847/1538-4357/aaf1a6}

\bibitem[{{Kauffmann} {et~al.}(2003){Kauffmann}, {Heckman}, {White}, {Charlot}, {Tremonti}, {Brinchmann}, {Bruzual}, {Peng}, {Seibert}, {Bernardi}, {Blanton}, {Brinkmann}, {Castander}, {Cs{\'a}bai}, {Fukugita}, {Ivezic}, {Munn}, {Nichol}, {Padmanabhan}, {Thakar}, {Weinberg}, \& {York}}]{Kauffmann_2003}
{Kauffmann}, G., {Heckman}, T.~M., {White}, S. D.~M., {et~al.} 2003, \mnras, 341, 33, \dodoi{10.1046/j.1365-8711.2003.06291.x}

\bibitem[{{Kennicutt} \& {Evans}(2012)}]{kennicutt_evans_2012}
{Kennicutt}, R.~C., \& {Evans}, N.~J. 2012, \araa, 50, 531, \dodoi{10.1146/annurev-astro-081811-125610}

\bibitem[{{Khaire} \& {Srianand}(2019)}]{Khaire_19}
{Khaire}, V., \& {Srianand}, R. 2019, \mnras, 484, 4174, \dodoi{10.1093/mnras/stz174}

\bibitem[{{Kormendy} \& {Ho}(2013)}]{K_ho_2013}
{Kormendy}, J., \& {Ho}, L.~C. 2013, \araa, 51, 511, \dodoi{10.1146/annurev-astro-082708-101811}

\bibitem[{{Kormendy} \& {Richstone}(1995)}]{kormendy_richstone_1995}
{Kormendy}, J., \& {Richstone}, D. 1995, \araa, 33, 581, \dodoi{10.1146/annurev.aa.33.090195.003053}

\bibitem[{{Koss} {et~al.}(2022){Koss}, {Trakhtenbrot}, {Ricci}, {Oh}, {Bauer}, {Stern}, {Caglar}, {den Brok}, {Mushotzky}, {Ricci}, {Mej{\'\i}a-Restrepo}, {Lamperti}, {Treister}, {B{\"a}r}, {Harrison}, {Powell}, {Privon}, {Riffel}, {Rojas}, {Schawinski}, \& {Urry}}]{Koss_2022}
{Koss}, M.~J., {Trakhtenbrot}, B., {Ricci}, C., {et~al.} 2022, \apjs, 261, 6, \dodoi{10.3847/1538-4365/ac650b}

\bibitem[{{Kroupa}(2001)}]{kroupa2001}
{Kroupa}, P. 2001, \mnras, 322, 231, \dodoi{10.1046/j.1365-8711.2001.04022.x}

\bibitem[{{L{\"a}sker} {et~al.}(2016){L{\"a}sker}, {Greene}, {Seth}, {van de Ven}, {Braatz}, {Henkel}, \& {Lo}}]{Lasker_2016}
{L{\"a}sker}, R., {Greene}, J.~E., {Seth}, A., {et~al.} 2016, \apj, 825, 3, \dodoi{10.3847/0004-637X/825/1/3}

\bibitem[{Lee(2020)}]{lee_2020}
Lee, L. 2020, NADA: Nondetects and Data Analysis for Environmental Data.
\newblock \url{https://CRAN.R-project.org/package=NADA}

\bibitem[{{Lehner} \& {Howk}(2011)}]{Lehner_howk_2011}
{Lehner}, N., \& {Howk}, J.~C. 2011, Science, 334, 955, \dodoi{10.1126/science.1209069}

\bibitem[{{Lehner} {et~al.}(2020){Lehner}, {Berek}, {Howk}, {Wakker}, {Tumlinson}, {Jenkins}, {Prochaska}, {Augustin}, {Ji}, {Faucher-Gigu{\`e}re}, {Hafen}, {Peeples}, {Barger}, {Berg}, {Bordoloi}, {Brown}, {Fox}, {Gilbert}, {Guhathakurta}, {Kalirai}, {Lockman}, {O'Meara}, {Pisano}, {Ribaudo}, \& {Werk}}]{Lehner_2020}
{Lehner}, N., {Berek}, S.~C., {Howk}, J.~C., {et~al.} 2020, \apj, 900, 9, \dodoi{10.3847/1538-4357/aba49c}

\bibitem[{{Lewis} {et~al.}(2015){Lewis}, {Dolphin}, {Dalcanton}, {Weisz}, {Williams}, {Bell}, {Seth}, {Simones}, {Skillman}, {Choi}, {Fouesneau}, {Guhathakurta}, {Johnson}, {Kalirai}, {Leroy}, {Monachesi}, {Rix}, \& {Schruba}}]{Lewis_2015}
{Lewis}, A.~R., {Dolphin}, A.~E., {Dalcanton}, J.~J., {et~al.} 2015, \apj, 805, 183, \dodoi{10.1088/0004-637X/805/2/183}

\bibitem[{{Magorrian} {et~al.}(1998){Magorrian}, {Tremaine}, {Richstone}, {Bender}, {Bower}, {Dressler}, {Faber}, {Gebhardt}, {Green}, {Grillmair}, {Kormendy}, \& {Lauer}}]{magorrian_1998}
{Magorrian}, J., {Tremaine}, S., {Richstone}, D., {et~al.} 1998, \aj, 115, 2285, \dodoi{10.1086/300353}

\bibitem[{Martin {et~al.}(2023)Martin, Hartikainen, Abril-Pla, Colin, Kumar, Rosheen~Naeem, Arroyuelo, Gautam, rpgoldman, Banerjea, Pasricha, Sanjay, Gruevski, Rochford, Axen, Mahweshwari, Matamoros, Phan, \& Zinkov}]{Martin_arviz_2023}
Martin, O.~A., Hartikainen, A., Abril-Pla, O., {et~al.} 2023, {ArviZ}, v0.17.0,  Zenodo, \dodoi{10.5281/zenodo.10436212}

\bibitem[{{Mathews} \& {Prochaska}(2017)}]{mathews_prochaska_2017}
{Mathews}, W.~G., \& {Prochaska}, J.~X. 2017, \apjl, 846, L24, \dodoi{10.3847/2041-8213/aa8861}

\bibitem[{{McConnell} \& {Ma}(2013)}]{mcconnell_2013}
{McConnell}, N.~J., \& {Ma}, C.-P. 2013, \apj, 764, 184, \dodoi{10.1088/0004-637X/764/2/184}

\bibitem[{{McQuinn} \& {Werk}(2018)}]{mcquinn_werk_2018}
{McQuinn}, M., \& {Werk}, J.~K. 2018, \apj, 852, 33, \dodoi{10.3847/1538-4357/aa9d3f}

\bibitem[{{Menon} {et~al.}(2015){Menon}, {Wesolowski}, {Zheng}, {Jetley}, {Kale}, {Quinn}, \& {Governato}}]{menon_2015}
{Menon}, H., {Wesolowski}, L., {Zheng}, G., {et~al.} 2015, Computational Astrophysics and Cosmology, 2, 1, \dodoi{10.1186/s40668-015-0007-9}

\bibitem[{{Miyoshi} {et~al.}(1995){Miyoshi}, {Moran}, {Herrnstein}, {Greenhill}, {Nakai}, {Diamond}, \& {Inoue}}]{miyoshi_1995}
{Miyoshi}, M., {Moran}, J., {Herrnstein}, J., {et~al.} 1995, \nat, 373, 127, \dodoi{10.1038/373127a0}

\bibitem[{{Molina} {et~al.}(2018){Molina}, {Eracleous}, {Barth}, {Maoz}, {Runnoe}, {Ho}, {Shields}, \& {Walsh}}]{Molina_2018}
{Molina}, M., {Eracleous}, M., {Barth}, A.~J., {et~al.} 2018, \apj, 864, 90, \dodoi{10.3847/1538-4357/aad5ed}

\bibitem[{{Monroe} {et~al.}(2016){Monroe}, {Prochaska}, {Tejos}, {Worseck}, {Hennawi}, {Schmidt}, {Tumlinson}, \& {Shen}}]{monroe_2016}
{Monroe}, T.~R., {Prochaska}, J.~X., {Tejos}, N., {et~al.} 2016, \aj, 152, 25, \dodoi{10.3847/0004-6256/152/1/25}

\bibitem[{{Nelson} {et~al.}(2018{\natexlab{a}}){Nelson}, {Pillepich}, {Springel}, {Weinberger}, {Hernquist}, {Pakmor}, {Genel}, {Torrey}, {Vogelsberger}, {Kauffmann}, {Marinacci}, \& {Naiman}}]{Nelson_2018}
{Nelson}, D., {Pillepich}, A., {Springel}, V., {et~al.} 2018{\natexlab{a}}, \mnras, 475, 624, \dodoi{10.1093/mnras/stx3040}

\bibitem[{{Nelson} {et~al.}(2018{\natexlab{b}}){Nelson}, {Kauffmann}, {Pillepich}, {Genel}, {Springel}, {Pakmor}, {Hernquist}, {Weinberger}, {Torrey}, {Vogelsberger}, \& {Marinacci}}]{Nelson_2018_oxygen}
{Nelson}, D., {Kauffmann}, G., {Pillepich}, A., {et~al.} 2018{\natexlab{b}}, \mnras, 477, 450, \dodoi{10.1093/mnras/sty656}

\bibitem[{{Nelson} {et~al.}(2019){Nelson}, {Pillepich}, {Springel}, {Pakmor}, {Weinberger}, {Genel}, {Torrey}, {Vogelsberger}, {Marinacci}, \& {Hernquist}}]{Nelson_2019b}
{Nelson}, D., {Pillepich}, A., {Springel}, V., {et~al.} 2019, \mnras, 490, 3234, \dodoi{10.1093/mnras/stz2306}

\bibitem[{{Oppenheimer}(2018)}]{oppy_2018_solo}
{Oppenheimer}, B.~D. 2018, \mnras, 480, 2963, \dodoi{10.1093/mnras/sty1918}

\bibitem[{{Oppenheimer} {et~al.}(2016){Oppenheimer}, {Crain}, {Schaye}, {Rahmati}, {Richings}, {Trayford}, {Tumlinson}, {Bower}, {Schaller}, \& {Theuns}}]{oppenheimer_2016}
{Oppenheimer}, B.~D., {Crain}, R.~A., {Schaye}, J., {et~al.} 2016, \mnras, 460, 2157, \dodoi{10.1093/mnras/stw1066}

\bibitem[{{Oppenheimer} {et~al.}(2020){Oppenheimer}, {Davies}, {Crain}, {Wijers}, {Schaye}, {Werk}, {Burchett}, {Trayford}, \& {Horton}}]{Oppenheimer_2020}
{Oppenheimer}, B.~D., {Davies}, J.~J., {Crain}, R.~A., {et~al.} 2020, \mnras, 491, 2939, \dodoi{10.1093/mnras/stz3124}

\bibitem[{{Ostriker} \& {McKee}(1988)}]{ostriker&mckee1988}
{Ostriker}, J.~P., \& {McKee}, C.~F. 1988, Reviews of Modern Physics, 60, 1, \dodoi{10.1103/RevModPhys.60.1}

\bibitem[{{Pakmor} \& {Springel}(2013)}]{pakmor13}
{Pakmor}, R., \& {Springel}, V. 2013, \mnras, 432, 176, \dodoi{10.1093/mnras/stt428}

\bibitem[{pandas~development team(2024)}]{pandas_2024}
pandas~development team, T. 2024, {pandas-dev/pandas: Pandas}, v2.2.1,  Zenodo, \dodoi{10.5281/zenodo.10697587}

\bibitem[{{Pandya} {et~al.}(2021){Pandya}, {Fielding}, {Angl{\'e}s-Alc{\'a}zar}, {Somerville}, {Bryan}, {Hayward}, {Stern}, {Kim}, {Quataert}, {Forbes}, {Faucher-Gigu{\`e}re}, {Feldmann}, {Hafen}, {Hopkins}, {Kere{\v{s}}}, {Murray}, \& {Wetzel}}]{pandya_2021}
{Pandya}, V., {Fielding}, D.~B., {Angl{\'e}s-Alc{\'a}zar}, D., {et~al.} 2021, \mnras, 508, 2979, \dodoi{10.1093/mnras/stab2714}

\bibitem[{{P{\^a}ris} {et~al.}(2018){P{\^a}ris}, {Petitjean}, {Aubourg}, {Myers}, {Streblyanska}, {Lyke}, {Anderson}, {Armengaud}, {Bautista}, {Blanton}, {Blomqvist}, {Brinkmann}, {Brownstein}, {Brandt}, {Burtin}, {Dawson}, {de la Torre}, {Georgakakis}, {Gil-Mar{\'\i}n}, {Green}, {Hall}, {Kneib}, {LaMassa}, {Le Goff}, {MacLeod}, {Mariappan}, {McGreer}, {Merloni}, {Noterdaeme}, {Palanque-Delabrouille}, {Percival}, {Ross}, {Rossi}, {Schneider}, {Seo}, {Tojeiro}, {Weaver}, {Weijmans}, {Y{\`e}che}, {Zarrouk}, \& {Zhao}}]{Paris_2018}
{P{\^a}ris}, I., {Petitjean}, P., {Aubourg}, {\'E}., {et~al.} 2018, \aap, 613, A51, \dodoi{10.1051/0004-6361/201732445}

\bibitem[{{Peeples} {et~al.}(2014){Peeples}, {Werk}, {Tumlinson}, {Oppenheimer}, {Prochaska}, {Katz}, \& {Weinberg}}]{Peeples_2014}
{Peeples}, M.~S., {Werk}, J.~K., {Tumlinson}, J., {et~al.} 2014, \apj, 786, 54, \dodoi{10.1088/0004-637X/786/1/54}

\bibitem[{Phan {et~al.}(2019)Phan, Pradhan, \& Jankowiak}]{phan2019composable}
Phan, D., Pradhan, N., \& Jankowiak, M. 2019, arXiv preprint arXiv:1912.11554

\bibitem[{{Pillepich} {et~al.}(2018){Pillepich}, {Springel}, {Nelson}, {Genel}, {Naiman}, {Pakmor}, {Hernquist}, {Torrey}, {Vogelsberger}, {Weinberger}, \& {Marinacci}}]{pillepich_2018}
{Pillepich}, A., {Springel}, V., {Nelson}, D., {et~al.} 2018, \mnras, 473, 4077, \dodoi{10.1093/mnras/stx2656}

\bibitem[{{Piotrowska} {et~al.}(2022){Piotrowska}, {Bluck}, {Maiolino}, \& {Peng}}]{Piotrowska_2022}
{Piotrowska}, J.~M., {Bluck}, A. F.~L., {Maiolino}, R., \& {Peng}, Y. 2022, \mnras, 512, 1052, \dodoi{10.1093/mnras/stab3673}

\bibitem[{{Planck Collaboration} {et~al.}(2013){Planck Collaboration}, {Ade}, {Aghanim}, {Arnaud}, {Ashdown}, {Atrio-Barandela}, {Aumont}, {Baccigalupi}, {Balbi}, {Banday}, {Barreiro}, {Barrena}, {Bartlett}, {Battaner}, {Benabed}, {Bernard}, {Bersanelli}, {Bikmaev}, {Bock}, {B{\"o}hringer}, {Bonaldi}, {Bond}, {Borrill}, {Bouchet}, {Bourdin}, {Burenin}, {Burigana}, {Butler}, {Cabella}, {Chamballu}, {Chary}, {Chiang}, {Chon}, {Christensen}, {Clements}, {Colafrancesco}, {Colombi}, {Colombo}, {Comis}, {Coulais}, {Crill}, {Cuttaia}, {Da Silva}, {Dahle}, {Davis}, {de Bernardis}, {de Gasperis}, {de Rosa}, {de Zotti}, {Delabrouille}, {D{\'e}mocl{\`e}s}, {Diego}, {Dole}, {Donzelli}, {Dor{\'e}}, {Douspis}, {Dupac}, {Efstathiou}, {En{\ss}lin}, {Finelli}, {Flores-Cacho}, {Forni}, {Frailis}, {Franceschi}, {Frommert}, {Galeotta}, {Ganga}, {G{\'e}nova-Santos}, {Giard}, {Giraud-H{\'e}raud}, {Gonz{\'a}lez-Nuevo}, {G{\'o}rski}, {Gregorio}, {Gruppuso}, {Hansen}, {Harrison}, {Hern{\'a}ndez-Monteagudo}, {Herranz}, {Hildebrandt},
  {Hivon}, {Hobson}, {Holmes}, {Hornstrup}, {Hovest}, {Huffenberger}, {Hurier}, {Jaffe}, {Jaffe}, {Jones}, {Juvela}, {Keih{\"a}nen}, {Keskitalo}, {Khamitov}, {Kisner}, {Kneissl}, {Knoche}, {Kunz}, {Kurki-Suonio}, {L{\"a}hteenm{\"a}ki}, {Lamarre}, {Lasenby}, {Lawrence}, {Le Jeune}, {Leonardi}, {Lilje}, {Linden-V{\o}rnle}, {L{\'o}pez-Caniego}, {Lubin}, {Luzzi}, {Mac{\'\i}as-P{\'e}rez}, {MacTavish}, {Maffei}, {Maino}, {Mandolesi}, {Maris}, {Marleau}, {Marshall}, {Mart{\'\i}nez-Gonz{\'a}lez}, {Masi}, {Massardi}, {Matarrese}, {Mazzotta}, {Mei}, {Melchiorri}, {Melin}, {Mendes}, {Mennella}, {Mitra}, {Miville-Desch{\^e}nes}, {Moneti}, {Montier}, {Morgante}, {Mortlock}, {Munshi}, {Murphy}, {Naselsky}, {Nati}, {Natoli}, {N{\o}rgaard-Nielsen}, {Noviello}, {Novikov}, {Novikov}, {Osborne}, {Oxborrow}, {Pajot}, {Paoletti}, {Perotto}, {Perrotta}, {Piacentini}, {Piat}, {Pierpaoli}, {Piffaretti}, {Plaszczynski}, {Pointecouteau}, {Polenta}, {Popa}, {Poutanen}, {Pratt}, {Prunet}, {Puget}, {Rachen}, {Rebolo}, {Reinecke},
  {Remazeilles}, {Renault}, {Ricciardi}, {Ristorcelli}, {Rocha}, {Roman}, {Rosset}, {Rossetti}, {Rubi{\~n}o-Mart{\'\i}n}, {Rusholme}, {Sandri}, {Savini}, {Scott}, {Spencer}, {Starck}, {Stolyarov}, {Sudiwala}, {Sunyaev}, {Sutton}, {Suur-Uski}, {Sygnet}, {Tauber}, {Terenzi}, {Toffolatti}, {Tomasi}, {Tristram}, {Valenziano}, {Van Tent}, {Vielva}, {Villa}, {Vittorio}, {Wade}, {Wandelt}, {Wang}, {Welikala}, {Weller}, {White}, {White}, {Yvon}, {Zacchei}, \& {Zonca}}]{PCXI_2013}
{Planck Collaboration}, {Ade}, P.~A.~R., {Aghanim}, N., {et~al.} 2013, \aap, 557, A52, \dodoi{10.1051/0004-6361/201220941}

\bibitem[{{Planck Collaboration} {et~al.}(2016){Planck Collaboration}, {Ade}, {Aghanim}, {Arnaud}, {Ashdown}, {Aumont}, {Baccigalupi}, {Banday}, {Barreiro}, {Bartlett}, {Bartolo}, {Battaner}, {Battye}, {Benabed}, {Beno{\^\i}t}, {Benoit-L{\'e}vy}, {Bernard}, {Bersanelli}, {Bielewicz}, {Bock}, {Bonaldi}, {Bonavera}, {Bond}, {Borrill}, {Bouchet}, {Boulanger}, {Bucher}, {Burigana}, {Butler}, {Calabrese}, {Cardoso}, {Catalano}, {Challinor}, {Chamballu}, {Chary}, {Chiang}, {Chluba}, {Christensen}, {Church}, {Clements}, {Colombi}, {Colombo}, {Combet}, {Coulais}, {Crill}, {Curto}, {Cuttaia}, {Danese}, {Davies}, {Davis}, {de Bernardis}, {de Rosa}, {de Zotti}, {Delabrouille}, {D{\'e}sert}, {Di Valentino}, {Dickinson}, {Diego}, {Dolag}, {Dole}, {Donzelli}, {Dor{\'e}}, {Douspis}, {Ducout}, {Dunkley}, {Dupac}, {Efstathiou}, {Elsner}, {En{\ss}lin}, {Eriksen}, {Farhang}, {Fergusson}, {Finelli}, {Forni}, {Frailis}, {Fraisse}, {Franceschi}, {Frejsel}, {Galeotta}, {Galli}, {Ganga}, {Gauthier}, {Gerbino}, {Ghosh}, {Giard},
  {Giraud-H{\'e}raud}, {Giusarma}, {Gjerl{\o}w}, {Gonz{\'a}lez-Nuevo}, {G{\'o}rski}, {Gratton}, {Gregorio}, {Gruppuso}, {Gudmundsson}, {Hamann}, {Hansen}, {Hanson}, {Harrison}, {Helou}, {Henrot-Versill{\'e}}, {Hern{\'a}ndez-Monteagudo}, {Herranz}, {Hildebrandt}, {Hivon}, {Hobson}, {Holmes}, {Hornstrup}, {Hovest}, {Huang}, {Huffenberger}, {Hurier}, {Jaffe}, {Jaffe}, {Jones}, {Juvela}, {Keih{\"a}nen}, {Keskitalo}, {Kisner}, {Kneissl}, {Knoche}, {Knox}, {Kunz}, {Kurki-Suonio}, {Lagache}, {L{\"a}hteenm{\"a}ki}, {Lamarre}, {Lasenby}, {Lattanzi}, {Lawrence}, {Leahy}, {Leonardi}, {Lesgourgues}, {Levrier}, {Lewis}, {Liguori}, {Lilje}, {Linden-V{\o}rnle}, {L{\'o}pez-Caniego}, {Lubin}, {Mac{\'\i}as-P{\'e}rez}, {Maggio}, {Maino}, {Mandolesi}, {Mangilli}, {Marchini}, {Maris}, {Martin}, {Martinelli}, {Mart{\'\i}nez-Gonz{\'a}lez}, {Masi}, {Matarrese}, {McGehee}, {Meinhold}, {Melchiorri}, {Melin}, {Mendes}, {Mennella}, {Migliaccio}, {Millea}, {Mitra}, {Miville-Desch{\^e}nes}, {Moneti}, {Montier}, {Morgante}, {Mortlock},
  {Moss}, {Munshi}, {Murphy}, {Naselsky}, {Nati}, {Natoli}, {Netterfield}, {N{\o}rgaard-Nielsen}, {Noviello}, {Novikov}, {Novikov}, {Oxborrow}, {Paci}, {Pagano}, {Pajot}, {Paladini}, {Paoletti}, {Partridge}, {Pasian}, {Patanchon}, {Pearson}, {Perdereau}, {Perotto}, {Perrotta}, {Pettorino}, {Piacentini}, {Piat}, {Pierpaoli}, {Pietrobon}, {Plaszczynski}, {Pointecouteau}, {Polenta}, {Popa}, {Pratt}, {Pr{\'e}zeau}, {Prunet}, {Puget}, {Rachen}, {Reach}, {Rebolo}, {Reinecke}, {Remazeilles}, {Renault}, {Renzi}, {Ristorcelli}, {Rocha}, {Rosset}, {Rossetti}, {Roudier}, {Rouill{\'e} d'Orfeuil}, {Rowan-Robinson}, {Rubi{\~n}o-Mart{\'\i}n}, {Rusholme}, {Said}, {Salvatelli}, {Salvati}, {Sandri}, {Santos}, {Savelainen}, {Savini}, {Scott}, {Seiffert}, {Serra}, {Shellard}, {Spencer}, {Spinelli}, {Stolyarov}, {Stompor}, {Sudiwala}, {Sunyaev}, {Sutton}, {Suur-Uski}, {Sygnet}, {Tauber}, {Terenzi}, {Toffolatti}, {Tomasi}, {Tristram}, {Trombetti}, {Tucci}, {Tuovinen}, {T{\"u}rler}, {Umana}, {Valenziano}, {Valiviita}, {Van Tent},
  {Vielva}, {Villa}, {Wade}, {Wandelt}, {Wehus}, {White}, {White}, {Wilkinson}, {Yvon}, {Zacchei}, \& {Zonca}}]{Plank2016}
---. 2016, \aap, 594, A13, \dodoi{10.1051/0004-6361/201525830}

\bibitem[{{Pontzen} {et~al.}(2013){Pontzen}, {Ro{\v s}kar}, {Stinson}, {Woods}, {Reed}, {Coles}, \& {Quinn}}]{pynbody}
{Pontzen}, A., {Ro{\v s}kar}, R., {Stinson}, G.~S., {et~al.} 2013, {pynbody: Astrophysics Simulation Analysis for Python}

\bibitem[{Prochaska {et~al.}(2017{\natexlab{a}})Prochaska, Tejos, Crighton, jnburchett, tiffanyhsyu, Tuo-Ji, marijana777, ktirimba, jhennawi, Cooke, O'Meara, \& Werk}]{prochaska_linetools_2017}
Prochaska, J.~X., Tejos, N., Crighton, N., {et~al.} 2017{\natexlab{a}}, {Linetools/Linetools: Third Minor Release}, v0.3,  Zenodo, \dodoi{10.5281/zenodo.1036773}

\bibitem[{Prochaska {et~al.}(2017{\natexlab{b}})Prochaska, Tejos, cwotta, jnburchett, Fumagalli, marijana777, O'Meara, Werk, Marc, mneeleman, \& kheegan}]{prochaska_pyigm_2017}
Prochaska, J.~X., Tejos, N., cwotta, {et~al.} 2017{\natexlab{b}}, {Pyigm/Pyigm: Initial release for publications}, v1.0,  Zenodo, \dodoi{10.5281/zenodo.1045479}

\bibitem[{{Reines} \& {Volonteri}(2015)}]{reines_2015}
{Reines}, A.~E., \& {Volonteri}, M. 2015, \apj, 813, 82, \dodoi{10.1088/0004-637X/813/2/82}

\bibitem[{{Rieke} {et~al.}(2009){Rieke}, {Alonso-Herrero}, {Weiner}, {P{\'e}rez-Gonz{\'a}lez}, {Blaylock}, {Donley}, \& {Marcillac}}]{Reike_2009}
{Rieke}, G.~H., {Alonso-Herrero}, A., {Weiner}, B.~J., {et~al.} 2009, \apj, 692, 556, \dodoi{10.1088/0004-637X/692/1/556}

\bibitem[{{Rosas-Guevara} {et~al.}(2016){Rosas-Guevara}, {Bower}, {Schaye}, {McAlpine}, {Dalla Vecchia}, {Frenk}, {Schaller}, \& {Theuns}}]{rosas_guevara_2016}
{Rosas-Guevara}, Y., {Bower}, R.~G., {Schaye}, J., {et~al.} 2016, \mnras, 462, 190, \dodoi{10.1093/mnras/stw1679}

\bibitem[{{Rosas-Guevara} {et~al.}(2015){Rosas-Guevara}, {Bower}, {Schaye}, {Furlong}, {Frenk}, {Booth}, {Crain}, {Dalla Vecchia}, {Schaller}, \& {Theuns}}]{rosas-guevara_2015}
{Rosas-Guevara}, Y.~M., {Bower}, R.~G., {Schaye}, J., {et~al.} 2015, \mnras, 454, 1038, \dodoi{10.1093/mnras/stv2056}

\bibitem[{{Saglia} {et~al.}(2016){Saglia}, {Opitsch}, {Erwin}, {Thomas}, {Beifiori}, {Fabricius}, {Mazzalay}, {Nowak}, {Rusli}, \& {Bender}}]{saglia_2016}
{Saglia}, R.~P., {Opitsch}, M., {Erwin}, P., {et~al.} 2016, \apj, 818, 47, \dodoi{10.3847/0004-637X/818/1/47}

\bibitem[{{Sanchez} {et~al.}(2021){Sanchez}, {Tremmel}, {Werk}, {Pontzen}, {Christensen}, {Quinn}, {Loebman}, \& {Cruz}}]{sanchez_2021}
{Sanchez}, N.~N., {Tremmel}, M., {Werk}, J.~K., {et~al.} 2021, \apj, 911, 116, \dodoi{10.3847/1538-4357/abeb15}

\bibitem[{{Sanchez} {et~al.}(2019){Sanchez}, {Werk}, {Tremmel}, {Pontzen}, {Christensen}, {Quinn}, \& {Cruz}}]{sanchez_2019}
{Sanchez}, N.~N., {Werk}, J.~K., {Tremmel}, M., {et~al.} 2019, \apj, 882, 8, \dodoi{10.3847/1538-4357/ab3045}

\bibitem[{{Sanchez} {et~al.}(in prep){Sanchez}, {Werk}, {Christensen}, {Telford}, {Tremmel}, {Quinn}, {Mead}, {Sharma}, \& {Brooks}}]{Sanchez_2023}
{Sanchez}, N.~N., {Werk}, J.~K., {Christensen}, C., {et~al.} in prep, arXiv e-prints, arXiv:2305.07672, \dodoi{10.48550/arXiv.2305.07672}

\bibitem[{{Schaye} \& {Dalla Vecchia}(2008)}]{Schaye_dalla_vecchia_2008}
{Schaye}, J., \& {Dalla Vecchia}, C. 2008, \mnras, 383, 1210, \dodoi{10.1111/j.1365-2966.2007.12639.x}

\bibitem[{{Schaye} {et~al.}(2015){Schaye}, {Crain}, {Bower}, {Furlong}, {Schaller}, {Theuns}, {Dalla Vecchia}, {Frenk}, {McCarthy}, {Helly}, {Jenkins}, {Rosas-Guevara}, {White}, {Baes}, {Booth}, {Camps}, {Navarro}, {Qu}, {Rahmati}, {Sawala}, {Thomas}, \& {Trayford}}]{schaye_2015}
{Schaye}, J., {Crain}, R.~A., {Bower}, R.~G., {et~al.} 2015, \mnras, 446, 521, \dodoi{10.1093/mnras/stu2058}

\bibitem[{{Sharma} {et~al.}(2020){Sharma}, {Brooks}, {Somerville}, {Tremmel}, {Bellovary}, {Wright}, \& {Quinn}}]{Sharma_2020}
{Sharma}, R.~S., {Brooks}, A.~M., {Somerville}, R.~S., {et~al.} 2020, \apj, 897, 103, \dodoi{10.3847/1538-4357/ab960e}

\bibitem[{{Sijacki} {et~al.}(2015){Sijacki}, {Vogelsberger}, {Genel}, {Springel}, {Torrey}, {Snyder}, {Nelson}, \& {Hernquist}}]{sijacki_2015}
{Sijacki}, D., {Vogelsberger}, M., {Genel}, S., {et~al.} 2015, \mnras, 452, 575, \dodoi{10.1093/mnras/stv1340}

\bibitem[{{Silk} \& {Rees}(1998)}]{Silk_rees_1998}
{Silk}, J., \& {Rees}, M.~J. 1998, \aap, 331, L1, \dodoi{10.48550/arXiv.astro-ph/9801013}

\bibitem[{{Springel}(2005)}]{Springel_2005}
{Springel}, V. 2005, \mnras, 364, 1105, \dodoi{10.1111/j.1365-2966.2005.09655.x}

\bibitem[{{Springel}(2010)}]{springel10}
---. 2010, \mnras, 401, 791, \dodoi{10.1111/j.1365-2966.2009.15715.x}

\bibitem[{{Stinson} {et~al.}(2012{\natexlab{a}}){Stinson}, {Brook}, {Prochaska}, {Hennawi}, {Shen}, {Wadsley}, {Pontzen}, {Couchman}, {Quinn}, {Macci{\`o}}, \& {Gibson}}]{stinson2012}
{Stinson}, G.~S., {Brook}, C., {Prochaska}, J.~X., {et~al.} 2012{\natexlab{a}}, \mnras, 425, 1270, \dodoi{10.1111/j.1365-2966.2012.21522.x}

\bibitem[{{Stinson} {et~al.}(2012{\natexlab{b}}){Stinson}, {Brook}, {Prochaska}, {Hennawi}, {Shen}, {Wadsley}, {Pontzen}, {Couchman}, {Quinn}, {Macci{\`o}}, \& {Gibson}}]{Stinson_2012}
---. 2012{\natexlab{b}}, \mnras, 425, 1270, \dodoi{10.1111/j.1365-2966.2012.21522.x}

\bibitem[{{Tchernyshyov} {et~al.}(2022){Tchernyshyov}, {Werk}, {Wilde}, {Prochaska}, {Tripp}, {Burchett}, {Bordoloi}, {Howk}, {Lehner}, {O'Meara}, {Tejos}, \& {Tumlinson}}]{Tchernyshyov2022}
{Tchernyshyov}, K., {Werk}, J.~K., {Wilde}, M.~C., {et~al.} 2022, arXiv e-prints, arXiv:2211.06436, \dodoi{10.48550/arXiv.2211.06436}

\bibitem[{{Tchernyshyov} {et~al.}(2023){Tchernyshyov}, {Werk}, {Wilde}, {Prochaska}, {Tripp}, {Burchett}, {Bordoloi}, {Howk}, {Lehner}, {O'Meara}, {Tejos}, \& {Tumlinson}}]{Tchernyshyov_2023}
---. 2023, \apj, 949, 41, \dodoi{10.3847/1538-4357/acc86a}

\bibitem[{{Terrazas} {et~al.}(2016){Terrazas}, {Bell}, {Henriques}, {White}, {Cattaneo}, \& {Woo}}]{Terrazas_2016}
{Terrazas}, B.~A., {Bell}, E.~F., {Henriques}, B. M.~B., {et~al.} 2016, \apjl, 830, L12, \dodoi{10.3847/2041-8205/830/1/L12}

\bibitem[{{Terrazas} {et~al.}(2017){Terrazas}, {Bell}, {Woo}, \& {Henriques}}]{Terrazas_2017}
{Terrazas}, B.~A., {Bell}, E.~F., {Woo}, J., \& {Henriques}, B. M.~B. 2017, \apj, 844, 170, \dodoi{10.3847/1538-4357/aa7d07}

\bibitem[{{Terrazas} {et~al.}(2020){Terrazas}, {Bell}, {Pillepich}, {Nelson}, {Somerville}, {Genel}, {Weinberger}, {Habouzit}, {Li}, {Hernquist}, \& {Vogelsberger}}]{terrazas20}
{Terrazas}, B.~A., {Bell}, E.~F., {Pillepich}, A., {et~al.} 2020, \mnras, 493, 1888, \dodoi{10.1093/mnras/staa374}

\bibitem[{{Thielemann} {et~al.}(1986){Thielemann}, {Nomoto}, \& {Yokoi}}]{thielemann1986}
{Thielemann}, F.~K., {Nomoto}, K., \& {Yokoi}, K. 1986, \aap, 158, 17

\bibitem[{{Tonry} {et~al.}(2001){Tonry}, {Dressler}, {Blakeslee}, {Ajhar}, {Fletcher}, {Luppino}, {Metzger}, \& {Moore}}]{Tonry_2001}
{Tonry}, J.~L., {Dressler}, A., {Blakeslee}, J.~P., {et~al.} 2001, \apj, 546, 681, \dodoi{10.1086/318301}

\bibitem[{{Tremmel} {et~al.}(2015){Tremmel}, {Governato}, {Volonteri}, \& {Quinn}}]{tremmel_2015}
{Tremmel}, M., {Governato}, F., {Volonteri}, M., \& {Quinn}, T.~R. 2015, \mnras, 451, 1868, \dodoi{10.1093/mnras/stv1060}

\bibitem[{{Tremmel} {et~al.}(2017){Tremmel}, {Karcher}, {Governato}, {Volonteri}, {Quinn}, {Pontzen}, {Anderson}, \& {Bellovary}}]{tremmel_2017}
{Tremmel}, M., {Karcher}, M., {Governato}, F., {et~al.} 2017, \mnras, 470, 1121, \dodoi{10.1093/mnras/stx1160}

\bibitem[{{Tremmel} {et~al.}(2019){Tremmel}, {Quinn}, {Ricarte}, {Babul}, {Chadayammuri}, {Natarajan}, {Nagai}, {Pontzen}, \& {Volonteri}}]{Tremmel_2019}
{Tremmel}, M., {Quinn}, T.~R., {Ricarte}, A., {et~al.} 2019, \mnras, 483, 3336, \dodoi{10.1093/mnras/sty3336}

\bibitem[{{Tumlinson} {et~al.}(2017){Tumlinson}, {Peeples}, \& {Werk}}]{tumlinson_2017}
{Tumlinson}, J., {Peeples}, M.~S., \& {Werk}, J.~K. 2017, \araa, 55, 389, \dodoi{10.1146/annurev-astro-091916-055240}

\bibitem[{{Tumlinson} {et~al.}(2011){Tumlinson}, {Thom}, {Werk}, {Prochaska}, {Tripp}, {Weinberg}, {Peeples}, {O'Meara}, {Oppenheimer}, {Meiring}, {Katz}, {Dav{\'e}}, {Ford}, \& {Sembach}}]{Tumlinson11}
{Tumlinson}, J., {Thom}, C., {Werk}, J.~K., {et~al.} 2011, Science, 334, 948, \dodoi{10.1126/science.1209840}

\bibitem[{{van den Bosch}(2016)}]{van_den_Bosch_2016}
{van den Bosch}, R. C.~E. 2016, \apj, 831, 134, \dodoi{10.3847/0004-637X/831/2/134}

\bibitem[{Virtanen {et~al.}(2020)Virtanen, Gommers, Oliphant, Haberland, Reddy, Cournapeau, Burovski, Peterson, Weckesser, Bright, {van der Walt}, Brett, Wilson, Millman, Mayorov, Nelson, Jones, Kern, Larson, Carey, Polat, Feng, Moore, {VanderPlas}, Laxalde, Perktold, Cimrman, Henriksen, Quintero, Harris, Archibald, Ribeiro, Pedregosa, {van Mulbregt}, \& {SciPy 1.0 Contributors}}]{2020SciPy-NMeth}
Virtanen, P., Gommers, R., Oliphant, T.~E., {et~al.} 2020, Nature Methods, 17, 261, \dodoi{10.1038/s41592-019-0686-2}

\bibitem[{{Voit} {et~al.}(2023){Voit}, {Oppenheimer}, {Bell}, {Terrazas}, \& {Donahue}}]{voit2023}
{Voit}, G.~M., {Oppenheimer}, B.~D., {Bell}, E.~F., {Terrazas}, B., \& {Donahue}, M. 2023, arXiv e-prints, arXiv:2309.14818, \dodoi{10.48550/arXiv.2309.14818}

\bibitem[{{Wadsley} {et~al.}(2017){Wadsley}, {Keller}, \& {Quinn}}]{wadsley17}
{Wadsley}, J.~W., {Keller}, B.~W., \& {Quinn}, T.~R. 2017, \mnras, 471, 2357, \dodoi{10.1093/mnras/stx1643}

\bibitem[{{Wadsley} {et~al.}(2004){Wadsley}, {Stadel}, \& {Quinn}}]{wadsley04}
{Wadsley}, J.~W., {Stadel}, J., \& {Quinn}, T. 2004, \na, 9, 137, \dodoi{10.1016/j.newast.2003.08.004}

\bibitem[{{Weinberger} {et~al.}(2017){Weinberger}, {Springel}, {Hernquist}, {Pillepich}, {Marinacci}, {Pakmor}, {Nelson}, {Genel}, {Vogelsberger}, {Naiman}, \& {Torrey}}]{weinberger17}
{Weinberger}, R., {Springel}, V., {Hernquist}, L., {et~al.} 2017, \mnras, 465, 3291, \dodoi{10.1093/mnras/stw2944}

\bibitem[{{Werk} {et~al.}(2012){Werk}, {Prochaska}, {Thom}, {Tumlinson}, {Tripp}, {O'Meara}, \& {Meiring}}]{Werk_2012}
{Werk}, J.~K., {Prochaska}, J.~X., {Thom}, C., {et~al.} 2012, \apjs, 198, 3, \dodoi{10.1088/0067-0049/198/1/3}

\bibitem[{{Werk} {et~al.}(2013){Werk}, {Prochaska}, {Thom}, {Tumlinson}, {Tripp}, {O'Meara}, \& {Peeples}}]{werk_2013}
---. 2013, \apjs, 204, 17, \dodoi{10.1088/0067-0049/204/2/17}

\bibitem[{{Werk} {et~al.}(2014){Werk}, {Prochaska}, {Tumlinson}, {Peeples}, {Tripp}, {Fox}, {Lehner}, {Thom}, {O'Meara}, {Ford}, {Bordoloi}, {Katz}, {Tejos}, {Oppenheimer}, {Dav{\'e}}, \& {Weinberg}}]{Werk_2014}
{Werk}, J.~K., {Prochaska}, J.~X., {Tumlinson}, J., {et~al.} 2014, \apj, 792, 8, \dodoi{10.1088/0004-637X/792/1/8}

\bibitem[{{Wiersma} {et~al.}(2009{\natexlab{a}}){Wiersma}, {Schaye}, \& {Smith}}]{Wiesrsma_Schaye_Smith_2009a}
{Wiersma}, R. P.~C., {Schaye}, J., \& {Smith}, B.~D. 2009{\natexlab{a}}, \mnras, 393, 99, \dodoi{10.1111/j.1365-2966.2008.14191.x}

\bibitem[{{Wiersma} {et~al.}(2009{\natexlab{b}}){Wiersma}, {Schaye}, {Theuns}, {Dalla Vecchia}, \& {Tornatore}}]{Wiersma_2009b}
{Wiersma}, R. P.~C., {Schaye}, J., {Theuns}, T., {Dalla Vecchia}, C., \& {Tornatore}, L. 2009{\natexlab{b}}, \mnras, 399, 574, \dodoi{10.1111/j.1365-2966.2009.15331.x}

\bibitem[{{Williams} {et~al.}(2017){Williams}, {Dolphin}, {Dalcanton}, {Weisz}, {Bell}, {Lewis}, {Rosenfield}, {Choi}, {Skillman}, \& {Monachesi}}]{Williams_2017}
{Williams}, B.~F., {Dolphin}, A.~E., {Dalcanton}, J.~J., {et~al.} 2017, \apj, 846, 145, \dodoi{10.3847/1538-4357/aa862a}

\bibitem[{{Woosley} \& {Weaver}(1995)}]{woosley&weaver1995}
{Woosley}, S.~E., \& {Weaver}, T.~A. 1995, \apjs, 101, 181, \dodoi{10.1086/192237}

\bibitem[{{Yoon} {et~al.}(2012){Yoon}, {Putman}, {Thom}, {Chen}, \& {Bryan}}]{yoon_2012}
{Yoon}, J.~H., {Putman}, M.~E., {Thom}, C., {Chen}, H.-W., \& {Bryan}, G.~L. 2012, \apj, 754, 84, \dodoi{10.1088/0004-637X/754/2/84}

\bibitem[{{Zahedy} {et~al.}(2019){Zahedy}, {Chen}, {Johnson}, {Pierce}, {Rauch}, {Huang}, {Weiner}, \& {Gauthier}}]{Zahedy_2019}
{Zahedy}, F.~S., {Chen}, H.-W., {Johnson}, S.~D., {et~al.} 2019, \mnras, 484, 2257, \dodoi{10.1093/mnras/sty3482}

\end{thebibliography}
\bibliographystyle{aasjournal}

%% This command is needed to show the entire author+affiliation list when
%% the collaboration and author truncation commands are used.  It has to
%% go at the end of the manuscript.
%\allauthors

%% Include this line if you are using the \added, \replaced, \deleted
%% commands to see a summary list of all changes at the end of the article.
%\listofchanges

\end{document}